\documentclass[a4paper,10pt]{article}
\pdfoutput=1 
\usepackage{jheppub}
\usepackage{mathrsfs}
\usepackage{amsmath}
\usepackage{amssymb}
\usepackage{amsfonts}
\usepackage[T1]{fontenc} % if needed
\usepackage{breqn}
\usepackage{subcaption}

\title{\boldmath Ensemble Dependent Holographic Phase Transitions in 4$D$ Dyonic AdS Black Holes}

\author[1,2]{ Abhishek Baruah }
\author[1,3]{ Prabwal Phukon}
 \affiliation[1]{Department Of Physics, Dibrugarh University, Dibrugarh,786004, Assam, India}
 \affiliation[2]{Department of Physics, Patkai Christian College, Chum$\ddot{o}$ukedima, 797013, Nagaland, India}
\affiliation[3]{Theoretical Physics Division, Centre for Atmospheric Studies, Dibrugarh University, Dibrugarh, 786004, Assam, India}

\emailAdd{rs{\_}abhishekbaruah@dibru.ac.in}
\emailAdd{prabwal@dibru.ac.in}

\abstract{This work studies the holographic thermodynamics of $4D$ dyonic Anti--de Sitter black holes within the AdS/CFT correspondence. By allowing variations of the cosmological constant $\Lambda$ and Newton's constant $G_N$, the gravitational thermodynamics is extended to include the CFT central charge $C$ and its conjugate chemical potential $\mu$, together with the standard thermodynamic pair $(T,S)$. A further conjugate pair $(p,\mathcal{V})$, interpreted as the CFT pressure and volume, is also introduced. The presence of both electric and magnetic charges, described by $(\tilde{\Phi}_e,\tilde{Q}_e)$ and $(\tilde{\Phi}_m,\tilde{Q}_m)$, leads to a significantly enriched phase structure. A systematic analysis of all sixteen thermodynamic ensembles reveals a strong dependence of phase behavior on ensemble choice. In the ensembles $(\tilde{Q}_m,\tilde{\Phi}_e,\mathcal{V},C)$ and $(\tilde{Q}_e,\tilde{\Phi}_m,\mathcal{V},C)$, the system exhibits Van der Waals transitions, superfluid $\lambda$ transitions, and Davies criticality, while the ensemble $(\tilde{Q}_m,\tilde{Q}_e,\mathcal{V},C)$ supports similar transitions except for the Davies transitions. When the chemical potential $\mu$ is fixed, only Davies--type transitions occur. Ensembles involving the CFT pressure $p$ show no critical behavior, although Davies transitions persist. Notably, the ensemble $(\tilde{\Phi}_m,\tilde{\Phi}_e,\mathcal{V},C)$ uniquely displays a confined/deconfined transition alongside Davies criticality. These results highlight the crucial role of dyonic charges and CFT variables in shaping holographic phase transitions.}

\begin{document} 
\maketitle
\flushbottom
\section{Introduction}
\label{sec:intro}
In the cosmic fabric of spacetime, where light bends and time warps, black holes are the cosmic enigmas and a source of gravitational singularity derived from general relativity. Black holes are being considered to mimic black bodies once the quantum effects are considered since they are taken as nature's ideal absorbers. For approximately half a century, thermodynamics of black holes has been the centre of extensive study. There is two main phases to thermodynamics of black hole. Bekenstein, Bardeen,  Carter, Hawking and others pioneered the first and initial stage \cite{a, aa1,b,bb1}.
 Conventional black hole thermodynamics (TBHT) is the formalism or concept developed at this phase. Black hole mass is regarded as the internal energy, and black hole temperature is determined by its angular momentum, mass, charge, in addition to other physical setup-related parameters. Contrary to a fluid like system, where entropy corresponds to volume, conjugate thermodynamic entropy in Einstein's gravity is directly proportional to the A, that is the horizon area. We set $\hbar=c=k_B=1$. We consider the Hawking temperature $T$ and Bekenstein-Hawking entropy $S$ of dyonic AdS black holes is given as \cite{0a,0b,0c,0d,0e}:-
\begin{equation}
\label{eq:T:S}
T=\frac{\kappa}{2 \pi},\qquad S=\frac{A}{4 G_N}
\end{equation}
where in \eqref{eq:T:S},  the surface gravity is $\kappa$ and $G_N$ is the Newton constant. One class of black holes that created immense expansion in the study of  thermodynamics of black holes is the asymptotically AdS (Anti-de-Sitter) black holes. Based on their ensembles and types (hairy, rotating, charged, uncharged, both charged and rotating) they are characterized by rich phase structures. Currently, there are other ways of studying the rich phase structures \cite{1a,1b,1c}, using certain tools like the thermodynamic geometry \cite{2a,2b,2c,2d,2e,2f,2g,2h} and the thermodynamic topology \cite{3a,3b,3c,3d,3e,3f,3g,3h,3i,3j,3k,3l,3m,3n}. Asymptotically, the  AdS (Anti-de-Sitter) black holes also known as holographic theory, which are analogous to thermal boundary states in the dual (CFT) conformal field theory \cite{c} greatly enhances the thermodynamics of black holes. The breakthrough of the Hawking-Page phase transition \cite{d} for the Anti-de-Sitter (AdS) black holes is another significant advance in this direction. This is a alteration of phase from a state of black hole to a completely  state of thermal gas, distinguished by diminishing Gibbs free energy.\\
The second phase is the (EPST) Extended Phase Space Thermodynamics \cite{y, yy,yy1,x,i,j1,yy2,yy3,yy4}. The formalism was introduced by scientists namely Kastor, Traschen and Ray and then it was further developed by more researchers \cite{p}. The main advancement at this stage is the addition of a new set of variables, like  the negative cosmological constant $\Lambda$ (which is proportional to thermodynamic pressure $P$), and its complimentary variable, the thermodynamic volume $V$. Interpreting mass parameter as the enthalpy which is the price paid for this modification in this formalism.
Asymptotically, the charged AdS (Anti-de-Sitter) black holes can achieve steady state (thermal equilibrium) with their Hawking radiation and is obtained to display a variety of intriguing behaviours in the phase, such as transition to radiation which is a first-order  that corresponds to the confinement or de confinement of the dual quark-gluon plasma \cite{e} and many more \cite{f,g,h,j,o,m,n,k,l}. AdS black holes can also be considered as heat engines thanks to the presence of the $(P, V)$ variables \cite{b1,bc1}. All these phenomena listed above have been found out via the concept of the thermodynamics of the extended phase space, where  pressure $P$ is identified with  $\Lambda$ i.e  the cosmological constant and is given as:-
\begin{equation}
P=-\frac{\Lambda}{8 \pi G_N}, \quad \Lambda=-\frac{(d-1)(d-2)}{2L^2}
\end{equation}

Visser \cite{u} added a new pair of complimentary thermodynamic variables to the research of Extended Phase Space Thermodynamics by taking the chemical potential $\mu$ and the central charge $C$.  This extension of the fundamental law of the thermodynamics of black holes, the investigation of the related criticality on central charge $C$  has recently come into existence \cite{v,vv}.\\
In the backdrop of the ADS/CFT correspondence, in the paper \cite{k2} they have examined the holographic counterpart of the extended thermodynamics of spherically symmetric, charged Anti-de-Sitter black holes by allowing variations of the Newton's constant $G_N$ and the cosmological constant $\Lambda$. Also the chemical potential $\mu$ and central charge $C$ are included in the dual CFT. In paper \cite{new,new1}, they have done the same analysis but with a rotating AdS black hole.
AdS black holes exhibit enriched phase behaviour because of the inclusion of a cosmological constant $\Lambda$ that is negative. There are some novel phase transitions are acquired by addressing the cosmological constant $\Lambda$ as an extra variable for the black holes.
Since Newton's constant $G_N$ can be considered as a coupling constant that can alter in the gravitational theories space and, fluctuates along the renormalisation group flow if quantum revisions are taken into account, it has recently been listed to the extended thermodynamic phase space (EPST) as a parameter that can be altered \cite{q,r,s,t}. Within the idea of AdS/CFT correspondence based on Visser's idea, a nascent formalism was given by Ghao and Zhao popularised as the Restricted Phase Space (RPS) Thermodynamics \cite{rps}. This formalism has the freedom from pressure $P$ and volume $V$ and interprets the mass as the internal energy. Several literature review have been done using this concept or formalism \cite{rps1,rps2,rps3,rps4,rps5}. \\
The basic thermodynamical characteristics of charged (electrically) black holes at arbitrary diameters are widely understood. However, due to the electromagnetic duality, a black hole can be created in the four space-time dimensions that can transport electric and magnetic charges. This sort of black holes are categorised as a dyonic black hole. In the literature \cite{l2} they explored a dual charged black hole in asymptotically Anti-de-Sitter spacetime.
A dyonic (dual-charge) black hole consists of the U(1) gauge and the graviton (metric) fields. This metric fulfils the Anti-de-Sitter boundary, and then the 2 alternative boundary conditions for the gauge field, the first is a mode that can be normalized and is associated with the vacuum expectation value (VEV) of the corresponding dual operator, and the second is a mode that cannot be normalized corresponding to the use of an  gauge field which is external which alters the theory at the boundary. The existence of a magnetic monopole adds complexity to the phase diagram of the Anti-de-Sitter spacetime black holes. Deformed dimensional field theories have recently received considerable attention in the presence of a holographic interpretation of certain phenomena such as superconductivity/superfluidity \cite{aa}, the Hall effect \cite{ab}, magnetohydrodynamics \cite{ac,ac1,ac2}, the Nernst effect \cite{ad} and many more \cite{z,a1,d1,e1,f1,c1}.\\
Black hole thermodynamics is highly influenced by the choice of ensemble. As we have seen in the case of CFT thermodynamics of charged AdS black holes, out of the eight possible ensembles in only three ensembles phase transitions are observed. To understand the intricate relationship between the nature of ensemble and phase transitions in holographic thermodynamics, it is pertinent that one explores black holes with more and diverse set of thermodynamic ensembles. The dual CFT corresponding to 4-D charged dyonic AdS black hole is enriched with 16 different ensembles. The presence of ensembles in addition to the ones available in the case of dual CFT of charged AdS black holes makes it a perfect choice to investigate the dependence of CFT phase structure on the nature of ensembles. 
The introduction of both the electric and magnetic charges together may introduce some new interactions which might not be present in purely electrical-charged systems. These combinations can change the thermodynamic potentials and their corresponding functions which leads to a combination of fixed for varying charges and potential both electric and magnetic. This can be referred to as mixed ensembles. A dual charged black hole is capable to show richer phase structure such as criticality depending on both the charges.\\

The primary motivation of this work is to explore phase transitions and critical phenomena within the holographic counterpart of thermodynamics in the extended phase space of dyonic Anti-de Sitter black holes. This framework reveals novel thermodynamic ensembles that do not appear in conventional black hole thermodynamics of dyonic AdS black holes.There are a total of 16 potential thermodynamic (grand) canonical and mixed ensembles in the CFT for the four conjugate quantities $(\tilde{\Phi}_e, \tilde{Q}_e)$, $(\tilde{\Phi}_m, \tilde{Q}_m)$, $(p, \mathcal{V})$ and $(\mu, C)$. Section \ref{sub:Thermodynamic ensemble}, we discover that seven of the ensembles— the ensembles at fixed $(\tilde{\Phi}_e, \tilde{Q}_m, \mathcal{V}, C)$, $(\tilde{\Phi}_e, \tilde{Q}_m, \mathcal{V}, \mu)$, $(\tilde{\Phi}_m, \tilde{Q}_e, \mathcal{V}, C)$, $(\tilde{\Phi}_m, \tilde{Q}_e, \mathcal{V}, \mu)$, $(\tilde{\Phi}_e, \tilde{\Phi}_m, \mathcal{V}, C)$, $(\tilde{Q}_e, \tilde{Q}_m, \mathcal{V}, C)$, $(\tilde{Q}_e, \tilde{Q}_m, \mathcal{V}, \mu)$ ensembles exhibits fascinating phase behavior or critical events. We evaluate the pertinent phase diagrams and graph the corresponding free energy, specific heat and coexistence plots in these ensembles with respect to temperature. In Section \ref{sec:other}, we plot the remaining free energy plots for the $p$ containing ensembles. In Section \ref{sec:Discussions} we look into conclusions and discussions.
\section{Holographic thermodynamics of dyonic  Anti-de-Sitter black holes}
\label{sec:Holographic thermodynamic}
The bulk thermodynamics of AdS black holes and the complimentary CFT are related by the AdS/CFT correspondence. For the extended thermodynamics of dual charge Anti-de-Sitter black holes the holographic dictionary is described and reviewed in this section. All charts are made for $d\rightarrow 3$, or the $AdS_4/CFT_3$ relationship because it gets too complicated to generalize in $d$-dimension.
\subsection{Extended phase space thermodynamics of Dyonic Anti-de-Sitter black holes}
\label{subsec:Extended thermodynamic}
The Reissner-Nordstr\"om action in the existence of a negative cosmological constant $\Lambda$ are given as:-
\begin{equation}
\label{eq:I:1}
I=\frac{1}{16 \pi G_4}\int d^4 x\sqrt{g}(-R+F^2-\frac{6}{L^2})
\end{equation}
The equations of motion are:-
\begin{equation}
\label{eq:EOM:1}
 R_{\mu \nu}-\frac{1}{2}g_{\mu\nu}R-\frac{3}{L^2}g_{\mu \nu}=2(F_{\mu \lambda}F_\nu^\lambda-\frac{1}{4}g_{\mu \nu} F_{\alpha \beta} F^{\alpha \beta}), \quad
 \nabla_\mu F^{\mu \nu}=0
\end{equation}
These equations have a static, spherically symmetric solution provided by:-
\begin{equation}
\label{eq:A}
A=\left(-\frac{q_e}{r}+\frac{q_e}{r_+}\right) dt+(q_m \cos\theta) d\phi
\end{equation}
and 
\begin{equation}
\label{eq:line:element}
ds^2=-f(r)dt^2+\frac{1}{f(r)}dr^2+r^2 d\theta^2+r^2\sin^2 \theta d\phi^2
\end{equation}
where the electromagnetic four-potential is $A$ and $f(r)$ is given as:- 
\begin{equation}
\label{eq:fr}
f(r_+)=\left(1+\frac{r_+^2}{L^2}-\frac{m}{r_+}+\frac{q^2_e+q^2_m}{r_+^2}\right)
\end{equation}
where in \eqref{eq:fr}, $q_e$ is the electric charge parameter, $q_m$ is the magnetic charge parameter, and $m$ is and ADM mass parameter of the blackhole. For $d=3$
The ADM mass and Bekenstein-Hawking entropy of the blackhole as:-
\begin{equation}
\label{eq:M:3}
M=\frac{(d-1)\Omega_{d-1}}{16 \pi G_N}m=\frac{m}{2 G_N}, \quad S=\frac{\Omega_{d-1} r_+^{d-1}}{4 G_N}=\frac{\pi r^2}{G_N}
\end{equation}
 The electric charge parameter $q_e$ and magnetic charge parameter $q_m$ can be connected to the $Q_e$ electric charge and $Q_m$ magnetic charge as:-
 \begin{equation}
 \label{eq:Qe:1}
 Q_e=\frac{(d-1)\Omega_{d-1}}{8 \pi G_N} \alpha q_e=\frac{q_e}{G_N}, \quad Q_m=\frac{(d-1)\Omega_{d-1}}{8 \pi G_N} \alpha q_m =\frac{q_m}{G_N}
 \end{equation}
 For $d=3$ and $\alpha=\sqrt{\frac{2(d-2)}{d-1}}=1$. We operate in an ensemble where $A_t$ has a constant asymptotic value, hence the potentials are:-
 \begin{equation}
 \label{:eq:Phie:Phim}
 \Phi_e=\frac{q_e}{r_+}, \quad \text{and} \quad \Phi_m=\frac{q_m}{r_+}
 \end{equation}
 With this decision, $\Phi_e$ and $\Phi_m$ stand for any hypothetical distinction between infinity and the outside horizon. The generalized Smarr formula relates the black hole parameters to one another as:-
 \begin{equation}
 \label{eq:M:6}
 M=\frac{d-1}{d-2} \frac{\kappa A}{8 \pi G_N}+\Phi_e Q_e+\Phi_m Q_m-\frac{1}{d-2}\frac{\Theta \Lambda}{4 \pi G_N}
 \end{equation}

Putting $f(r_+)=0$ in \eqref{eq:fr} we can calculate the mass as:-
\begin{equation}
\label{eq:fr:1}
 m= \frac{\frac{r_+^4}{L^2}+q^2_e+q^2_m+r_+^2}{r_+}
\end{equation}
We do certain calculations for 4 different kinds of variable combinations which will be modified later.
\subsubsection{For $\Phi_e$ and $Q_m$}
\label{subsubsec:phie:qm}
Here the quantities $\Phi_e$ and $Q_m$ are fixed. Hence $\Phi_e=\frac{q_e}{r_+}$ and
 Hawking temperature of the dyonic AdS black hole is:-
 \begin{equation}
 \label{eq:T:1}
 \begin{split}
  & T_1=\frac{\kappa}{2 \pi}=\frac{1}{4 \pi r_+}\left(1+\frac{3 r_+^2}{L^2}-\Phi_e^2-\frac{q_m^2}{r_+^2}\right) 
 \end{split}
 \end{equation} 
\subsubsection{For $\Phi_m$ and $Q_e$}
\label{subsubsec:phim:qe}
Here the quantities $\Phi_m$ and $Q_e$ as fixed. Hence $\Phi_m=\frac{q_m}{r_{+}}$. Temperature is given as:-
 \begin{equation}
 \label{eq:T:2}
   T_2=\frac{\kappa}{2\pi}= \frac{1}{4 \pi r_+}\left(1+\frac{3 r_+^2}{L^2}-\Phi_m^2-\frac{q_e^2}{r_+^2}\right)
 \end{equation} 
 \subsubsection{For $\Phi_e$ and $\Phi_m$}
 \label{subsubsec:phie:phim}
Here the quantities $\Phi_m$ and $\Phi_e$ as fixed. Hence $\Phi_e=\frac{q_e}{r_{+}}$, $\Phi_m=\frac{q_m}{r_+}$.
The temperature is given as:-
 \begin{equation}
 \label{eq:T:3}
 T_3=\frac{\kappa}{2 \pi}=\frac{1}{4 \pi r_+}\left[1+\frac{3 r_+^2}{L^2}-\Phi_e^2-\Phi_m^2\right]
 \end{equation} 
 \subsubsection{For $Q_e$ and $Q_m$}
 \label{subsubsec:qe:qm}
Here $Q_m$ and $Q_e$ are fixed quantities. 
Hence the temperature is given as:-
 \begin{equation}
 \label{eq:T:4}
   T_4=\frac{\kappa}{2 \pi}\frac{1}{4 \pi r_+}\left(1+\frac{3 r_+^2}{L^2}-\frac{q_e^2}{r_+^2}-\frac{q_m^2}{r_+^2}\right)
 \end{equation} 
 
 By relating to a scaling argument in \cite{h1,i1}, first law mechanics for the charged dyonic AdS black hole can be stretched to the generalized Smarr formula. The $L$ which is the AdS radius is being scaled with the black hole parameters as $M\sim L^{d-2}, A \sim L^{d-1}, Q \sim L^{d-2}$ and also $\Lambda \sim L^{-2}$, similarly with Newton's constant of gravitation, we get the scaling as $M, Q \sim G_N^{-1}$.
First law of gravitation for the  charged dyonic Anti-de-Sitter black holes stretches to incorporate alterations of the theoretical variables of the $G_N$ gravitational constant and $\Lambda$ cosmological constant are:-
 \begin{equation}
 \label{eq:dM:5}
dM=\frac{\kappa}{8 \pi G_N}dA+\Phi_e dQ_e+\Phi_m dQ_m+\frac{\Theta}{8 \pi G_N} d\Lambda-(M-\Phi_e Q_e-\Phi_m Q_m) \frac{d G_N}{G_N} 
 \end{equation}
 The integration of the $\Lambda$ cosmological constant variation for first rule has been extensively examined in the literature, as was indicated in the introduction. The fluctuation of Newton's constant has previously been discussed in the Anti-de-Sitter black holes extended fundamental and the fundamental law of holographic entanglement entropy in \cite{k1,l1}.
 \subsection{Thermodynamics of CFT }
 \label{subsec:CFT thermodynamic}
 The Smarr formula \eqref{eq:M:6} similar for Anti-de-Sitter black holes was demonstrated in \cite{m1}  that has a $p\mathcal{V}$ term and lacking $\mu C$ term. The dual Euler equation for charged dyonic AdS black holes is:-

 \begin{equation}
 \label{eq:E:1}
 E=TS+\tilde{\Phi}_e \tilde{Q}_e+\tilde{\Phi}_m \tilde{Q}_m+\mu C 
 \end{equation}
Here $\mu$ is chemical potential conjugate to $C$ which is the central charge and $\tilde{\Phi}_e$, is the CFT electric potential and $\tilde{Q}_e$ is the CFT electric charge and $\tilde{\Phi}_m$ is the CFT magnetic potential and, $\tilde{Q}_m$ is the CFT magnetic charge. The extended first law in \eqref{eq:dM:5}, which incorporates both a  $pd \mathcal{V}$ and $\mu dC$ terms , was aligned with the expanded version of the CFT thermodynamics first law as:-
\begin{equation}
\label{eq:dE:2}
dE=TdS+\tilde{\Phi}_e d \tilde{Q}_e+\tilde{\Phi}_m d \tilde{Q}_m-p d\mathcal{V}+\mu dC
\end{equation} 
 where in \eqref{eq:dE:2},  $\mathcal{V}$ is the volume of the CFT and $p$ is the field theory pressure.
In AdS/CFT correspondence, the bulk theory's gravitational coupling constants and the AdS radius are related to the CFT's central charge. According to this dictionary, Einstein's gravity reads as:-
 \begin{equation}
 \label{eq:C:1}
 C=\frac{\Omega_{d-1} L^{d-1}}{16 \pi G_N}
 \end{equation}
 The volume of space $\mathcal{V}$ of the geometry where the CFT is placed is another CFT characteristic that is mentioned in the thermodynamic first law. The CFT is frequently placed on a sphere of $L$ AdS radius, resulting in a volume of $\mathcal{V} = \Omega_{d-1} L^{d-1}$. We make a distinction between the CFT's volume of space $\mathcal{V}$ and its $C$ central charge. Consequently, we choose to make the bulk $L$ curvature radius different from the boundary $R$ curvature radius, and the volume is written as:-
 \begin{equation}
 \label{eq:V:1}
 \mathcal{V}=\Omega_{d-1} R^{d-1}
 \end{equation}
This sphere volume of radius $R$ in the  AdS bulk which is in $(d - 1)$ dimensions and $\mathbf{R} \times S^{d-1}$, which relates to the 'area' in the CFT boundary which is in $(d + 1)$ -dimensions.
The dual field theory in AdS/CFT is located on the asymptotically AdS spacetime's conformal field theory boundary.
  By using the Gubser-Klebanov-Polyakov-Witten (GKPW) prescription, the AdS metric $g_{\text{AdS}}$ is related to the  CFT metric $g_{\text{CFT}}$  when taken or utilised on the asymptotic boundary edge tending to a Weyl rescalling \cite{s1,t1} as:-
 \begin{equation}
 \label{eq:gcft}
 g_{CFT}=\lim_{r \to \infty}(\lambda(x) g_{AdS})
\end{equation}  
where $\lambda(x)$ is a freely chosen Weyl rescale factor and $r\to \infty$ denotes spatial infinity. As $r\to \infty$, the asymptotically AdS spacetime line element approaches as:- 
 \begin{equation}
 \label{eq:line:element:2}
 ds^2=-\frac{r^2}{L^2} dt^2 +\frac{L^2}{r^2} dr^2+r^2 d\Omega_{d-1}^2
 \end{equation}
  The model choice of the Weyl rescale factor $\lambda=L/r$ the line element of the edge CFT is $ds^2=-dt^2+L^2 d\Omega^2$. 
 In this instance, the curvature radius of the boundary and the bulk curvature radius are the same. Instead, we'll assume that the Weyl rescale factor is $\lambda= R/r$, which places the CFT on a sphere with $R$ as constant radius. As a result, line element of the CFT's becomes with d=3 as:-
 \begin{equation}
 \label{eq:line:element:3}
 ds^2=-\frac{R^2}{L^2}dt^2+R^2 d \Omega_2^2
 \end{equation}
 For the $S$ entropy, $E$ energy, $T$ temperature, $\tilde{\Phi}_m$ magnetic potential, $\tilde{Q}_m$ magnetic charge, $\tilde{\Phi} _e$ electric potential, $\tilde{Q}_e$ electric charge,  the holographic dictionary can be written for the choice of the above CFT metric \cite{u}:-
 \begin{equation}
 \label{eq:holodic:2}
 \begin{split}
& S=\frac{A}{4 G_N},\quad E=M\frac{L}{R},\quad T=\frac{\kappa}{2 \pi} \frac{L}{R},\quad \tilde{\Phi}_e=\frac{\Phi_e}{L} \frac{L}{R},\quad 
  \tilde{Q}_e=Q_e L,\quad \tilde{\Phi}_m=\frac{\Phi_m}{L} \frac{L}{R},\quad \tilde{Q}_m=Q_m L
 \end{split}
 \end{equation}
 Due to a factor $R/L$ difference between the bulk and the boundary CFT time in \eqref{eq:holodic:2}, the  rescale factor $L/R$ occurs in the dictionaries for the temperature, energy, and electric and magnetic potential \cite{u1}. As a result, the dictionaries for energy, temperature, and potential are different from the holographic dictionaries by a factor of $L/R$.
 The vacuum state energy vanishes in the holographic dictionary above. The counterterm subtraction approach \cite{v1,w1}, is used used to calculate the energy of the vacuum for a CFT. $E=(M + E_0)L/R$ is the proper dictionary for the rescaled energy that results via this approach, where the energy of the vacuum $E_0 \sim C$ corresponds to the $C$ central charge. Since the Casimir energy has little bearing on (holographic) thermodynamics, hence disregarded.\\
 The generalized Smarr formula \eqref{eq:M:6}  replaces $\Theta$ which is the quantity conjugate to $\Lambda$ in the first law extension \eqref{eq:dM:5} and plug in $d\Lambda/\Lambda =- 2dL/L$, which is  vital in the matching of the bulk and edge/ boundary first laws \cite{u}. Then we can write the extended first law as:-
 \begin{equation}
 \label{eq:dM:6}
 \begin{split}
& d\left(M\frac{L}{R}\right)=\frac{\kappa}{2 \pi}\frac{L}{R} d\left(\frac{A}{4 G_N}\right)+\frac{\Phi_e}{R} d(Q_e L)+\frac{\Phi_m}{R} d(Q_m L)-\frac{M}{2}\frac{L}{R}\frac{dR^2}{R^2}\\
&  +\left(M\frac{L}{R}-\frac{\kappa A}{8\pi G_N}\frac{L}{R}-\frac{\Phi_e}{R} Q_e L-\frac{\Phi_e}{R} Q_e L\right)\frac{d(L^2/G_N)}{L^2/G_N}
\end{split}
\end{equation}  
The appearance of $M$ on both sides of the extended first law equation specifically in the context of generalized Smarr formula and first law extensions for black holes in AdS spacetimes, is fundamentally linked to the inclusion of additional terms that account for the variation of thermodynamic-like variables beyond the standard mass, and charges This particular formulation takes into account the effects of varying cosmological constant $\Lambda$ or related parameters such as the AdS length scale $L$, which directly interacts with the mass $M$ of the black hole.\\
 Dyonic AdS black holes extended first law is then implied by the holographic dictionary \eqref{eq:C:1}, \eqref{eq:V:1}, and \eqref{eq:holodic:2} corresponds to the following thermodynamic first law in the CFT as:-
\begin{equation}
\label{eq:dE:3}
dE=TdS+\tilde{\Phi}_e d\tilde{Q}_e+\tilde{\Phi}_m d\tilde{Q}_m-pd\mathcal{V}+\mu dC
\end{equation} 
 The chemical potential $\mu$ linked with the $C$ central charge and the CFT field pressure $p$ are found to be set via a detail comparison of the bulk and boundary first laws, which are given in equations \eqref{eq:dM:6} and \eqref{eq:dE:3}, respectively as:-
 \begin{equation}
 \label{eq:mu:p:1}
 \mu=\frac{1}{C}(E-TS-\tilde{\Phi}_e \tilde{Q}_e-\tilde{\Phi}_m \tilde{Q}_m), \qquad p=\frac{1}{2}\frac{E}{\mathcal{V}}
 \end{equation}
 The first equation is the Euler equation given in \eqref{eq:E:1} with constant magnetic and electric charges and the later equation is the CFT equation of state (EOS).  The first equation also implies that the $C$  (or $N^2$) is proportional to the free energy in the grand canonical ensemble i.e. $F_{11}=E-TS-\tilde{\Phi}_e \tilde{Q}_e-\tilde{\Phi}_m \tilde{Q}_m=\mu C$.\\
 At last, we provide clear expressions for the thermodynamic variables of charged dyonic black holes dual to CFT states. We introduce certain dimensional variables:-
 \begin{equation}
 \label{eq:dim:para}
 x=\frac{r_+}{L}, \quad y=\frac{q_e}{L},\quad y=\Phi_e \quad z=\frac{q_m}{L}, \quad z=\Phi_m
 \end{equation}
using the holographic dictionary, the entropy \eqref{eq:M:3}, electric and magnetic charge \eqref{eq:Qe:1} and \eqref{:eq:Phie:Phim} for the electric and  magnetic potential can be related to the variables in the CFT \eqref{eq:holodic:2} to get:-
 \begin{equation}
 \label{eq:holodic:3}
 S=4 \pi  C x^2,\quad \tilde{Q}_e=4 C y,\quad \tilde{\Phi}_e=\frac{y}{R x},\quad \tilde{Q}_m=4 C z,\quad \tilde{\Phi}_m=\frac{z}{R x}
 \end{equation}
 where  $C$ is  central charge and $R$ is boundary radius.

 \subsection{ Energy ($E$), temperature ($T$) and chemical potential ($\mu$) for $(\tilde{\Phi}_e,\tilde{Q}_m)$}
 \label{subsec:ADM1}
 Using \eqref{eq:M:3}, \eqref{eq:Qe:1}, \eqref{eq:fr:1} the ADM mass $M$, the Hawking's temperature and the chemical potential $\mu$ using \eqref{eq:T:1} can be related by using \eqref{eq:holodic:2}, \eqref{eq:holodic:3} to the CFT boundary energy $E$ and temperature $T$ as:-
 \begin{equation}
 \label{eq:E1:T1}
 E=\frac{2 c \left(x^2 \left(y^2+1\right)+x^4+z^2\right)}{R x},\qquad T=-\frac{x^2 \left(y^2-1\right)-3 x^4+z^2}{4 \pi  R x^3}
 \end{equation}
 The chemical potential $\mu$ can be written in $x$, $y$, $z$ and $R$ by using \eqref{eq:mu:p:1}, \eqref{eq:holodic:3} and \eqref{eq:E1:T1} we get:-
 \begin{equation}
 \label{eq:mu1}
 \mu=-\frac{-x^2 \left(3 y^2+1\right)+x^4+4 y^2+z^2}{R x}
 \end{equation}
 
 \subsection{Energy ($E$), Temperature ($T$) and chemical potential ($\mu$) for $(\tilde{\Phi}_m, \tilde{Q}_e)$}
 \label{subsec:ADM2}
 Now for $\tilde{\Phi}_m$ and $\tilde{Q}_e$, using \eqref{eq:M:3}, \eqref{eq:Qe:1}, \eqref{eq:fr:1}, the ADM mass $M$ and the Hawking temperature using \eqref{eq:T:2} can be linked to the CFT boundary energy $E$ and temperature $T$ by using \eqref{eq:holodic:2} and \eqref{eq:holodic:3} as:-
 \begin{equation}
 \label{eq:E2:T2}
 E=\frac{2 c \left(x^4+x^2 \left(z^2+1\right)+y^2\right)}{R x}, \qquad T=-\frac{-3 x^4+x^2 \left(z^2-1\right)+y^2}{4 \pi  R x^3}
 \end{equation}
 And the chemical potential $\mu$ can be written in terms of $x$, $y$, $z$ and $R$ by using \eqref{eq:mu:p:1}, \eqref{eq:holodic:3} and \eqref{eq:E2:T2} and obtain:-
 \begin{equation}
 \label{eq:mu2}
 \mu=-\frac{x^4-x^2 \left(3 z^2+1\right)+y^2+4 z^2}{R x}
 \end{equation}
 \subsection{Energy ($E$), Temperature ($T$) and chemical potential ($\mu$) for $(\tilde{\Phi}_e, \tilde{\Phi}_m)$}
 \label{subsec:ADM3}
 Now for $\tilde{\Phi}_e$ and $\tilde{\Phi}_m$, using  \eqref{eq:M:3}, \eqref{eq:Qe:1}, \eqref{eq:fr:1}, the ADM mass $M$ and the Hawking temperature using \eqref{eq:T:3} can be co-related to the CFT boundary temperature and CFT boundary energy by using \eqref{eq:holodic:2} and \eqref{eq:holodic:3} as:-
 \begin{equation}
 \label{eq:E3:T3}
E=\frac{2 C x \left(x^2+y^2+z^2+1\right)}{R},  \qquad T=-\frac{-3 x^2+y^2+z^2-1}{4 \pi  R x}
 \end{equation}
 The chemical potential $\mu$ can be written in terms of $x$, $y$, $z$ and $R$ by using \eqref{eq:mu:p:1}, \eqref{eq:holodic:3} and \eqref{eq:E3:T3} and obtain:-
 \begin{equation}
 \label{eq:mu3}
 \mu=\frac{-x^4+x^2 \left(3 y^2+3 z^2+1\right)-4 \left(y^2+z^2\right)}{R x}
 \end{equation}
 \subsection{Energy ($E$), Temperature ($T$) and chemical potential ($\mu$) for $(\tilde{Q}_e,\tilde{Q}_m)$}
 \label{subsec:ADM4}
 Now, for $\tilde{Q}_e$ and $\tilde{Q}_m$, using \eqref{eq:M:3}, \eqref{eq:Qe:1}, \eqref{eq:fr:1}, the  ADM mass $M$ and the Hawking's temperature using \eqref{eq:T:4} can be corelated to the CFT boundary temperature and CFT energy at the boundary by using \eqref{eq:holodic:2} and \eqref{eq:holodic:3} as:-
\begin{equation} 
\label{eq:E4:T4}
 E=\frac{2 C \left(x^4+x^2+y^2+z^2\right)}{R x},\qquad T=-\frac{-3 x^4-x^2+y^2+z^2}{4 \pi  R x^3}
 \end{equation}
 The chemical potential $\mu$ can be written in terms of $x$, $y$, $z$ and $R$ by using \eqref{eq:mu:p:1}, \eqref{eq:holodic:3} and \eqref{eq:E4:T4} and obtain:-
 \begin{equation}
 \label{eq:mu4}
 \mu=-\frac{x^4-x^2+y^2+z^2}{R x}
 \end{equation}
 \section{Different types of ensembles in the CFT thermodynamics}
 \label{sub:Thermodynamic ensemble}
 We want to examine the different ensembles of the CFT thermodynamics using the holographic AdS/CFT dictionary. There are five complimentary variable groups in the CFT thermodynamics depiction for dyonic AdS black holes, namely $(\tilde{\Phi}_e,\tilde{Q}_e)$, $(\tilde{\Phi}_m,\tilde{Q}_m)$, $(T, S)$, $(p,\mathcal{V})$ and $(\mu, C)$. For the dyonic AdS black holes, the ensembles are:-
 \begin{subequations}
 \label{free34}
 \begin{align}
 & \text{fixed}\quad (\tilde{Q}_m, \tilde{\Phi}_e, p,C): \qquad F_1\equiv E-TS-\tilde{\Phi}_e \tilde{Q}_e +p\mathcal{V}=\tilde{\Phi}_m \tilde{Q}_m+\mu C+p\mathcal{V}\\
 & \text{fixed}\quad(\tilde{Q}_m, \tilde{\Phi}_e, p, \mu):\qquad F_2\equiv E-TS-\tilde{\Phi}_e \tilde{Q}_e +p\mathcal{V}-\mu C=\tilde{\Phi}_m \tilde{Q}_m+p\mathcal{V}\\
 & \text{fixed} \quad (\tilde{Q}_m, \tilde{\Phi}_e,\mathcal{V},C):\qquad F_3\equiv E-TS-\tilde{\Phi}_e \tilde{Q}_e =\tilde{\Phi}_m \tilde{Q}_m+\mu C\\
 & \text{fixed} \quad (\tilde{Q}_m, \tilde{\Phi}_e, \mathcal{V},\mu):\qquad F_4\equiv E-TS-\tilde{\Phi}_e \tilde{Q}_e -\mu C=\tilde{\Phi}_m \tilde{Q}_m\\
 & \text{fixed}\quad (\tilde{\Phi}_m, \tilde{Q}_e,p, C): \qquad F_5 \equiv E-TS+\tilde{\Phi}_m \tilde{Q}_m+p \mathcal{V}=\tilde{\Phi}_e \tilde{Q}_e + \mu C+p\mathcal{V}\\
 & \text{fixed}\quad (\tilde{\Phi}_m, \tilde{Q}_e,p,\mu):\qquad F_6 \equiv E-TS-\tilde{\Phi}_m \tilde{Q}_m+p \mathcal{V}-\mu C=\tilde{\Phi}_e \tilde{Q}_e +p\mathcal{V}\\
 & \text{fixed}\quad (\tilde{\Phi}_m, \tilde{Q}_e,\mathcal{V}, C):\qquad F_7 \equiv E-TS-\tilde{\Phi}_m \tilde{Q}_m=\tilde{\Phi}_e \tilde{Q}_e + \mu C\\
 & \text{fixed}\quad (\tilde{\Phi}_m, \tilde{Q}_e,\mathcal{V}, \mu):\qquad F_8 \equiv E-TS-\tilde{\Phi}_m \tilde{Q}_m-\mu C=\tilde{\Phi}_e \tilde{Q}_e \\
 & \text{fixed}\quad (\tilde{\Phi}_m, \tilde{\Phi}_e,p, C): \qquad F_9 \equiv E-TS-\tilde{\Phi}_m \tilde{Q}_m-\tilde{\Phi}_e \tilde{Q}_e+p \mathcal{V}=  \mu C+p\mathcal{V}\\
 & \text{fixed}\quad(\tilde{\Phi}_m, \tilde{\Phi}_e,p, \mu):\qquad F_{10} \equiv E-TS-\tilde{\Phi}_m \tilde{Q}_m-\tilde{\Phi}_e \tilde{Q}_e-\mu C+p \mathcal{V}=p\mathcal{V}\\
 & \text{fixed} \quad(\tilde{\Phi}_m, \tilde{\Phi}_e,\mathcal{V}, C):\qquad F_{11} \equiv E-TS-\tilde{\Phi}_m \tilde{Q}_m-\tilde{\Phi}_e \tilde{Q}_e=  \mu C\\
 & \text{fixed}\quad(\tilde{\Phi}_m, \tilde{\Phi}_e,\mathcal{V}, \mu):\qquad F_{12} \equiv E-TS-\tilde{\Phi}_m \tilde{Q}_m-\tilde{\Phi}_e \tilde{Q}_e+\mu C=0\\
 & \text{fixed} \quad (\tilde{Q}_m, \tilde{Q}_e,p, C):\qquad F_{13} \equiv E-TS+p \mathcal{V}= +\tilde{\Phi}_m \tilde{Q}_m+\tilde{\Phi}_e \tilde{Q}_e+ \mu C+p\mathcal{V}\\
 & \text{fixed} \quad (\tilde{Q}_m, \tilde{Q}_e,p,\mu):\qquad F_{14} \equiv E-TS+p \mathcal{V}+\mu C= +\tilde{\Phi}_m \tilde{Q}_m+\tilde{\Phi}_e \tilde{Q}_e+p\mathcal{V}\\
 & \text{fixed} \quad (\tilde{Q}_m, \tilde{Q}_e,\mathcal{V}, C):\qquad F_{15} \equiv E-TS= \tilde{\Phi}_m \tilde{Q}_m+\tilde{\Phi}_e \tilde{Q}_e+ \mu C\\
 & \text{fixed} \quad (\tilde{Q}_m, \tilde{Q}_e,\mathcal{V}, \mu):\qquad F_{16} \equiv E-TS-\mu C= \tilde{\Phi}_m \tilde{Q}_m+\tilde{\Phi}_e \tilde{Q}_e
 \end{align}
 \end{subequations}
In the subsections carried on, we study the behaviours of different phases seen in the list ensembles mentioned above.

 \subsection{For the fixed $(\tilde{Q}_m, \tilde{\Phi}_e, \mathcal{V},C)$ ensemble}
  \label{subsec:Ensemble3}
\begin{figure}[htp]
\centering
\begin{subfigure}[b]{0.3\textwidth} 
\includegraphics[width=1\textwidth]{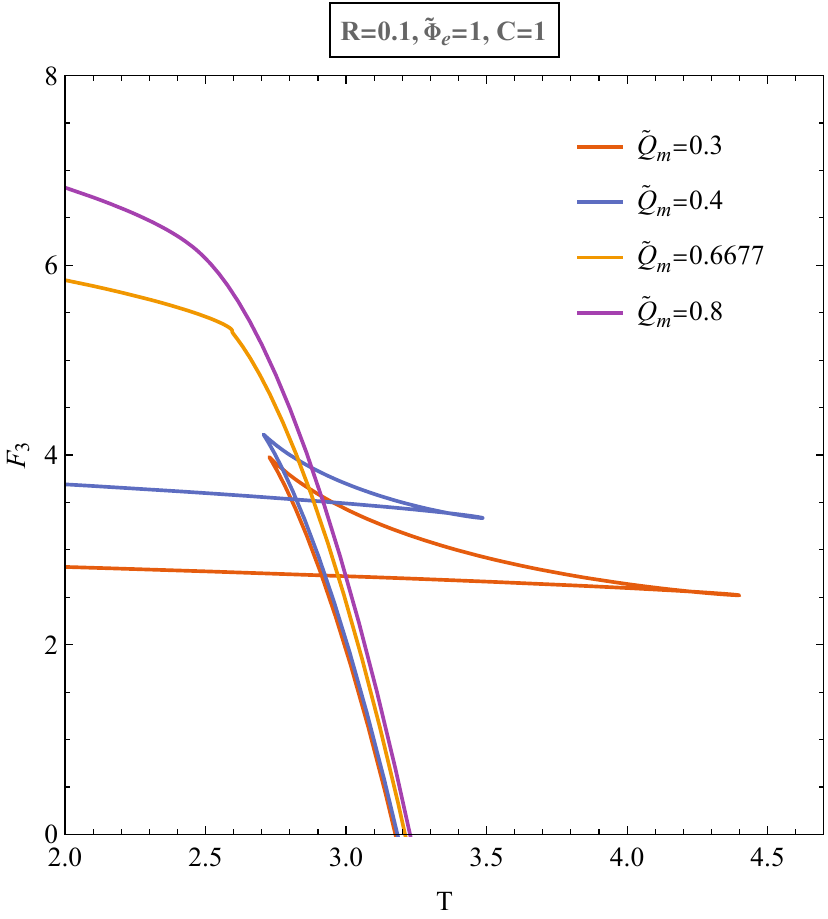}
\caption{}
\label{Fig:2}
\end{subfigure}
\begin{subfigure}[b]{0.3\textwidth}
\includegraphics[width=1\textwidth]{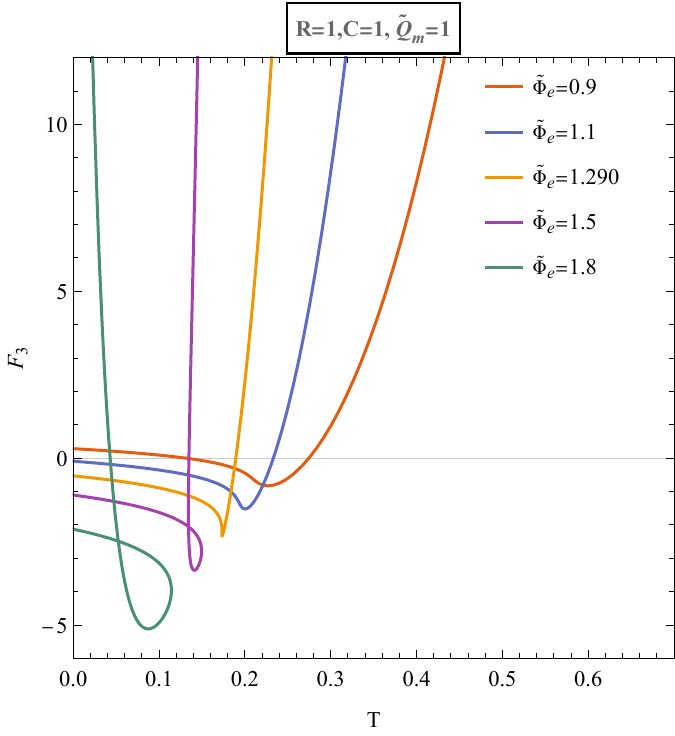}
\caption{}
\label{Fig:3}
\end{subfigure}
\begin{subfigure}[b]{0.3\textwidth} 
\includegraphics[width=1\textwidth]{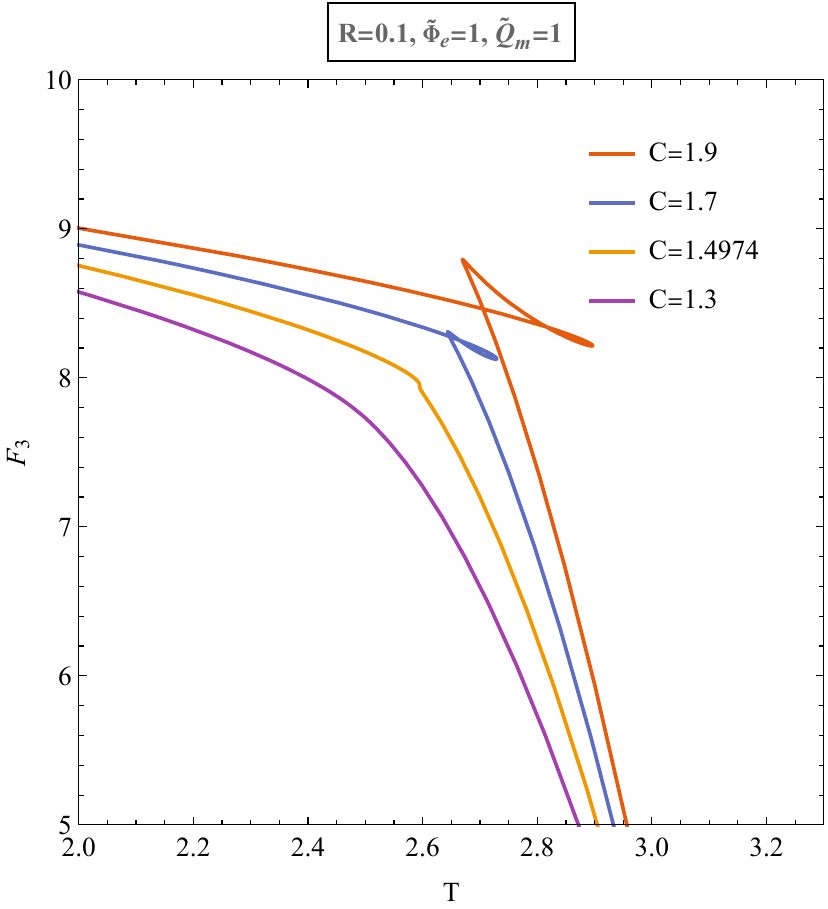}
\caption{}
\label{Fig:4}
\end{subfigure}
    \caption{Free energy $F_3$ vs. temperature $T$ plot for the fixed $(\tilde{\Phi}_e,\tilde{Q}_m,R,C)$ ensemble in $(D=4/d=3)$.  }
\label{fig:en1}
\end{figure}

\begin{figure}[htp]
\centering
\begin{subfigure}[b]{0.3\textwidth} 
\includegraphics[width=1\textwidth]{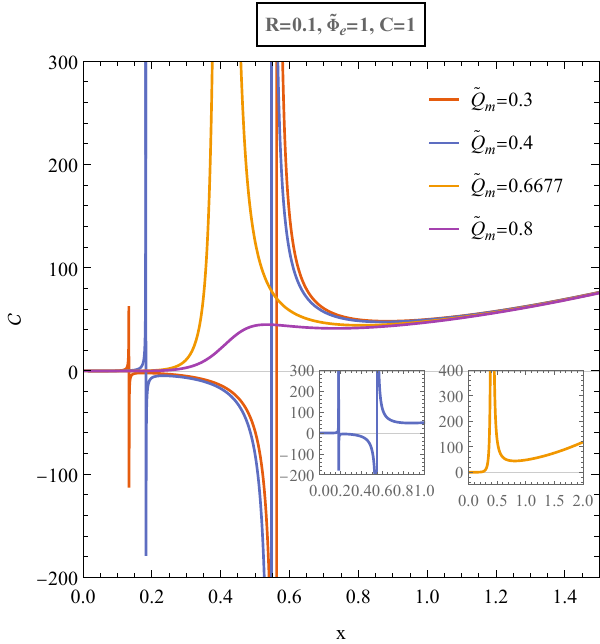}
\caption{}
\label{Fig:Sfig1}
\end{subfigure}
\begin{subfigure}[b]{0.3\textwidth}
\includegraphics[width=1\textwidth]{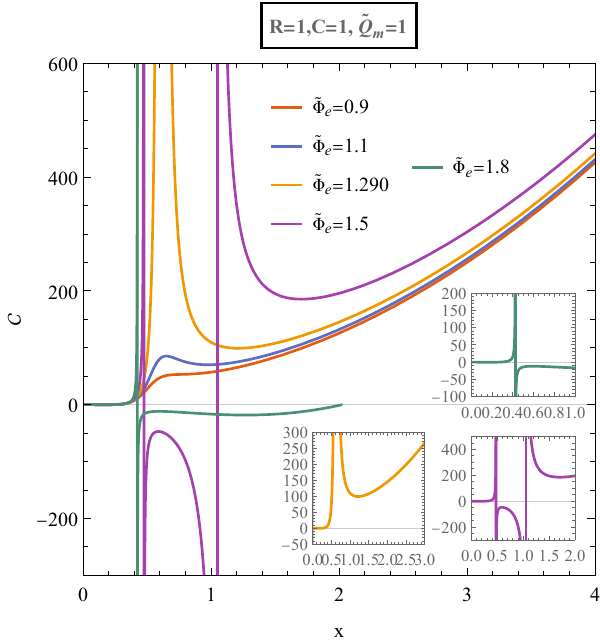}
\caption{}
\label{Fig:Sfig2}
\end{subfigure}
\begin{subfigure}[b]{0.3\textwidth} 
\includegraphics[width=1\textwidth]{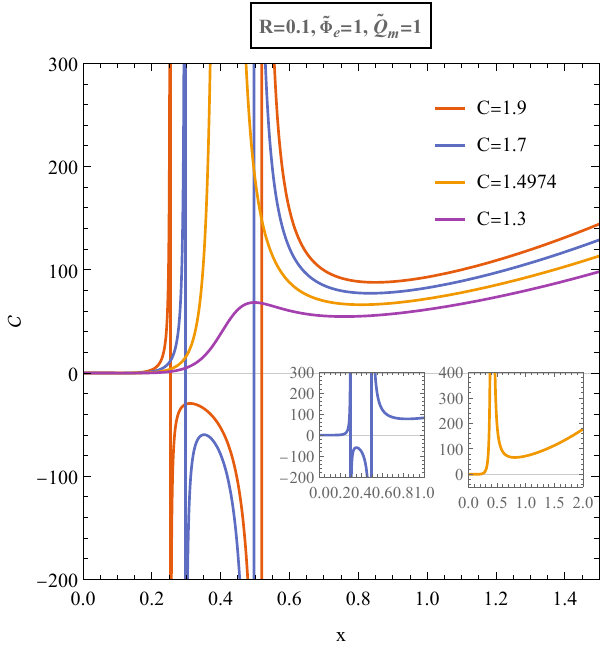}
\caption{}
\label{Fig:Sfig3}
\end{subfigure}
    \caption{Specific Heat $\mathcal{C}$ vs. $x$ plot for the fixed $(\tilde{\Phi}_e,\tilde{Q}_m,R,C)$ ensemble in $(D=4/d=3)$.  }
\label{fig:en1a}
\end{figure}

 Fixing the magnetic charge $\tilde{Q}_m$, an electric potential $\tilde{\Phi}_e$, volume of space $\mathcal{V}$ and CFT central charge $C$. The free energy is:-
 \begin{equation}
 \label{eq:F3:1}
 F_3 \equiv E-TS- \tilde{\Phi}_e \tilde{Q}_e=-\frac{C \left(x^4-x^2+y^2-3 z^2\right)}{R x}
 \end{equation}
The first law CFT given in \eqref{eq:dE:3}, by differentiating of $F_3$ we get
  \begin{equation}
  \label{eq:dF3}
  \begin{split}
  & dF_3=dE-TdS-SdT-\tilde{\Phi}_e d\tilde{Q}_e-\tilde{Q}_e d \tilde{\Phi}_e\\
  & =-SdT+\tilde{\Phi}_m d\tilde{Q}_m-\tilde{Q}_e d \tilde{\Phi}_e-p d\mathcal{V}+\mu C
  \end{split}
  \end{equation}
 Hence, we can see in \eqref{eq:dF3}, that $F_3$ is static for the fixed $(T,\tilde{Q}_m,\tilde{\Phi}_e,\mathcal{V},C)$.
 Let us explore how the free energy $F_3$ \eqref{eq:F3:1} behaves w.r.t to the temperature $T$ \eqref{eq:T:1} for different fixed $(T,\tilde{Q}_m,\tilde{\Phi}_e, \mathcal{V},C)$ values. We preferable to grasp the idea that $F_3$ and $T$ are the functions of $(T,\tilde{Q}_m,\tilde{\Phi}_e, \mathcal{V}, C, x)$.  The volume $\mathcal{V}$ is fixed by fixing the radius $R$. We write the temperature and free energy as below after substituting the parameters $y=\tilde{\Phi}_e R x$ and $z=\frac{\tilde{Q}_m}{4 C}$ for the $\tilde{\Phi}_e$ as electric potential and $\tilde{Q}_m$ as magnetic charge we get:-
 \begin{equation}
 \begin{split}
 \label{eq:F3:2}
 & F_3=\frac{3 \tilde{Q}_m^2}{16 C R x}-\frac{C x \left(R^2 \tilde{\Phi}_e^2+x^2-1\right)}{R}, \quad
  T=-\frac{\frac{\tilde{Q}_m^2}{16 C^2}+x^2 \left(R^2 \tilde{\Phi}_e^2-3 x^2-1\right)}{4 \pi  R x^3}
 \end{split}                
 \end{equation}
As a result, we may parametrically display $F(T)$ free energy and temperature using $x$ as a condition for fixed $(T,\tilde{Q}_m,\tilde{\Phi}_e,\mathcal{V},C)$. Since scale invariance has fixed the dependence of $R$, changing the value of $R$ will only cause the plots to rescale each other. When we plot $F_3(T)$ parametrically using \eqref{eq:F3:2} for various values of $\tilde{Q}_m$ in Figure \ref{Fig:2}, \ref{Fig:Sfig1}, for $\tilde{\Phi}_e$ in Figure \ref{Fig:3}, \ref{Fig:Sfig2} and for $C$ in Figure \ref{Fig:4}, \ref{Fig:Sfig3}.

In Figure \ref{Fig:2} we plot for various values of $\tilde{Q}_m$ where for subcritical values of $\tilde{Q}_m$ (red, blue) we get a Van der Waals type phase transition of order one which looks like a swallow-tail. For same values we see three branches in \ref{Fig:Sfig1} namely small (stable)- intermediate (unstable)- large (stable) entropy branches. At the critical point $\tilde{Q}_m$=0.6677 (orange) we see a kink like structure and is actually a superfulid $\lambda$ phase transition and is of order two In Figure \ref{Fig:Sfig1} it looks like a discontinuity in that point. For supercritical values (purple) we see no phase transition and a smooth monotonous plot is seen. \\
In figure \ref{Fig:3}, \ref{Fig:Sfig2} we plot for various values of $\tilde{\Phi}_e$ where for sub critical values (red, blue) we do not get any phase transition and a smooth monotonous plot is seen but at the critical point $\tilde{\Phi}_e$=1.290 (orange), we get the superfluid $\lambda$ phase transition of order two where we see two stable branches (small-large) and for super critical values of $\tilde{\Phi}_e$ we get the Van der Waals type phase transition signalling three branches namely small (stable)-intermediate (unstable)-large(stable) branches. At a certain super critical value $\tilde{\Phi}_e$=1.8 we get a Davies type phase transition showing only two small (stable)-large (unstable) branches seen in Figure \ref{Fig:Sfig2}.\\
\begin{figure}[htp]
\centering
\begin{subfigure}[b]{0.3\textwidth} 
\includegraphics[width=1\textwidth]{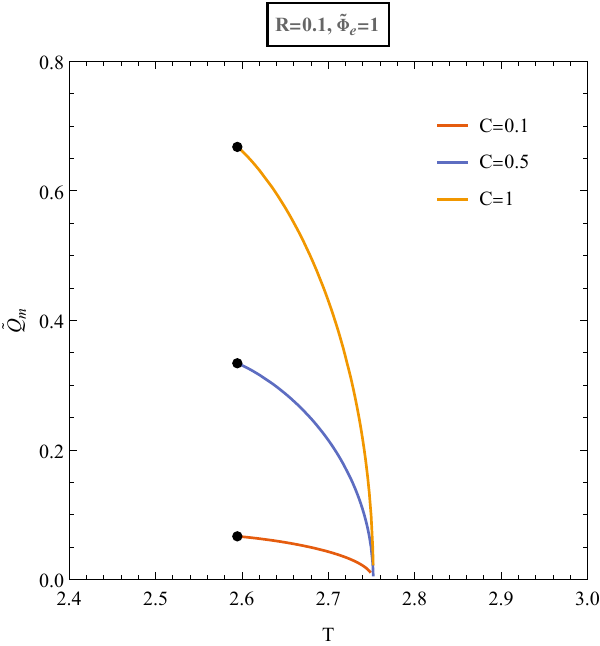}
\caption{}
\label{Fig:coexist1}
\end{subfigure}
\begin{subfigure}[b]{0.3\textwidth}
\includegraphics[width=1\textwidth]{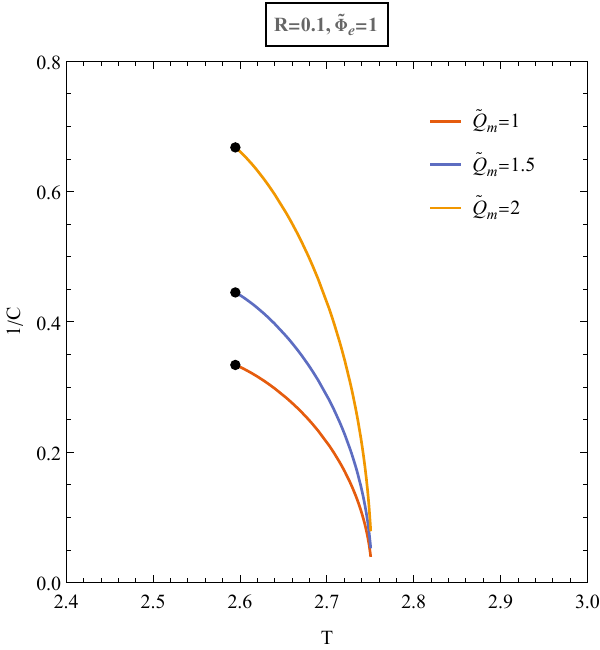}
\caption{}
\label{Fig:coexist2}
\end{subfigure}
\begin{subfigure}[b]{0.3\textwidth}
\includegraphics[width=1\textwidth]{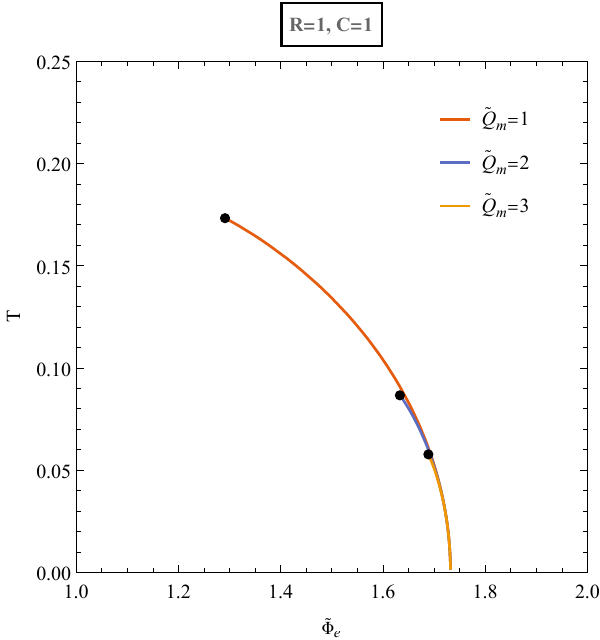}
\caption{}
\label{Fig:coexist3}
\end{subfigure}
    \caption{Coexistence plots}
\label{fig:Coexistenceplots}
\end{figure}
Plotting for various values of $C$ in Figure \ref{Fig:4} we see Van der Waals type phase transition where we see small (stable)-intermediate (unstable)-large (stable) branches in Figure \ref{Fig:Sfig3} for super crititcal values (red, blue). At the critical point $C$=1.4974 (orange) we see a superfluid $\lambda$ phase transition and for sub critical values we see only a monotonous plot (purple).\\
In Figure \ref{fig:Coexistenceplots} we plot the coexistence lines for the low and high entropy regimes for the CFT on the $\tilde{Q}_m-T$, $\tilde{\Phi}_e-T$ and $1/C-T$ phase plots. The coexistence line delineates the two phases on the planes, and the CFT experiences a first-order phase transition when it intersects this line. In these phase diagrams, the phase with lower entropy is found to the left of the coexistence line (for any given value of $C$, $R$, and $\tilde{\Phi}_e$ respectively), whereas the phase with higher entropy is situated to the right. The critical points are represented by black dots on the diagrams. Beyond these critical points, the CFT does not exhibit distinct phases.

 \subsection{For the fixed $(\tilde{Q}_m, \tilde{\Phi}_e, \mathcal{V},\mu)$ ensemble}
  \label{subsec:Ensemble4}
In this ensemble we explore the results of fixing the corresponding $\mu$ as the chemical potential. The ensemble's proper free energy $F_4$ depends on $(\tilde{Q}_m, \tilde{\Phi}_e, \mathcal{V}, \mu)$.
 \begin{equation}
 \label{eq:F4:1}
 F_4\equiv E-TS-\tilde{\Phi}_e \tilde{Q}_e -\mu C= \tilde{\Phi}_m \tilde{Q}_m
 \end{equation}
By differentiating $F_4$, we get:-
\begin{equation}
\label{eq:dF4}
\begin{split}
& dF_4=dE-TdS-SdT-\tilde{\Phi}_e d\tilde{Q}_e- \tilde{Q}_e d\tilde{\Phi}_e-\mu dC-C d\mu\\
& dF_4=-SdT+\tilde{\Phi}_m d \tilde{Q}_m-\tilde{Q}_e d \tilde{\Phi}_e-p d \mathcal{V}-C d \mu
\end{split}
\end{equation} 
where we put $dE$ from equation \eqref{eq:dE:3} and simplify.
 Therefore, the free energy $F_4$ depends on $(T, \tilde{Q}_m, \tilde{\Phi}_e, \mathcal{V},\mu)$ can be written as $
 F_4=\tilde{Q}_m \frac{z}{R x}$
 where we have put $\tilde{\Phi}_m=\frac{z}{R x}$. Now from \eqref{eq:mu1} we solve it in terms of $z$ to get $z= \sqrt{-\mu R x+x^2 \left(3 y^2+1\right)-x^4-4 y^2}$ and put it in the above free energy equation and temperature \eqref{eq:E1:T1} we get:-
 \begin{equation}
 \label{eq:F4:2}
 F_4=\frac{\tilde{Q}_m \sqrt{-\mu R x+x^2 \left(3 y^2+1\right)-x^4-4 y^2}}{R x}, \quad T=\frac{\mu R x+4 \left(-x^2 y^2+x^4+y^2\right)}{4 \pi  R x^3}
 \end{equation}
Representing the free energy $F_4$ \eqref{eq:F4:2} and $T$ for temperature  as functions $F_4=F_4(\tilde{Q}_m,\tilde{\Phi}_e, R, \mu , x)$
 and $T=T(\tilde{\Phi}_e, R,\mu, x)$ using lastly $y=\tilde{\Phi}_e R x$, we get:-
 \begin{equation}
 \label{eq:F4:3}
 \begin{split}
 & F_4=\frac{|\tilde{Q}_m| \sqrt{x \left(x^3 \left(3 R^2 \tilde{\Phi}_e^2-1\right)-4 R^2 \tilde{\Phi}_e^2 x-\mu R+x\right)}}{R x}, \quad
 T=\frac{x^3 \left(4-4 R^2 \tilde{\Phi}_e^2\right)+4 R^2 \tilde{\Phi}_e^2 x+\mu R}{4 \pi  R x^2}
 \end{split}
 \end{equation}
 We observe that the absolute value of $\tilde{Q}_m$ is there because the free energy cannot be negative.
\begin{figure}[htp]
\centering
\begin{subfigure}[b]{0.3\textwidth} 
\includegraphics[width=1\textwidth]{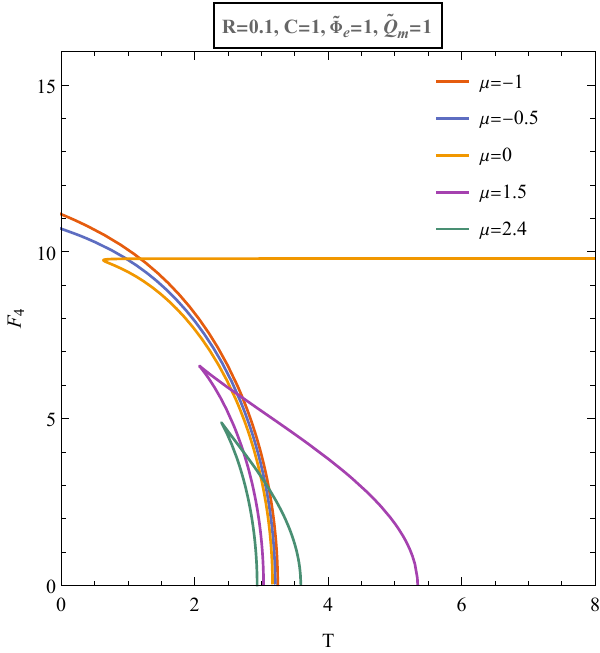}
\caption{}
\label{Fig:5}
\end{subfigure}
\begin{subfigure}[b]{0.3\textwidth}
\includegraphics[width=1\textwidth]{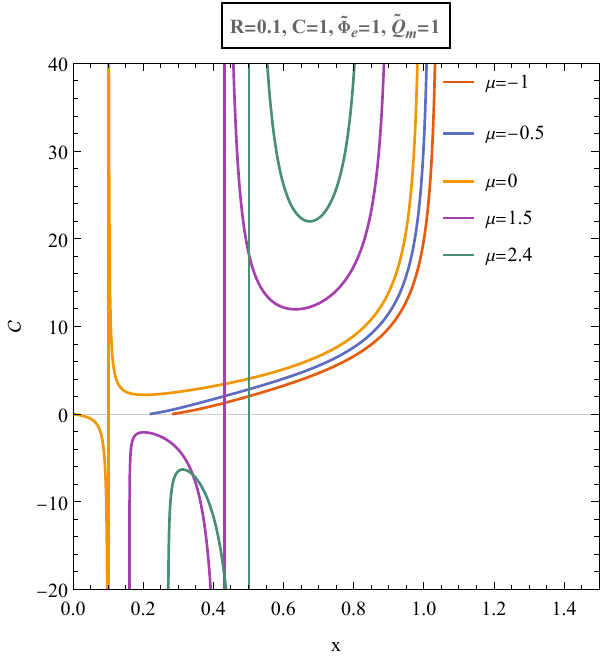}
\caption{}
\label{Fig:SfigF4}
\end{subfigure}
    \caption{ Plotting free energy $F_4$ vs $T$ as temperature and specific heat $\mathcal{C}$ vs x, we get the phase diagram for fixed $(\tilde{Q}_m, \tilde{\Phi}_e, R, \mu)$ ensemble. }
\label{fig:en1a}
\end{figure} 
 We can display $F_4(T)$ for given values of $(\tilde {Q}_m, \tilde{\Phi}_e,R,\mu)$ shown in Figure \ref{Fig:5}. At $\mu=0$, the graph exhibits a qualitative change that can be seen (orange curve). The free energy $F_4$ is single-valued for the temperature function for $\mu < 0$ (red, blue), which translates to a single phase transition in the CFT and it can also be see in Figure \ref{Fig:SfigF4}. The free energy curve for $0<\mu<\mu_{coin}$ (purple and green) has branches namely two that split $F_4=0$ at $T_1$ and $T_2$ temperatures, where $T_2 \geq T_1$. To solve this we take the two positive roots $x_1$, $x_2$ which are given in Appendix \ref{Appendix}. Putting the value of the roots $x_1$ and $x_2$ in the temperature equation \eqref{eq:F4:2} we get the two temperatures namely $T_1$ and $T_2$ written as:-
\begin{equation}
T_1=\frac{144 \left(B_1\right) \left(R \tilde{\Phi }_e-3 R^3 \tilde{\Phi }_e^3\right){}^2 \left(A_1\right){}^{2/3}+432 R \mu  \left(1-3 R^2 \tilde{\Phi }_e^2\right){}^3 \left(A_1\right)-R^2 \tilde{\Phi }_e^2 \left(B_1\right){}^3}{12 \pi  R \left(1-3 R^2 \tilde{\Phi }_e^2\right) \sqrt[3]{A_1} \left(B_1\right){}^2}
\end{equation} 
\begin{equation}
T_2=\frac{144 \left(B_2\right) \left(R \tilde{\Phi }_e-3 R^3 \tilde{\Phi }_e^3\right){}^2 \left(A_1\right){}^{2/3}+432 R \mu  \left(1-3 R^2 \tilde{\Phi }_e^2\right){}^3 \left(A\right)-R^2 \tilde{\Phi }_e^2 \left(B_2\right){}^3+\left(B_2\right){}^3}{12 \pi  R \left(1-3 R^2 \tilde{\Phi }_e^2\right) \sqrt[3]{A_1} \left(B_2\right){}^2}
\end{equation}
where $A_1$ $B_1$ and $B_2$ above  are given in Appendix \ref{Appendix}. Further if we put $\mu=0, \tilde{\Phi}_e=1$ (orange) we get $
 T_1=3.16698$, $ T_2=\infty$. The two temperatures $T_1$ and $T_2$ are same at the coincidence point, where for the plot $\mu_{coin}=3.849$ so that the coincidence temperature comes out to be $T_{coin}=2.75664$. We see that we do not get any plots of $F_4(T)$ when $\mu>\mu_{coin}$. So hence we only get the plots when $\mu < \mu_{coin}$. In Fig \ref{Fig:5} using the purple curve for reference, the temperature $T_2$ starts increasing with increasing value of x uptill $T_0$ after which it starts decreasing. The plot between $T_0$ and $T_2$ is low entropy branch and between $T_1$ and $T_0$ is high entropy branch which can also be seen in \ref{Fig:SfigF4}. The stationary point $T_0$ can be obtained by solving $(\frac{dT}{dx}=0)$ for $x$ and putting it back in $T$ to get $T_0$. The value of the turning point of the purple plot $T_0=2.07463$ and this is the point where we see a Davies type phase transition and the expression is in Appendix \ref{Appendix}.\\ 
When we analyze the phase transitions within this ensemble, there is a single transition for $\mu < 0$. However, within the range $0 \leq \mu < \mu_{\text{coin}}$, a phase shift occurs between the branches of low and high entropy. In this interval, the high-entropy state is thermodynamically favored because it has the lowest free energy for temperatures $T_0 < T < T_1$. This high-entropy phase concludes at $T = T_1$. For temperatures $T_1 < T < T_2$, the lower entropy state prevails and dominates the thermodynamic ensemble.\\

\subsection{For the fixed $(\tilde{Q}_e,\tilde{\Phi}_m,\mathcal{V},C)$ ensemble}
 \label{subsec:Ensemble7}
\begin{figure}[htp]
\centering
\begin{subfigure}[b]{0.3\textwidth} 
\includegraphics[width=1\textwidth]{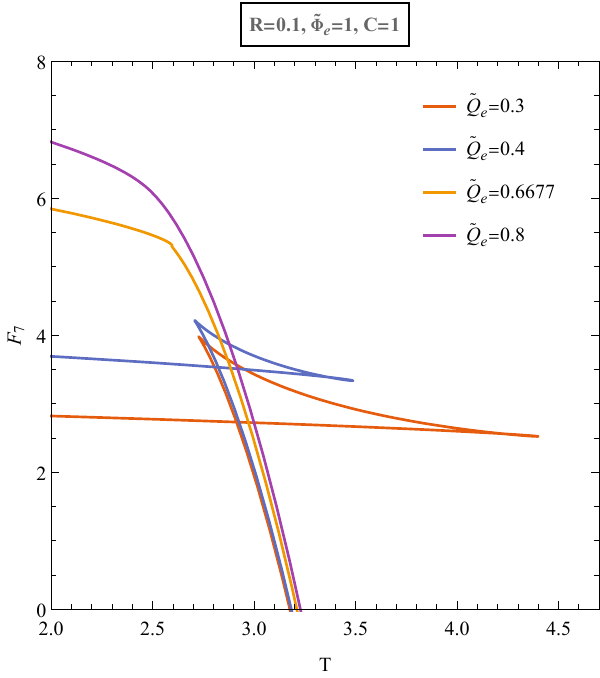}
\caption{}
\label{Fig:8}
\end{subfigure}
\begin{subfigure}[b]{0.3\textwidth}
\includegraphics[width=1\textwidth]{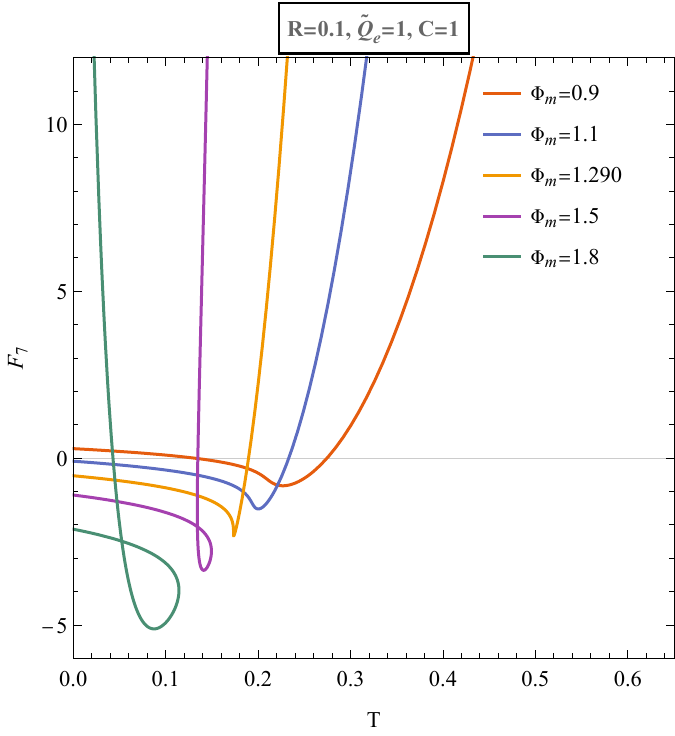}
\caption{}
\label{Fig:7}
\end{subfigure}
\begin{subfigure}[b]{0.3\textwidth} 
\includegraphics[width=1\textwidth]{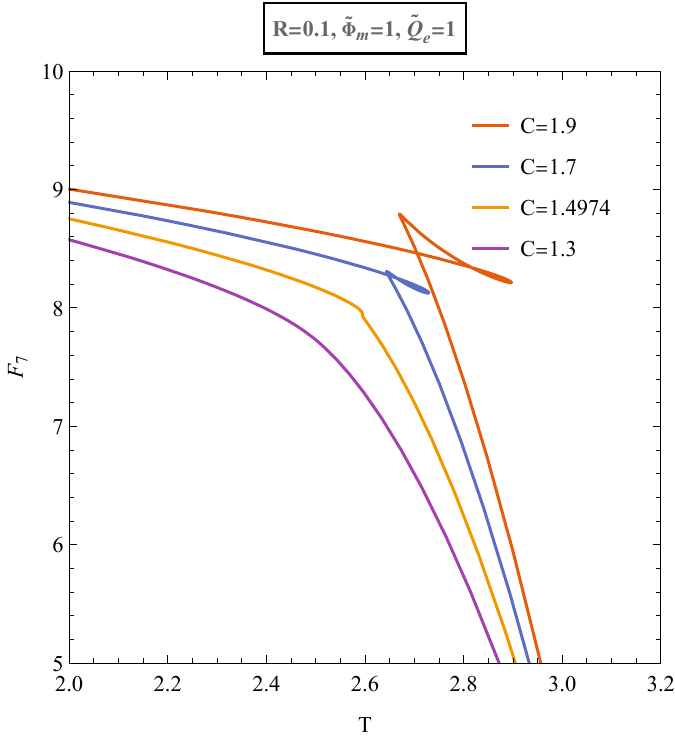}
\caption{}
\label{Fig:9}
\end{subfigure}
    \caption{Free energy $F_7$ vs. temperature $T$ plot for the fixed $(\tilde{\Phi}_m,\tilde{Q}_e,R,C)$ ensemble in $(D=4/d=3)$.  }
\label{fig:en3}
\end{figure}

 We take electric charge $\tilde{\Phi}_e$, magnetic potential $\tilde{\Phi}_m$, CFT volume $\mathcal{V}$ and CFT central charge $C$. Hence, now the free  energy for this ensemble is calculated as:-
\begin{equation}
\label{eq:F7:1}
F_7=E-TS- \tilde{\Phi}_m \tilde{Q}_m=\frac{c \left(-x^4+x^2 \left(3 z^2+1\right)+3 y^2-4 z^2\right)}{R x}
\end{equation}
By using the first CFT law \eqref{eq:dE:3} the differential of $F_7$ gives
\begin{equation}
\label{eq:dF7}
\begin{split}
& dF_7=dE-TdS-SdT-\tilde{\Phi}_m d\tilde{Q}_m-\tilde{Q}_m d\tilde{\Phi}_m\\
& dF_7=-SdT+\tilde{\Phi}_e d\tilde{Q}_e-\tilde{Q}_m d\tilde{\Phi}_m-p d\mathcal{V}+\mu dC
\end{split}
\end{equation}
Hence, from \eqref{eq:dF7} we can infer that $F_7$ is seen to be stationary at fixed $(T,\tilde{Q}_e, \tilde{\Phi}_m, \mathcal{V},C)$. Let us now study the performance of the Helmholtz free energy $F_7$ w.r.t to the temperature $T$ given for different values of the $(T, \tilde{Q}_e, \tilde{\Phi}_m, \mathcal{V}, C)$. The free energy and the temperature are written by simplifying the above equation by putting the values of $y=\frac{\tilde{Q}_e}{4 C}$ and $z=\tilde{\Phi}_m R x$ in \eqref{eq:F7:1} and \eqref{eq:E2:T2}  we get:-
\begin{equation}
\label{eq:F7:2}
F_7=\frac{3 \tilde{Q}_e^2}{16 c R x}+\frac{c \left(R^2 \tilde{\Phi}_m^2 \left(3 x^2-4\right) x-x^3+x\right)}{R}, \quad
T=-\frac{\frac{\tilde{Q}_e^2}{16 c^2}+x^4 \left(R^2 \tilde{\Phi}_m^2-3\right)-x^2}{4 \pi  R x^3}
\end{equation}
  By using the free energy and temperature expression \eqref{eq:F7:2}, we can plot the parametric plot between $F_7$ and $T$ and are seen in  Figure \ref{fig:en3}.

\begin{figure}[htp]
\centering
\begin{subfigure}[b]{0.3\textwidth} 
\includegraphics[width=1\textwidth]{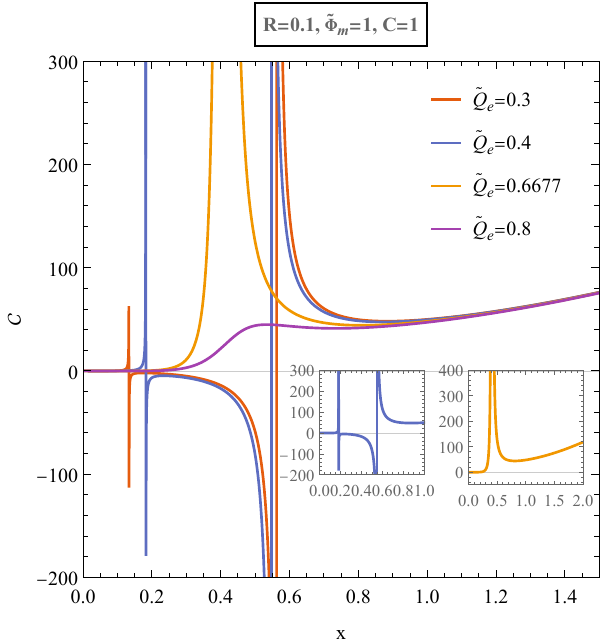}
\caption{}
\label{Fig:Sfig43}
\end{subfigure}
\begin{subfigure}[b]{0.3\textwidth}
\includegraphics[width=1\textwidth]{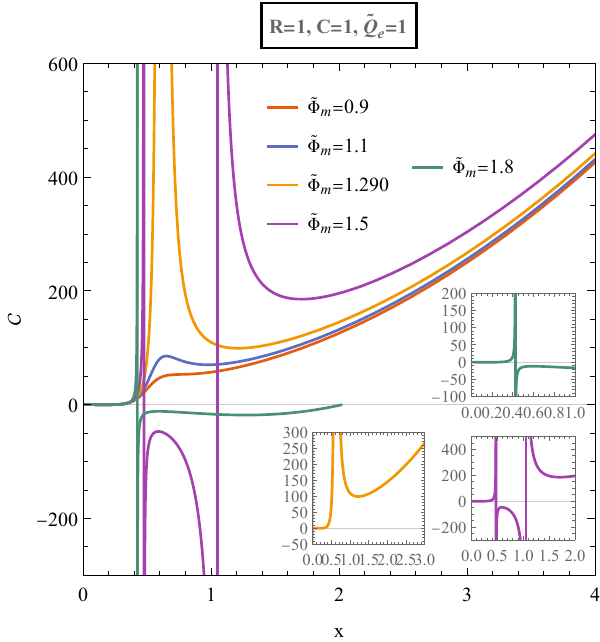}
\caption{}
\label{Fig:Sfig44}
\end{subfigure}
\begin{subfigure}[b]{0.3\textwidth} 
\includegraphics[width=1\textwidth]{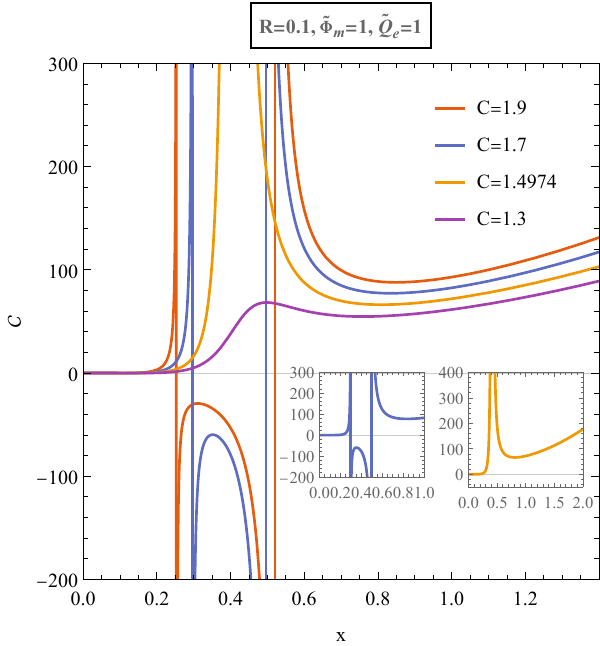}
\caption{}
\label{Fig:Sfig45}
\end{subfigure}
    \caption{Specific Heat $\mathcal{C}$ vs. $x$ plot for the fixed $(\tilde{\Phi}_e,\tilde{Q}_m,R,C)$ ensemble in $(D=4/d=3)$.  }
\label{fig:en3a}
\end{figure}  
In Figure \ref{Fig:8} and \ref{Fig:Sfig43} we plot for various values of $\tilde{Q}_e$ keeping other variables constant. We see for sub critical values (red, blue) of electric charge we get Van der Waals type phase transition showing swallow-tail and three branches in both plots effectively showing small (stable)-intermediate (unstable)-large (stable) entropy branches. At the critical value $\tilde{Q}_e=0.6677$ (orange) we see superfluid $\lambda$ phase transition showing two stable branches of small-large entropy branches and lastly for super critical values (purple) we get no phase transition. \\
\begin{figure}[htp]
\centering
\begin{subfigure}[b]{0.3\textwidth} 
\includegraphics[width=1\textwidth]{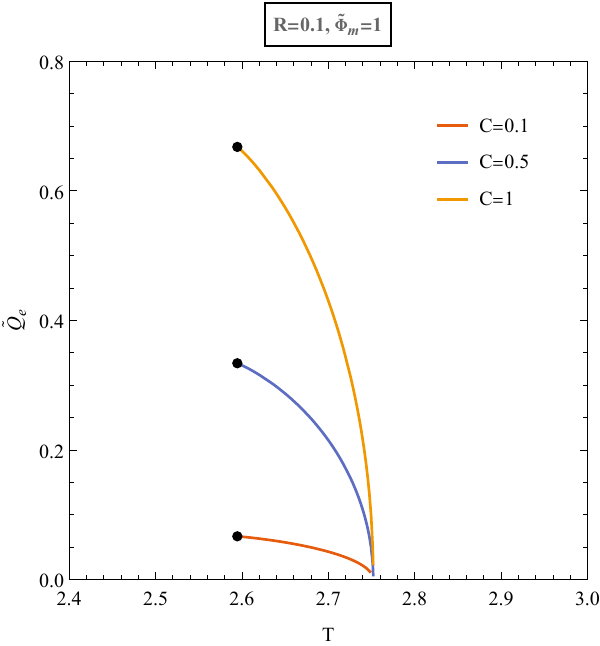}
\caption{}
\label{Fig:coexist4}
\end{subfigure}
\begin{subfigure}[b]{0.3\textwidth}
\includegraphics[width=1\textwidth]{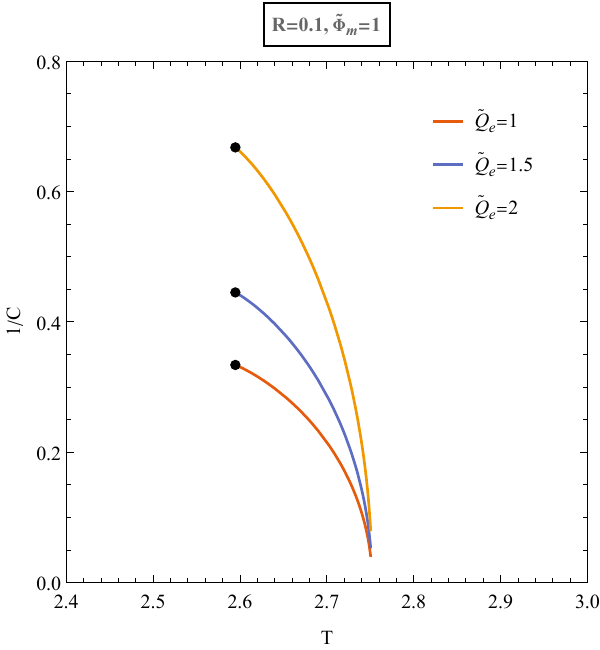}
\caption{}
\label{Fig:coexist5}
\end{subfigure}
\begin{subfigure}[b]{0.3\textwidth}
\includegraphics[width=1\textwidth]{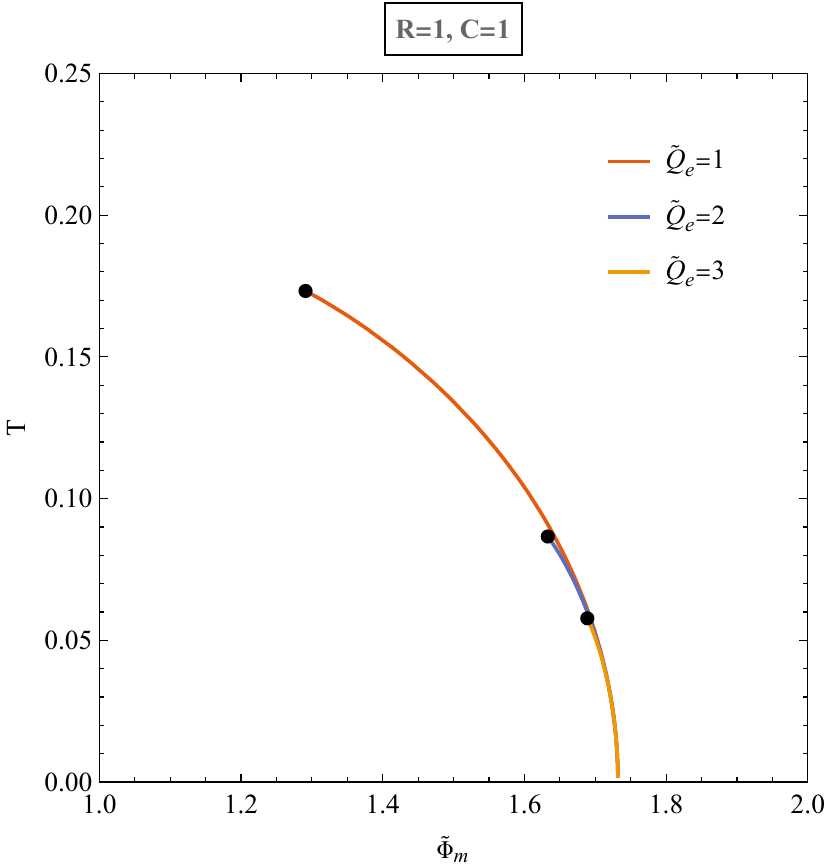}
\caption{}
\label{Fig:coexist6}
\end{subfigure}
    \caption{Coexistence plots}
\label{fig:Coexistenceplots1}
\end{figure}
In Figure \ref{Fig:7} and \ref{Fig:Sfig44}, we plot for various values of $\tilde{\Phi}_m$ where for sub critical values (red, blue) we see no transition but at the critical value $\tilde{\Phi}_m=1.290$ (orange) we see a superfluid $\lambda$ phase transiton which shows two stable branch of small-large entropy and last for super critical values (purple) we see a Van der Waals type phase transition showing three branches namely small (stable)-intermediate (unstable)-large (stable) entropy branches. Interestingly, at a certain supercritical value $\tilde{\Phi}_m=1.8$ (green), we see a Davies type phase transition showing small (stable)-large (unstable) entropy branch. Lastly plotting for various values of $C$ in \ref{Fig:9} and \ref{Fig:Sfig45} we see Van der Waals type phase transition for super critical values (red, blue) of $C$ showing swallow-tail and three branches namely small (stable)-intermediate (unstable)-large (stable) branches. At the critical point $C=1.4974$ (orange) we see a superfluid $\lambda$ phase transition and lastly for sub critical values (purple) we see no phase transition and just a monotonous plot.\\
In Figure \ref{fig:Coexistenceplots1} we plot the coexistence lines for the low and high entropy regimes for the CFT on the $\tilde{Q}_e-T$, $\tilde{\Phi}_m-T$ and $1/C-T$ phase plots. The coexistence line delineates the two phases on the planes, and the CFT experiences a first-order phase transition when it intersects this line. In these phase diagrams, the phase with lower entropy is found to the left of the coexistence line (for any given value of $C$, $R$, and $\tilde{\Phi}_m$ respectively), whereas the phase with higher entropy is situated to the right. The critical points are represented by black dots on the diagrams. Beyond these critical points, the CFT does not exhibit distinct phases.

\subsection{For the fixed $(\tilde{Q}_e,\tilde{\Phi}_m, \mathcal{V},\mu)$ ensemble}
 \label{subsec:Ensemble8}
 We study what happens when the appropriate chemical potential $\mu$ is fixed. The correct free energy of the ensemble$ (\tilde{Q}_e, \tilde{\Phi}_m, \mathcal{V}, \mu)$ is:-
 \begin{equation}
 \label{eq:F8:1}
 F_8=E-TS-\tilde{\Phi}_m \tilde{Q}_m-\mu C=\frac{4 C y^2}{R x}
\end{equation}
By differentiating $F_8$ and putting \eqref{eq:dE:3} we get:-
\begin{equation}
\label{eq:dF8}
    \begin{split}
        & dF_8=dE-TdS-SdT-\tilde{\Phi}_m d\tilde{Q}_m-\tilde{Q}_m d\tilde{\Phi}_m-\mu dC-C d\mu\\
        & dF_8=-SdT-\tilde{Q}_m d\tilde{\Phi}_m+\tilde{\Phi}_e d\tilde{Q}_e-C d\mu+ p d\mathcal{V}
    \end{split}
\end{equation}
 Hence for the fixed value of $(T,\tilde{\Phi}_m, 
 \tilde{Q}_e, \mathcal{V}, \mu)$ the free energy $F_8$ is stationary. Then we put the value of $C=\frac{\tilde{Q}_e}{4 y}$ we get $F_8=\frac{\tilde{Q}_e y}{R x}$.
Solving $\mu$ \eqref{eq:mu2} in terms of $y$ we get 
$y= \sqrt{-\mu R x-x^4+x^2 \left(3 z^2+1\right)-4 z^2}$
 and put it on the free energy equation $F_8=\frac{\tilde{Q}_e y}{R x}$ and temperature  \eqref{eq:E2:T2} and lastly putting $z=\tilde{\Phi}_m R x$ we get:-
\begin{equation}
\label{eq:F8:2}
F_8=\frac{|\tilde{Q}_e| \sqrt{x \left(x^3 \left(3 R^2 \tilde{\Phi}_m^2-1\right)-4 R^2 \tilde{\Phi}_m^2 x-\mu R+x\right)}}{R x}, \quad
T=\frac{4 \left(-R^2 \tilde{\Phi}_m^2 x^4+R^2 \tilde{\Phi}_m^2 x^2+x^4\right)+\mu R x}{4 \pi  R x^3}
\end{equation}
\begin{figure}[htp]
\centering
\begin{subfigure}[b]{0.3\textwidth} 
\includegraphics[width=1\textwidth]{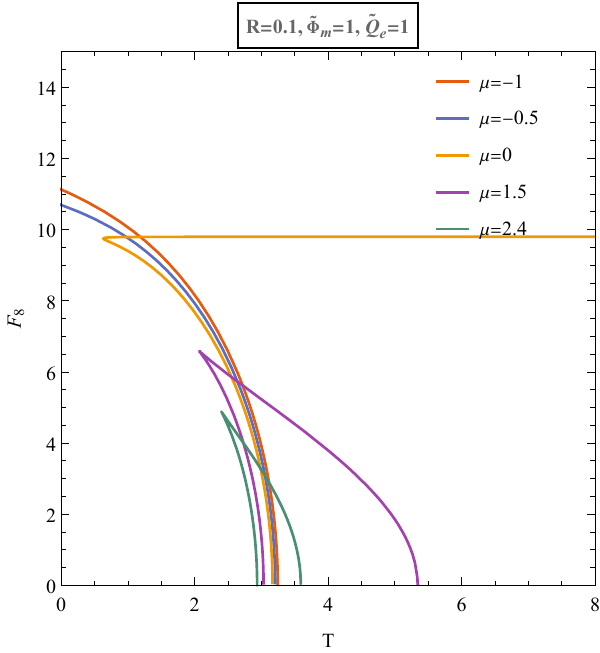}
\caption{}
\label{Fig:10}
\end{subfigure}
\begin{subfigure}[b]{0.3\textwidth}
\includegraphics[width=1\textwidth]{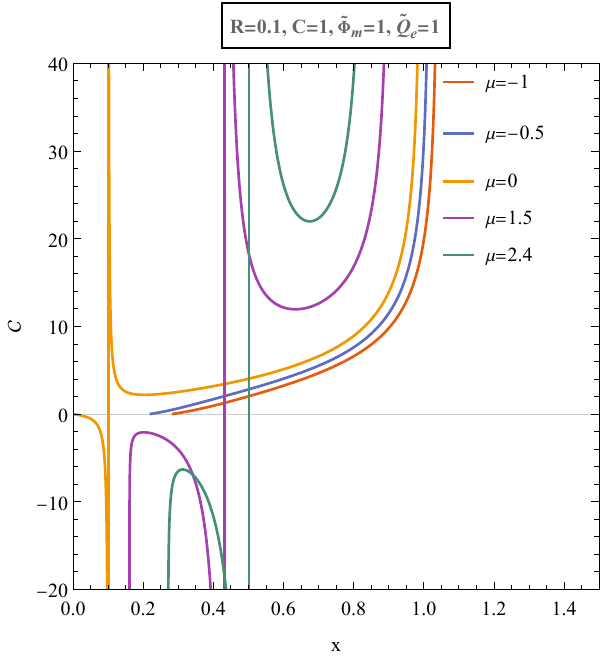}
\caption{}
\label{Fig:SfigS42}
\end{subfigure}
    \caption{ Plotting free energy $F_8$ vs $T$ as temperature and specific heat $\mathcal{C}$ vs x, we get the phase diagram for fixed $(\tilde{Q}_e, \tilde{\Phi}_m, R, \mu)$ ensemble. }
\label{fig:en4a}
\end{figure} 
The free energy $F_8$ and temperature $T$ in \eqref{eq:F8:2}  as a function of $F_8= F_8(\tilde{\Phi}_m, \tilde{Q}_e, R, \mu, x)$ and temperature $T=T(\tilde{\Phi}_m, R, \mu, x)$.
We can display the parametric plot for $F_8$ for different values of $\mu$ keeping rest parameters constant shown in the Figure \ref{Fig:10} using equation \eqref{eq:F8:2}. At $\mu= 0$ (orange) a qualitative change can be seen with the plot in the temperature. For $\mu < 0$ we can see that there is a single-value for the free energy which is a function of temperature. For $0 \leq \mu < \mu_{coin}$ we see that the free energy graph has branches namely two lines at temperatures $T_1$ and $T_2$ where $T_2 \ge T_1$ which is calculated from the roots of the function solving $F_8=0$. The two roots are $x_3$ and $x_4$ which are given in Appendix  \ref{Appendix}. Putting the value of the roots $x_1$ and $x_2$ in the temperature equation \eqref{eq:F8:2}  we get the two temperatures namely $T_1$ and $T_2$ as:-
\begin{equation}
T_1=\frac{144 \left(A_4\right) \left(R \tilde{\Phi }_m-3 R^3 \tilde{\Phi }_m^3\right){}^2 \left(B_4\right){}^{2/3}+432 R \text{$\mu $} \left(1-3 R^2 \tilde{\Phi }_m^2\right){}^3 \left(B_4\right)-R^2 \tilde{\Phi }_m^2 \left(A_4\right){}^3+\left(A_4\right){}^3}{B_4}
\end{equation}
\begin{equation}
T_2=
\frac{144 \left(A_5\right) \left(R \tilde{\Phi }_m-3 R^3 \tilde{\Phi }_m^3\right){}^2 \left(B_5\right)^{2/3}+432 R \text{$\mu $} \left(1-3 R^2 \tilde{\Phi }_m^2\right){}^3 \left(B_5\right)-R^2 \tilde{\Phi }_m^2 \left(A_5\right){}^3+\left(A_5\right){}^3}{B_5}
\end{equation}
where $A_4$, $B_4$, $A_5$, $B_5$ are in Appendix \ref{Appendix}. Further if we put $\mu = 0$, $\tilde{\Phi}_m=1$ we get $T_1=3.16698$ and  $T_2=\infty$. Further carrying on we notice that the two temperatures $T_1$ and $T_2$ are the same at the coincidence point, $\mu_{coin} =3.849$ for the Figure \ref{Fig:10} so that the coincidence point is given as $T_{coin} =2.75664$. We have not obtained any plot for $F_8(T)$ when $\mu>\mu_{coin}$. So only the plots when $\mu<\mu_{coin}$ are obtained.\\
In Figure \ref{Fig:10}, using the purple curve, we see that $T_2$ starts decreasing when $x$ increases untill the turning point $T_0$ after which it starts increasing. The plot between $T_0$ and $T_2$ is ow entropy branch and the plot between $T_1$ and $T_0$ is high entropy branch which also be seen in Figure \ref{Fig:SfigS42} as small (unstable)-large(stable) entropy branch. We solve $\left(\frac{dT}{dx}\right)_\mu=0$ for $x=x_{01}$ and putting it back in \eqref{eq:F8:2} we get $T_{01}=2.07463$ (purple). This is also the point where we see a Davies type phase transition.
The expression for $x_{01}$ and $T_{01}$ are in Appendix \ref{Appendix}.

 \subsection{For the fixed $(\tilde{\Phi}_e,\tilde{\Phi}_m,\mathcal{V},C)$ ensemble}
  \label{subsec:Ensemble11}
We fix the potentials $\tilde{\Phi}_e$, $\tilde{\Phi}_m$ as well as the central charge $C$ making it an ensemble such like $(\tilde{\Phi}_e,\tilde{\Phi}_m,\mathcal{V},C)$. Therefore the Gibbs energy namely as $F_{11}$ can be calculated as follows:-
 \begin{equation}
 \label{eq:F11:1}
 F_{11}=E-TS-\tilde{\Phi}_e \tilde{Q}_e-\tilde{\Phi}_m \tilde{Q}_m=\mu C
 \end{equation}
  The CFT first law in \eqref{eq:dE:3} and by differentiating $F_{11}$ is given as:-
 \begin{equation}
 \label{eq:dF11}
 \begin{split}
 & dF_{11}=dE-TdS-SdT-\tilde{\Phi}_e d\tilde{Q}_e-\tilde{Q}_e d\tilde{\Phi}_m-\tilde{\Phi}_m d\tilde{Q}_m-\tilde{Q}_m d\tilde{\Phi}_m\\
 & dF_{11}=-SdT-\tilde{Q}_e d\tilde{\Phi}_e-\tilde{Q}_m d\tilde{\Phi}_m+p\mathcal{V}+\mu dC
 \end{split}
 \end{equation}
 From equation \eqref{eq:dF11} we get that $F_{11}$ is stationary for $(\tilde{\Phi}_e, \tilde{\Phi}_m, \mathcal{V}, C)$. Writing the free energy $F_{11}=\mu C$ by putting the value of $\mu$ from \eqref{eq:mu3} and also the temperature $T$ from \eqref{eq:E3:T3} in the language of $\tilde{\Phi}_e$ and $\tilde{\Phi}_m$ using the formulas $y=\tilde{\Phi}_e Rx$ and $z=\tilde{\Phi}_mRx$. Hence we get the free energy $F_{11}$ and temperaure $T$ as:-
 \begin{equation}
 \label{eq:F11:2}
 F_{11}=\frac{C x \left(-4 R^2 \left({\tilde{\Phi}_e}^2+{\tilde{\Phi} _m}^2\right)+x^2 \left(3 R^2 \left({\tilde{\Phi}_e}^2+{\tilde{\Phi}_m}^2\right)-1\right)+1\right)}{R}, \quad
 T=\frac{1-x^2 \left(R^2 \left(\tilde{\Phi}_e^2+\tilde{\Phi}_m^2\right)-3\right)}{4 \pi  R x}
 \end{equation}
 We plot the parametric plot between Gibbs free energy $F_{11}$ and Temperature $T$ keeping $C$ and $\tilde{\Phi}_m$ constant and varying $\tilde{\Phi}_e$, we get a plot and it is given below:-
 \begin{figure}[htp]
\centering
\begin{subfigure}[b]{0.3\textwidth} 
\includegraphics[width=1\textwidth]{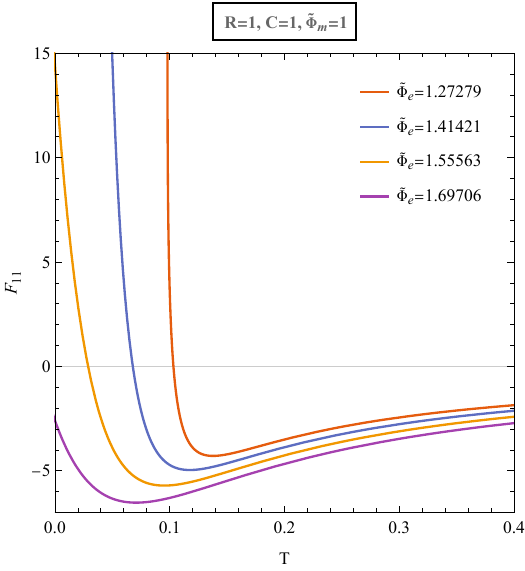}
\caption{}
\label{Fig:F11T}
\end{subfigure}
\begin{subfigure}[b]{0.3\textwidth}
\includegraphics[width=1\textwidth]{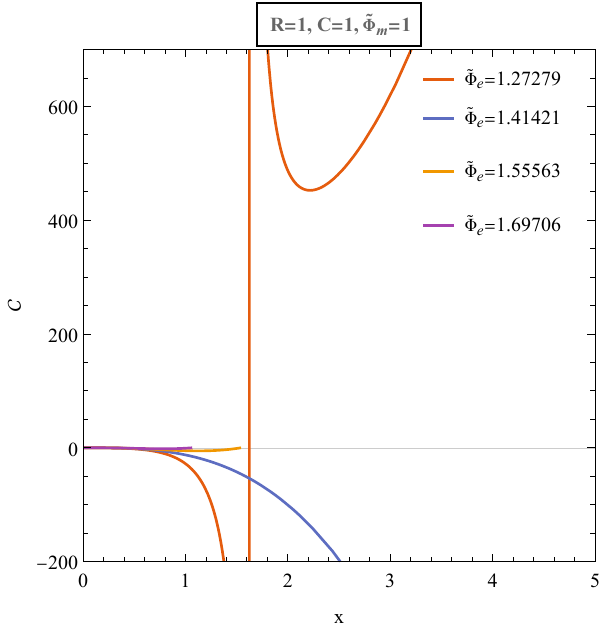}
\caption{}
\label{Fig:jfig1}
\end{subfigure}
\begin{subfigure}[b]{0.3\textwidth}
\includegraphics[width=1\textwidth]{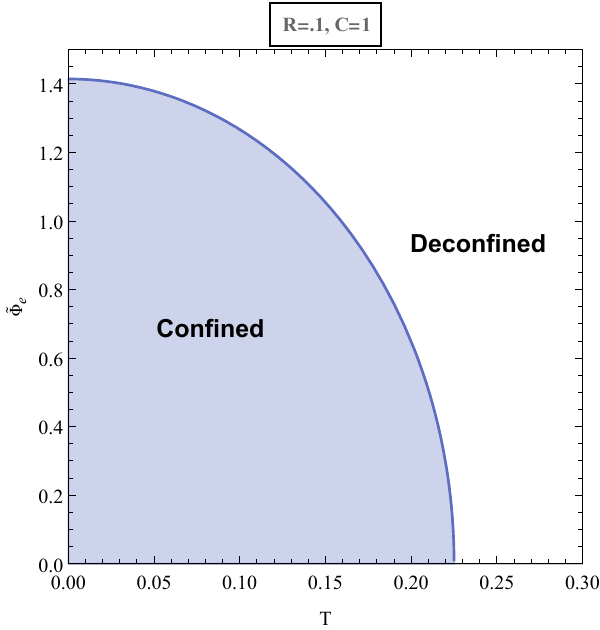}
\caption{}
\label{Fig:PhieT}
\end{subfigure}
    \caption{ Plotting free energy $F_{11}$ vs $T$ as temperature, specific heat $\mathcal{C}$ vs x and coexistence plots we get the phase diagram for fixed $(\tilde{\Phi}_e, \tilde{\Phi}_m, R, C)$ ensemble. }
\label{fig:en11}
\end{figure} 

 As seen in figure \ref{fig:en11}, this enables us to parametrically depict the free energy $F_{11}(T)$ using $x$ parameter for the ensemble at fixed $(\tilde{\Phi}_e,\tilde{\Phi}_m, R, C)$. Different behaviour is portrayed on the $F_{11}-T$ plot crossing and staying below a specific crucial value of potential $\tilde{\Phi}_m=1.55563$. While $F_{11} \le 0$ and the graph crosses the $F_{11}$ axis for $\tilde{\Phi}_m \ge\tilde{\Phi}_m^{crit}$ (blue, yellow, purple), we obtain a single value of free energy with the temperature as its function. In contrast, the free energy curve for $\tilde{\Phi}_m<\tilde{\Phi}_m^{crit}$ (red) has an higher and lower branch that seems to converge at a turning point or cusp, representing, small black holes corresponding to low entropy  and the big black holes corresponding to high entropy  states. The temperature reaches a minimum value for the function of $x$ by solving $\left(\frac{dT}{dx}\right)=0$ getting $ x=1/\sqrt{-R^2 \tilde{\Phi}_e^2-R^2 \tilde{\Phi}_m^2+3}
$. Putting the value of $x$ in \eqref{eq:F11:2} we get the minimum attained  temperature as:-
\begin{equation}
T_{min}=\frac{\sqrt{3-R^2 \left(\tilde{\Phi}_e^2+\tilde{\Phi}_m^2\right)}}{2 \pi  R}
\end{equation}
 For the red plot, the turning point is $x=1.62221$ and the minimum temperature to be achieved is $T_{min}=0.0981097$. We see a change of phase of the order of one known as (de)confined which is indicated by the lower branch's free energy changing sign at $F_{11}=0$. While the "confined" state is thermodynamically favoured when $F_{11}>0$, the huge entropy "deconfined" state dominates the ensemble at $F_{11}<0$. These phase transitions between (de)confined and huge black holes in the AdS space and the AdS space-time with thermal radiation \cite{d,e} have some similarity to a generalized transition of the Hawking Page phase. Although the Hawking page transition was first identified for charged AdS Schwarzschild black holes \cite{d}, it also occurs for charged black holes in the AdS spacetime and is dependent on the electric potential \cite{f}. At the Figure \ref{Fig:jfig1} for the subcritical value at at the turning point $x=1.62221$, we see a Davies type phase transition on top of the (de)confined type transition. At \ref{Fig:PhieT}, we draw the coexistence line $\tilde{\Phi}_e-T$ plane along which the first order phase transition for (de)confined phase occur. We note that at $T=0$, the transition occurs at $\tilde{\Phi}_e=1.41421$ which is the critical value. 

 \subsection{For the fixed $(\tilde{Q}_e,\tilde{Q}_m,\mathcal{V},C)$ ensemble}
  \label{subsec:Ensemble15}
We fixed both the electric charge $\tilde{Q}_e$, magnetic charge $\tilde{Q}_m$, CFT volume $\mathcal{V}$ and central charge $C$. The free energy is written as:-
 \begin{equation}
 \label{eq:F15:1}
 F_{15}\equiv E-TS=\tilde{\Phi}_e\tilde{Q}_e+\tilde{\Phi}_m\tilde{Q}_m+\mu C
 \end{equation}
By differentiating $F_{15}$ and using \eqref{eq:dE:3} it can be expressed as:-
 \begin{equation}
 \label{eq:dF15}
     \begin{split}
         &dF_{15}=dE-TdS-SdT
        =-SdT+\tilde{\phi}_e d\tilde{Q}_e+\tilde{\Phi}_m d\tilde{Q}_m+p d\mathcal{V}-\mu dC
     \end{split}
 \end{equation}
We can see that the free energy $F_{15}$ is static for the fixed $(T,\tilde{Q}_e, \tilde{Q}_m,\mathcal{V}, C)$ ensembles. We write equation $F_{15}$ \eqref{eq:F15:1} and temperature $T$ by using \eqref{eq:E4:T4}, \eqref{eq:S:1}, by taking values from the relation $y=\frac{\tilde{Q}_e}{4 C}$ and $z=\frac{\tilde{Q}_m}{4 C}$, so we get:-
 \begin{equation}
 \label{eq:F15:2}
 F_{15}=\frac{3 \left(\tilde{Q}_e^2+\tilde{Q}_m^2\right)}{16 C R x}+\frac{C \left(x-x^3\right)}{R}, \quad
T= -\frac{-16 C^2 x^2 \left(3 x^2+1\right)+\tilde{Q}_e^2+\tilde{Q}_m^2}{64 \pi  C^2 R x^3}
 \end{equation}
As a result, using $x$ as a variable for the fixed $(T, \tilde{Q}_m, \tilde{Q}_e, \mathcal{V}, C)$, we parametrically show $F_{15}(T)$ in Figure \ref{fig:en7}. 
\begin{figure}[htp]
\centering
\begin{subfigure}[b]{0.3\textwidth} 
\includegraphics[width=1\textwidth]{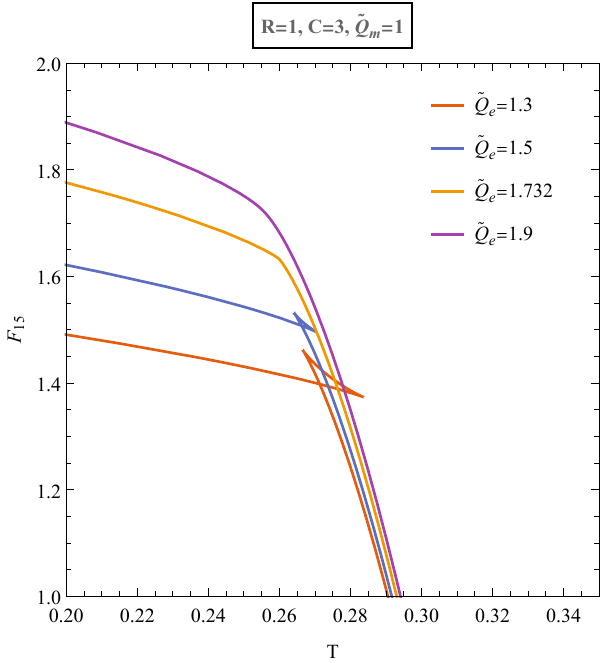}
\caption{}
\label{Fig:16}
\end{subfigure}
\begin{subfigure}[b]{0.3\textwidth}
\includegraphics[width=1\textwidth]{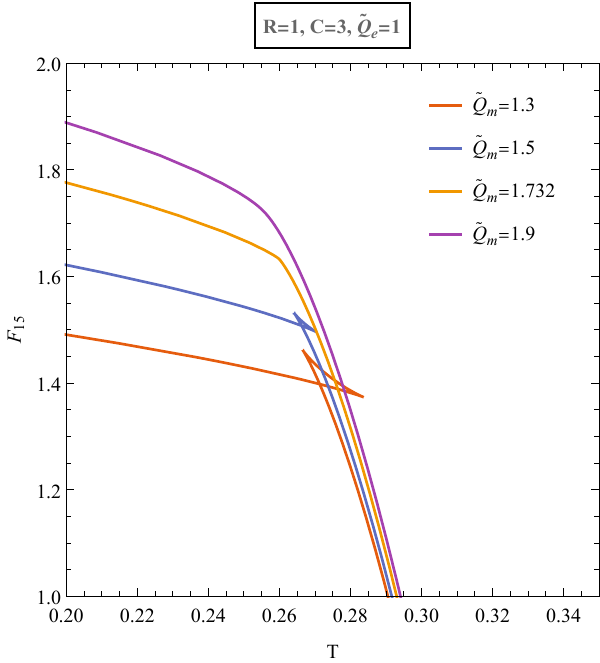}
\caption{}
\label{Fig:17}
\end{subfigure}
\begin{subfigure}[b]{0.3\textwidth} 
\includegraphics[width=1\textwidth]{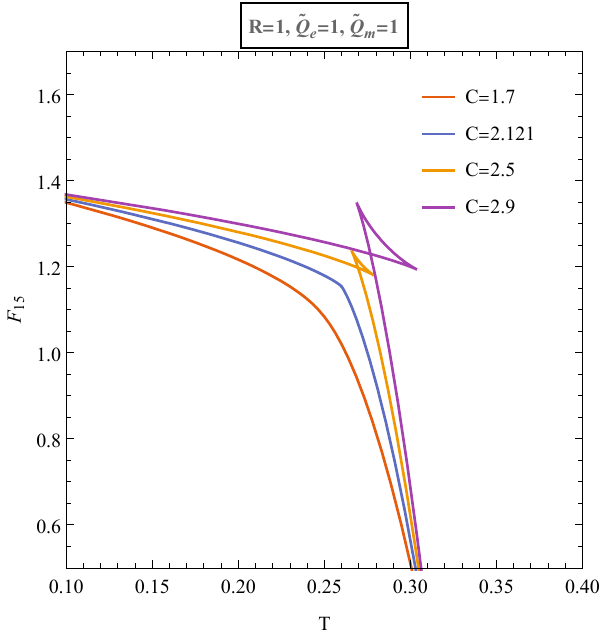}
\caption{}
\label{Fig:18}
\end{subfigure}
    \caption{Free energy $F_{15}$ vs. temperature $T$ plot for the fixed $(\tilde{Q}_e,\tilde{Q}_m,R,C)$ ensemble in $(D=4/d=3)$.  }
\label{fig:en7}
\end{figure}

\begin{figure}[htp]
\centering
\begin{subfigure}[b]{0.3\textwidth} 
\includegraphics[width=1\textwidth]{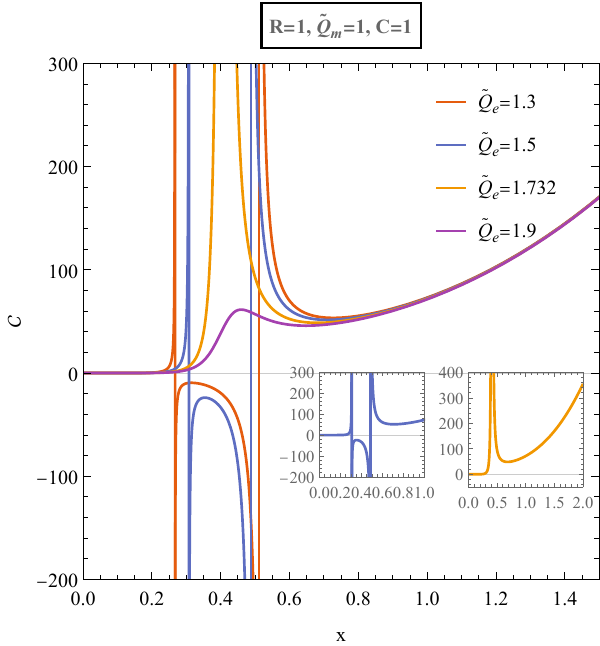}
\caption{}
\label{Fig:Vfig1}
\end{subfigure}
\begin{subfigure}[b]{0.3\textwidth}
\includegraphics[width=1\textwidth]{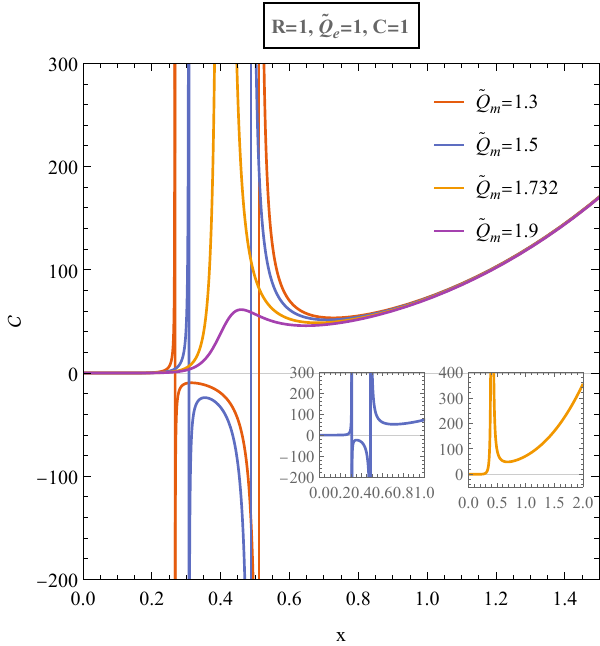}
\caption{}
\label{Fig:Vfig2}
\end{subfigure}
\begin{subfigure}[b]{0.3\textwidth} 
\includegraphics[width=1\textwidth]{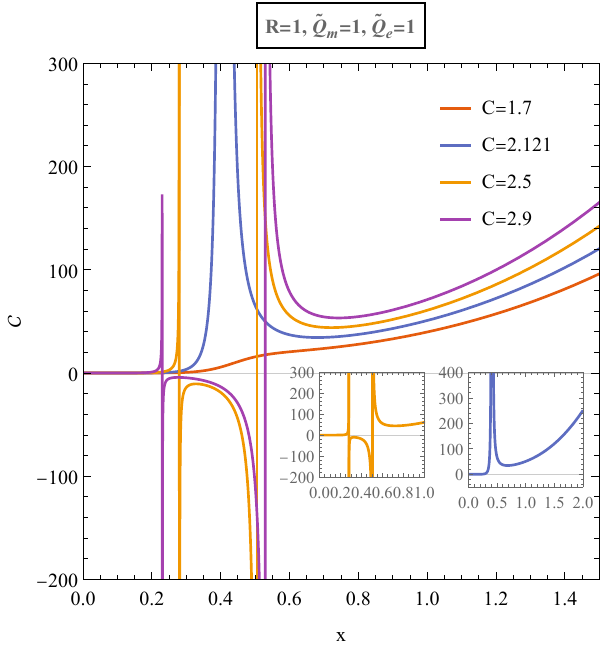}
\caption{}
\label{Fig:Vfig3}
\end{subfigure}
    \caption{Specific heat $\mathcal{C}$ vs. $x$ plot for the fixed $(\tilde{Q}_e,\tilde{Q}_m,R,C)$ ensemble in $(D=4/d=3)$.  }
\label{fig:en7}
\end{figure}
In Figure \ref{Fig:16} and \ref{Fig:Vfig1}, we plot for various values of electric charge $\mathcal{Q}_e$ keeping rest parameters constant. For sub critical values of $\tilde{Q}_e$ (red, blue) we see Van der Waals type phase transition of order one which looks like a swallowtail having three branches namely small (stable)-intermediate (unstable)-large(stable) branches where stabilities are seen in Figure \ref{Fig:Vfig1}. At the critical point $\tilde{Q}_e=1.732$ (orange) we see a kink in the $F_{15}(T)$ plot and inverse $\lambda$ like structure signalling a superfluid $\lambda$ phase transition of order two. For supercritical values of $\tilde{Q}_e$ we see no phase transition.\\
In Figure \ref{Fig:17} and \ref{Fig:Vfig2} we plot for various values of magnetic charge where for sub critical values of $\tilde{Q}_m$ (red, blue) we see Van der Waals type phase transition showing three branches namely small (stable) -intermediate (unstable)-large (stable) branches. At the critical point $\tilde{Q}_e=1.732$ (orange) we see a superfluid $\lambda$ phase transition of order two and at super critical values we get no phase transitions (purple).\\
\begin{figure}[htp]
\centering
\begin{subfigure}[b]{0.3\textwidth} 
\includegraphics[width=1\textwidth]{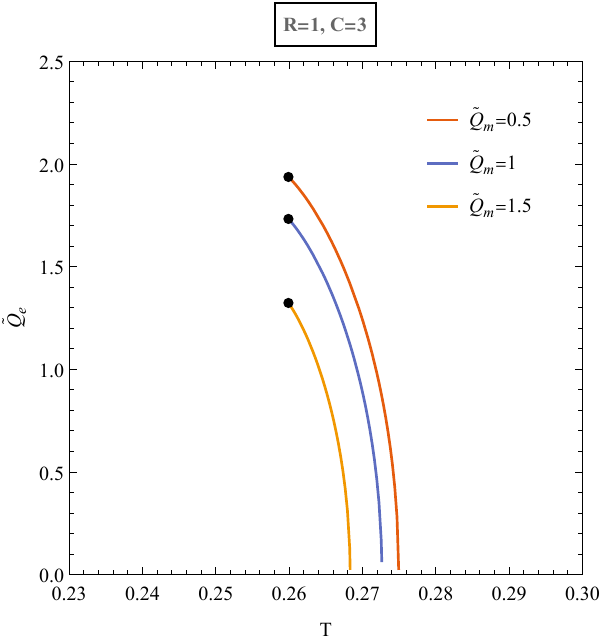}
\caption{}
\label{Fig:QeTcoexist1}
\end{subfigure}
\begin{subfigure}[b]{0.3\textwidth}
\includegraphics[width=1\textwidth]{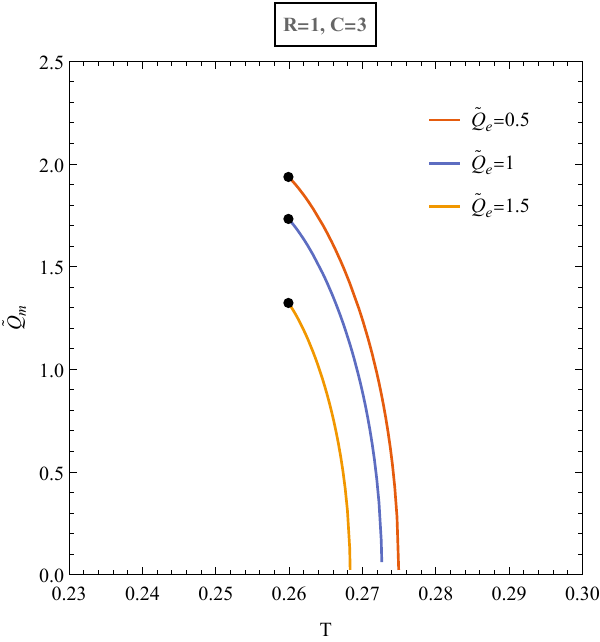}
\caption{}
\label{Fig:QmTcoexist1}
\end{subfigure}
\begin{subfigure}[b]{0.3\textwidth} 
\includegraphics[width=1\textwidth]{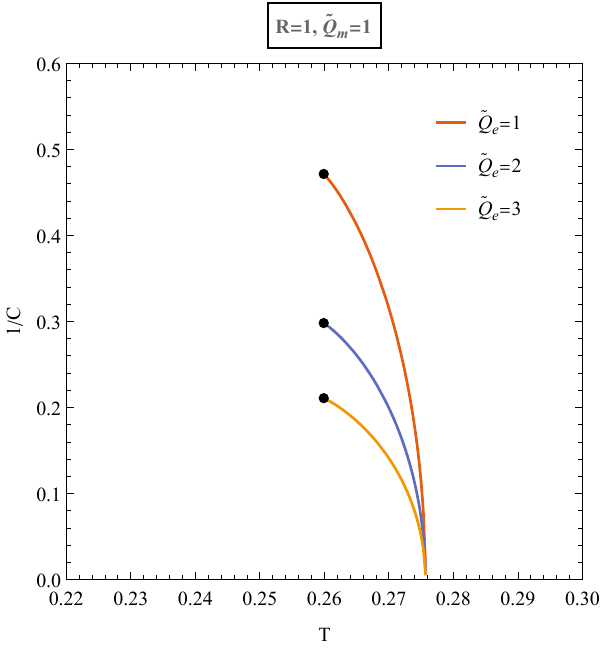}
\caption{}
\label{Fig:1byctcoexist2}
\end{subfigure}
    \caption{Coexistence Plots  }
\label{fig:coexistenceplots2}
\end{figure}
Lastly, for various values of $C$ we see in Figure \ref{Fig:18} and \ref{Fig:Vfig3} that super critical values (purple, orange) we see Van der Waals type phase transition showing three branches namely small (stable)-intermediate (unstable)-large (stable) branches. At the critical value $C=2.121$ (blue) we see a superfluid $\lambda$ phase transition of order two and for sub critical values of $C$ (red) we see no phase transition.\\
In Figure \ref{fig:coexistenceplots2} we plot the coexistence lines for the low and high entropy regimes for the CFT on the $\tilde{Q}_e-T$, $\tilde{Q}_m-T$ and $1/C-T$ phase plots. The coexistence line delineates the two phases on the planes, and the CFT experiences a first-order phase transition when it intersects this line. In these phase diagrams, the phase with lower entropy is found to the left of the coexistence line, whereas the phase with higher entropy is situated to the right. The critical points are represented by black dots on the diagrams. Beyond these critical points, the CFT does not exhibit distinct phases.

 \subsection{For the fixed $(\tilde{Q}_e, \tilde{Q}_m,\mathcal{V},\mu)$ ensemble}
  \label{subsec:Ensemble16}
 We try to fix the electric charge $\tilde{Q}_e$ and magnetic charges $\tilde{Q}_m$ as well as the chemical potential $\mu$ and obtaining the ensemble $(\tilde{Q}_e, \tilde{Q}_m,\mathcal{V},\mu)$. Hence, the Gibbs energy is obtained as below:-
 \begin{equation}
 \label{eq:F16:1}
 F_{16}\equiv E-TS-\mu C=\tilde{\Phi}_e \tilde{Q}_e+\tilde{\Phi}_m \tilde{Q}_m=\frac{4 C \left(y^2+z^2\right)}{R x}
 \end{equation}
After differentiating $F_{16}$ we can write it as:-
 \begin{equation}
 \label{eq:dF16}
     \begin{split}
        & dF_{16}=dE-TdS-SdT-\mu dC-C d\mu
        =-SdT-C d\mu +\tilde{\Phi}_e d\tilde{Q}_e+\tilde{\Phi}_m d\tilde{Q}_m-p \mathcal{V}
     \end{split}
 \end{equation}
 Expressing the free energy $F_{16}$ with respect to the required ensemble, we need to incorporate the chemical potential $\mu$ from \eqref{eq:mu4} by solving it for $C$ where $C=\frac{\sqrt{-\tilde{Q}_e^2-\tilde{Q}_m^2}}{4 \sqrt{\mu  R x+x^4-x^2}}$ where $y= \frac{\tilde{Q}_e}{4 C},z= \frac{\tilde{Q}_m}{4 C}$ in the $\mu$ equation and  then putting it free energy $F_{16}$ \eqref{eq:dF16} and temperature \eqref{eq:E4:T4}  we get:-
 \begin{equation}
 \label{eq:F16:2}
F_{16}=-\frac{\sqrt{-\tilde{Q}_e^2-\tilde{Q}_m^2} \sqrt{x \left(\mu  R+x^3-x\right)}}{R x}, \quad
T=\frac{\mu +\frac{4 x^3}{R}}{4 \pi  x^2}
\end{equation}

\begin{figure}
\centering
\begin{subfigure}[b]{0.3\textwidth} 
\includegraphics[width=1\textwidth]{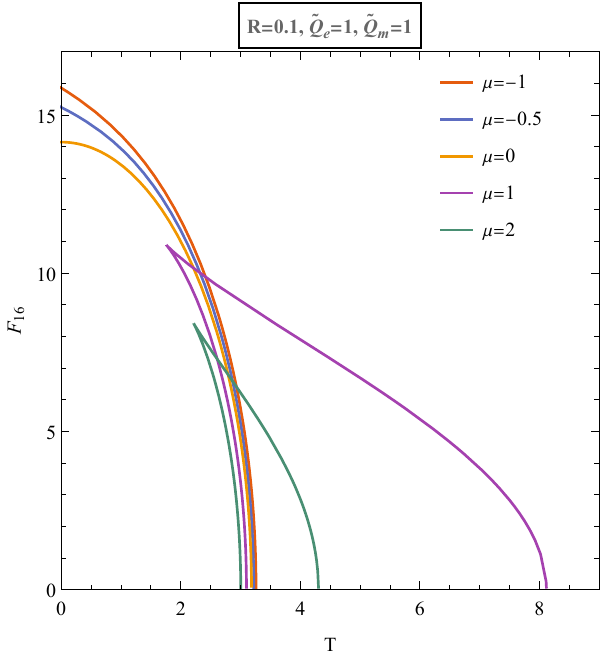}
\caption{}
\end{subfigure}
\begin{subfigure}[b]{0.3\textwidth}
\includegraphics[width=1\textwidth]{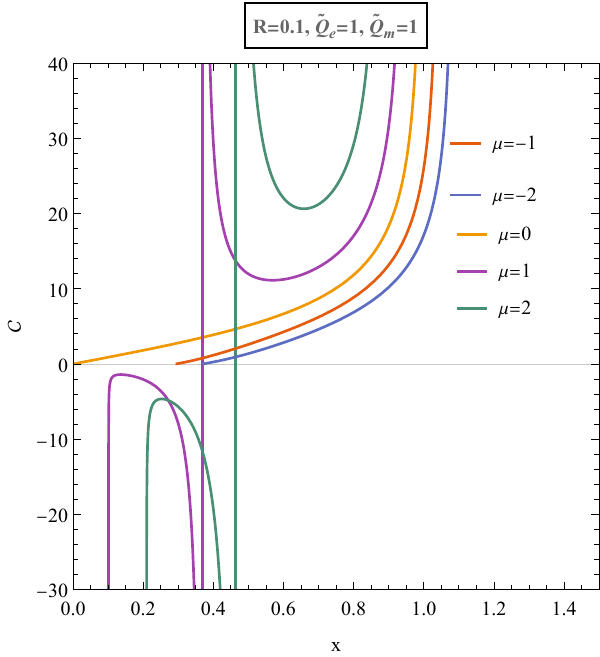}
\caption{}
\end{subfigure}
    \caption{Free energy $F_{15}$ vs. temperature $T$ plot for the fixed $(\tilde{Q}_e,\tilde{Q}_m,R,C)$ ensemble in $(D=4/d=3)$.  }
 \label{fig:en:161}
\end{figure}

The free energy $F_{16}$ and the temperature T given in \eqref{eq:F16:2} is shown as a function of $F_{16}=F_{16}(\tilde{Q}_e, \tilde{Q}_m, R, \mu, x)$ and temperature $T=T(\mu, R, x)$. We can show the parametric plot for $F_{16}$ for the different values of chemical potential $\mu$ keeping the rest parameters same as shown in the Figure \ref{fig:en:161} using equation \eqref{eq:F16:2}.

At $\mu=0$ (orange) we can see a qualitative change which can be seen both in the $F_{16}-T$ plot and $C-x$ plots. For $\mu<0$ we see that there is a single-valued for the free energy which is the function of temperature. For $0 \leq \mu<\mu_{coin}$ we get two branches namely two lines at temperatures $T_1$ and $T_2$ where $T_2 \leq T_1$ which is calculated by solving $F_{16}=0$. We get the two roots $x_5$ and $x_6$ which are given in Appendix \ref{Appendix}. Puting the roots back in the temperature we get the two temperatures as:-
 \begin{equation}
 T_1=\frac{6 \sqrt[3]{6} \left(\sqrt{81 \mu ^2 R^2-12}-9 \mu  R\right)^{2/3} \left(\mu +\frac{\left(i \sqrt[3]{2} \left(\sqrt{3}+i\right) \left(\sqrt{81 \mu ^2 R^2-12}-9 \mu  R\right)^{2/3}-2 \sqrt[3]{3} \left(1+i \sqrt{3}\right)\right)^3}{72 R \left(\sqrt{81 \mu ^2 R^2-12}-9 \mu  R\right)}\right)}{\pi  \left(2 \left(\sqrt[3]{3}+i 3^{5/6}\right)-i \sqrt[3]{2} \left(\sqrt{3}+i\right) \left(\sqrt{81 \mu ^2 R^2-12}-9 \mu  R\right)^{2/3}\right)^2}
 \end{equation}
 \begin{equation}
 T_2=\frac{6 \sqrt[3]{6} \left(\sqrt{81 \mu ^2 R^2-12}-9 \mu  R\right)^{2/3} \left(\mu +\frac{\left(2 i \sqrt[3]{3} \left(\sqrt{3}+i\right)-\sqrt[3]{2} \left(1+i \sqrt{3}\right) \left(\sqrt{81 \mu ^2 R^2-12}-9 \mu  R\right)^{2/3}\right)^3}{72 R \left(\sqrt{81 \mu ^2 R^2-12}-9 \mu  R\right)}\right)}{\pi  \left(2 i \sqrt[3]{3} \left(\sqrt{3}+i\right)-\sqrt[3]{2} \left(1+i \sqrt{3}\right) \left(\sqrt{81 \mu ^2 R^2-12}-9 \mu  R\right)^{2/3}\right)^2}
 \end{equation}
Putting  $\mu=0$ we get $T_1=3.1831$ and $T_2=0$ (orange). At the coincidence point $\mu_{coin}=3.849$ the two temperatures $T_1$ and $T_2$ are the same so the temperature at the coincidence point is given as $T_0=2.75664$. We do not see any plot for $\mu>\mu_{coin}$. So only the plots when $\mu<\mu_{coin}$ are obtained.\\
In Figure \ref{fig:en:161}, using the purple curve, we see that $T_2$ starts decreasing when x increases untill it reaches the turning point $T_0$ after which it starts increasing. The plot between $T_0$ and $T_2$ is low entropy branch and the branch between $T_1$ and $T_0$ is high enropy which is seen in the Figure as small (unstable)- large (stable) entropy branch. We solve
$\left(\frac{dT}{dx}\right)_\mu=0$ for the value of $x=x_{02}$ to get temperature $T_{02}$ which ia given as:-
\begin{equation}
T_{02}=\frac{3 \sqrt[3]{\mu}}{2 \sqrt[3]{2} \pi  R^{2/3}}, \quad x_{02}=\frac{\sqrt[3]{\mu} \sqrt[3]{R}}{\sqrt[3]{2}}
\end{equation}
For the purple plot the turning point is $T_{02}=1.759$. This is also the point where we see a Davies type phase transition.
\begin{figure}
\centering
\begin{subfigure}[b]{0.2\textwidth} 
\includegraphics[width=1\textwidth]{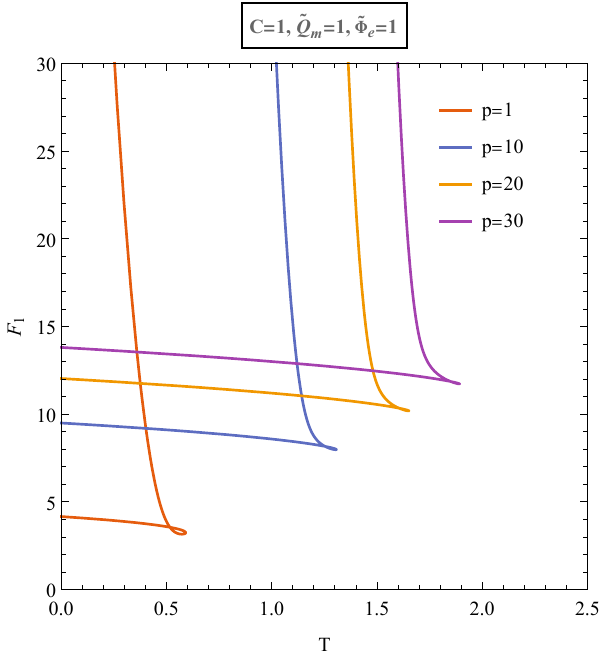}
\caption{}
\label{fig:plot1}
\end{subfigure}
\begin{subfigure}[b]{0.21\textwidth}
\includegraphics[width=1\textwidth]{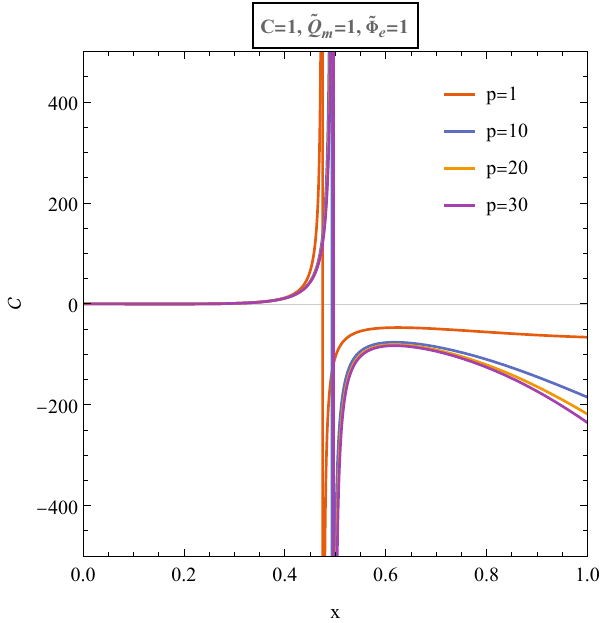}
\caption{}
\label{fig:plot1aa}
\end{subfigure}
\begin{subfigure}[b]{0.2\textwidth} 
\includegraphics[width=1\textwidth]{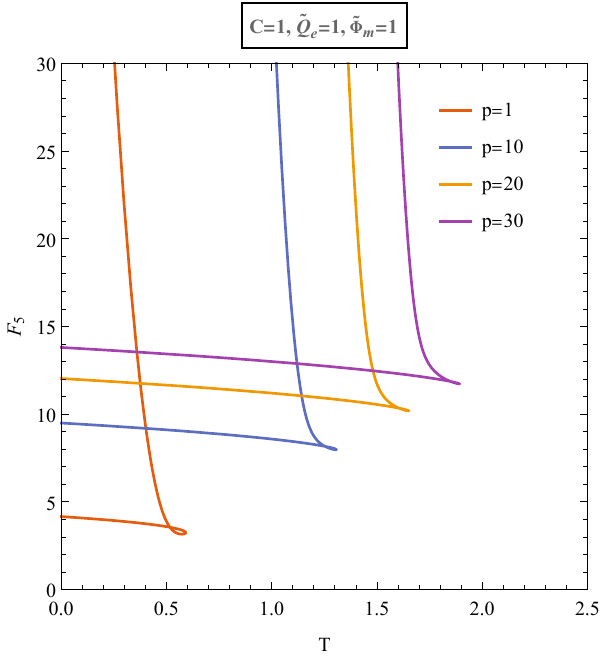}
\caption{}
\label{fig:plot2}
\end{subfigure}
\begin{subfigure}[b]{0.21\textwidth}
\includegraphics[width=1\textwidth]{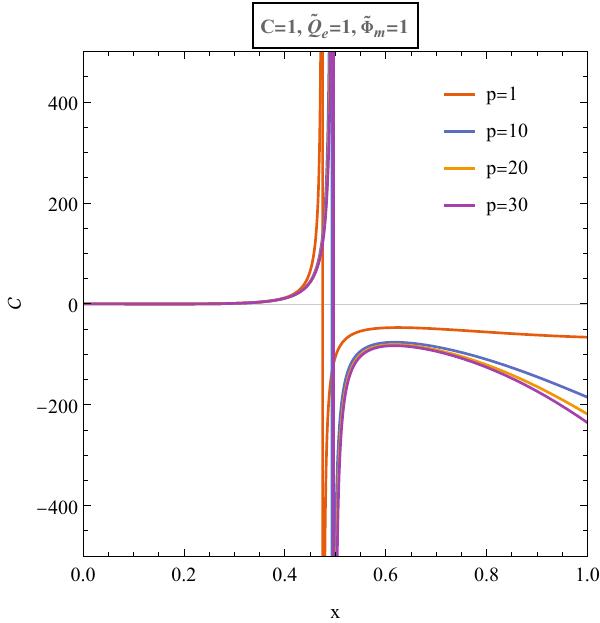}
\caption{}
\label{fig:plot1bb}
\end{subfigure}
\begin{subfigure}[b]{0.2\textwidth} 
\includegraphics[width=1\textwidth]{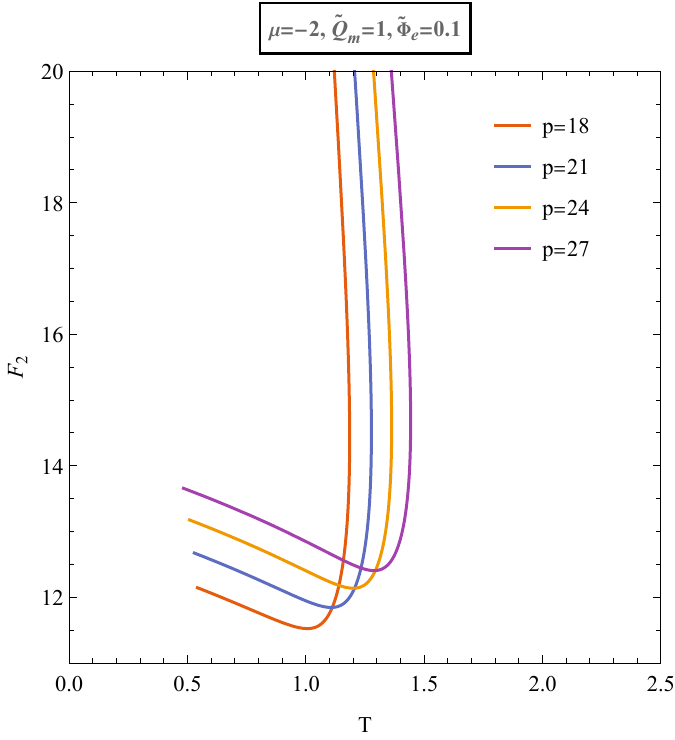}
\caption{}
\label{fig:plot3}
\end{subfigure}
\begin{subfigure}[b]{0.21\textwidth}
\includegraphics[width=1\textwidth]{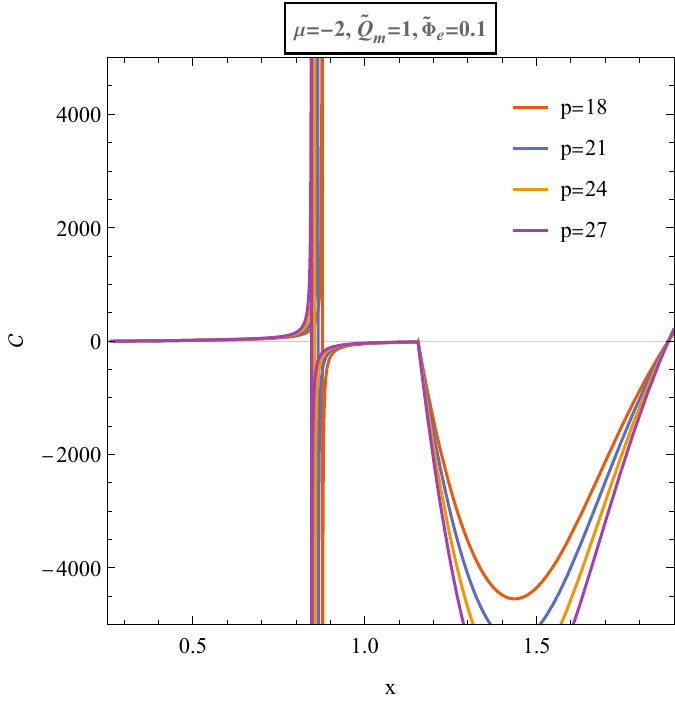}
\caption{}
\label{fig:plot1cc}
\end{subfigure}
\begin{subfigure}[b]{0.2\textwidth} 
\includegraphics[width=1\textwidth]{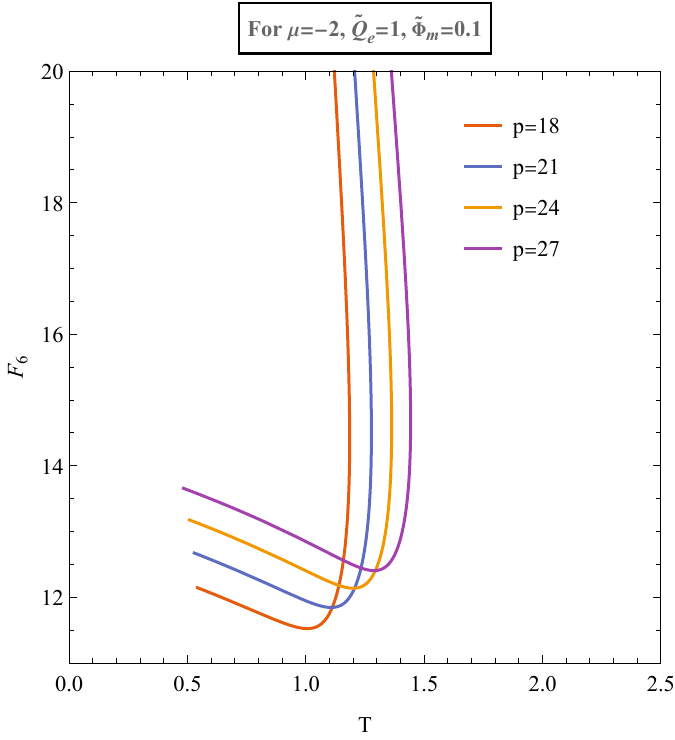}
\caption{}
\label{fig:plot4}
\end{subfigure}
\begin{subfigure}[b]{0.21\textwidth}
\includegraphics[width=1\textwidth]{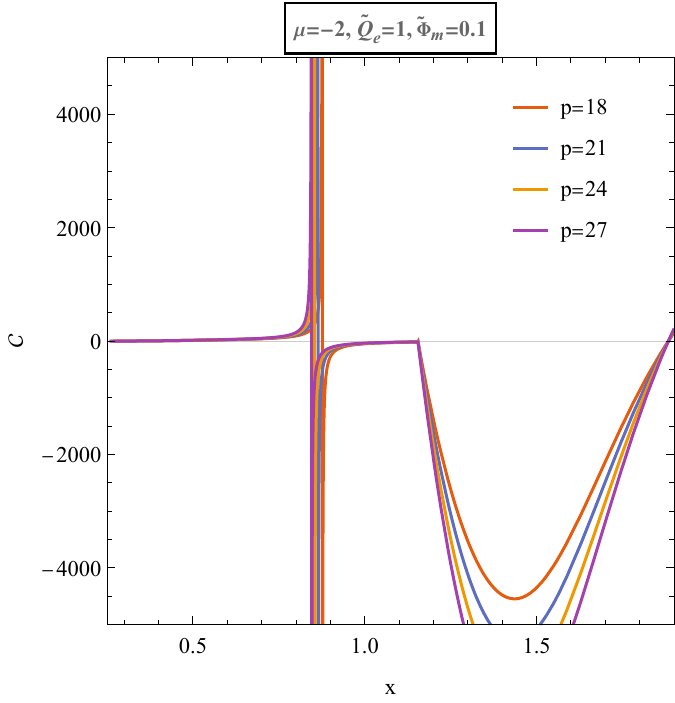}
\caption{}
\label{fig:plot1dd}
\end{subfigure}
\begin{subfigure}[b]{0.2\textwidth} 
\includegraphics[width=1\textwidth]{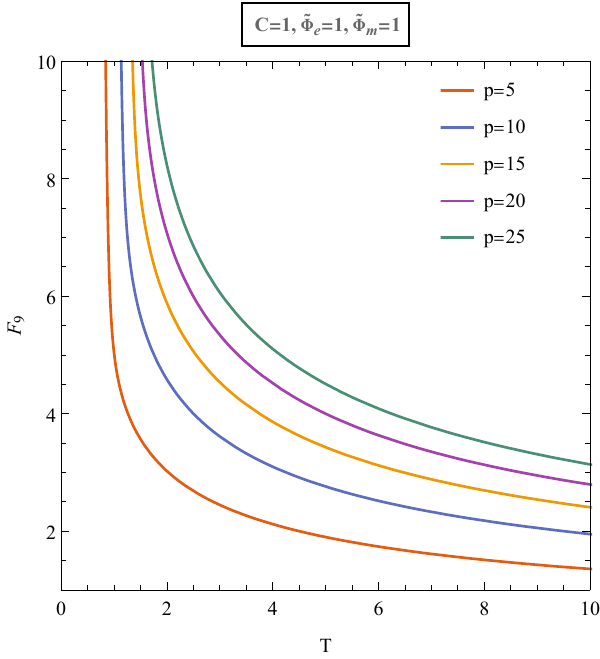}
\caption{}
\label{fig:plot5}
\end{subfigure}
\begin{subfigure}[b]{0.21\textwidth}
\includegraphics[width=1\textwidth]{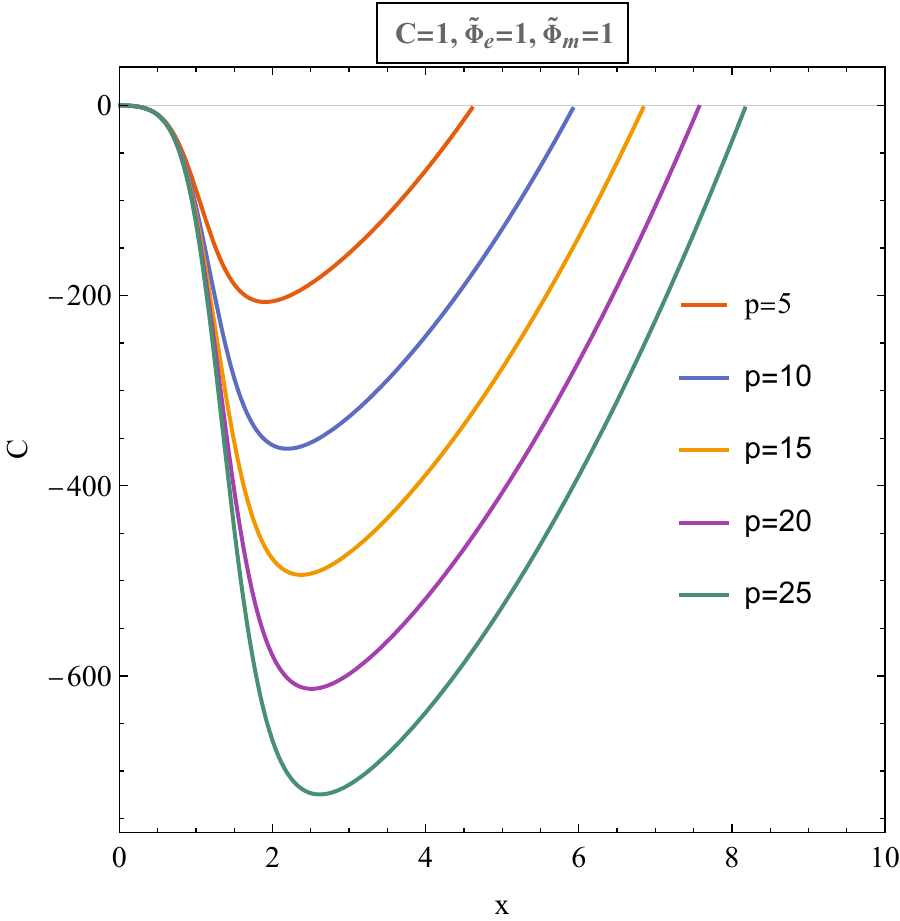}
\caption{}
\label{fig:plot1ee}
\end{subfigure}
\begin{subfigure}[b]{0.2\textwidth} 
\includegraphics[width=1\textwidth]{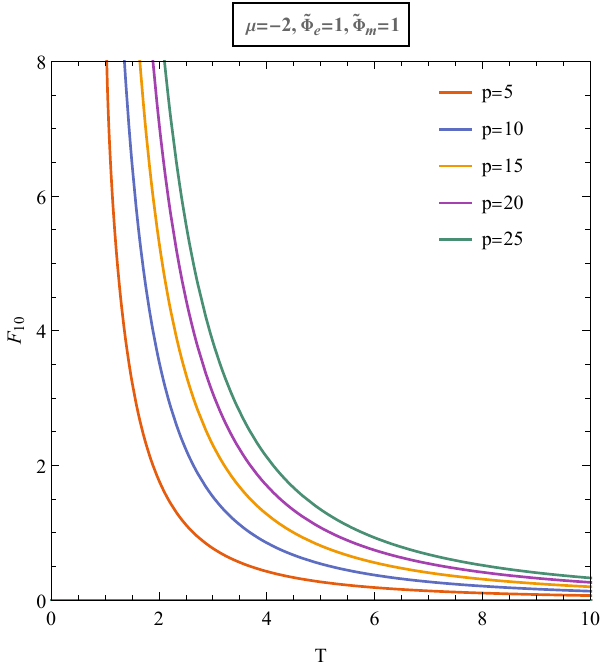}
\caption{}
\label{fig:plot6}
\end{subfigure}
\begin{subfigure}[b]{0.21\textwidth}
\includegraphics[width=1\textwidth]{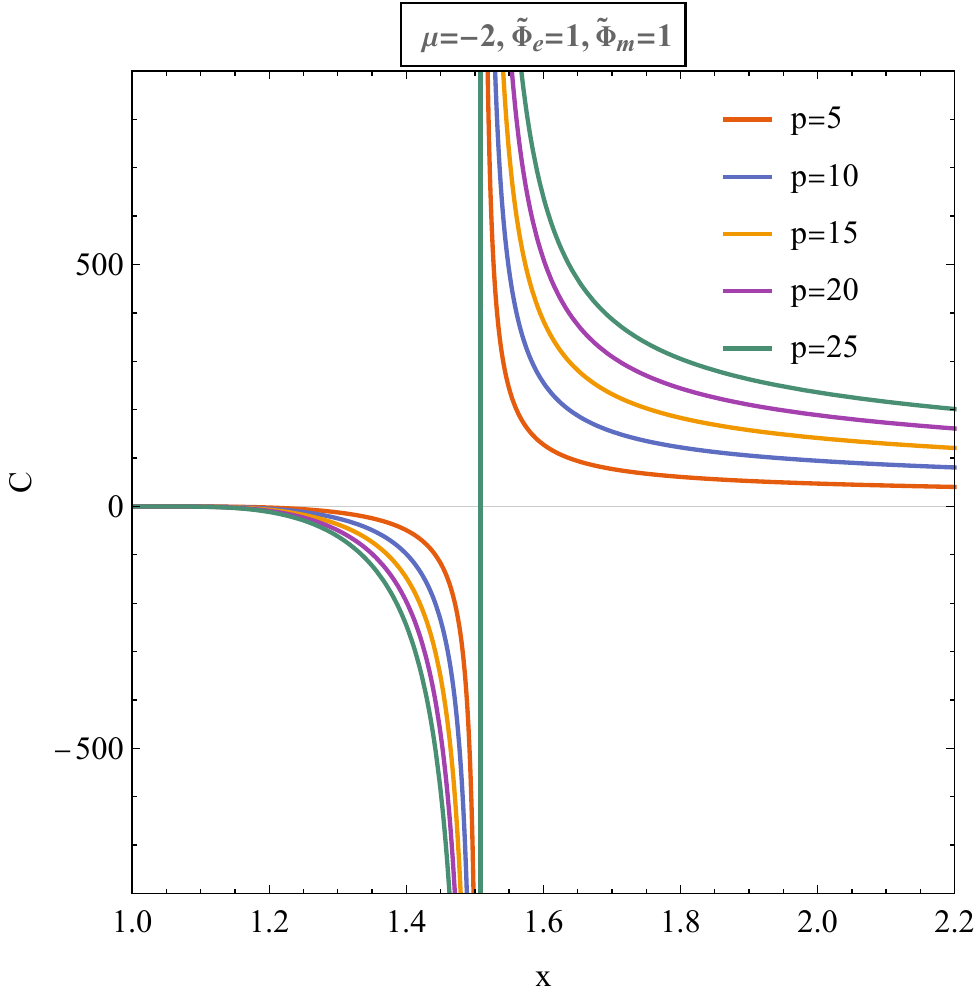}
\caption{}
\label{fig:plot1ff}
\end{subfigure}
\begin{subfigure}[b]{0.2\textwidth} 
\includegraphics[width=1\textwidth,]{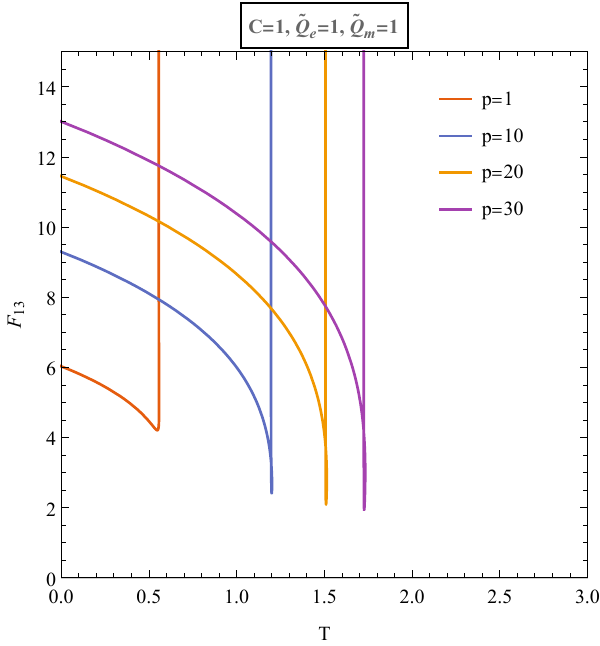}
\caption{}
\label{fig:plot7}
\end{subfigure}
\begin{subfigure}[b]{0.21\textwidth} 
\includegraphics[width=1\textwidth,]{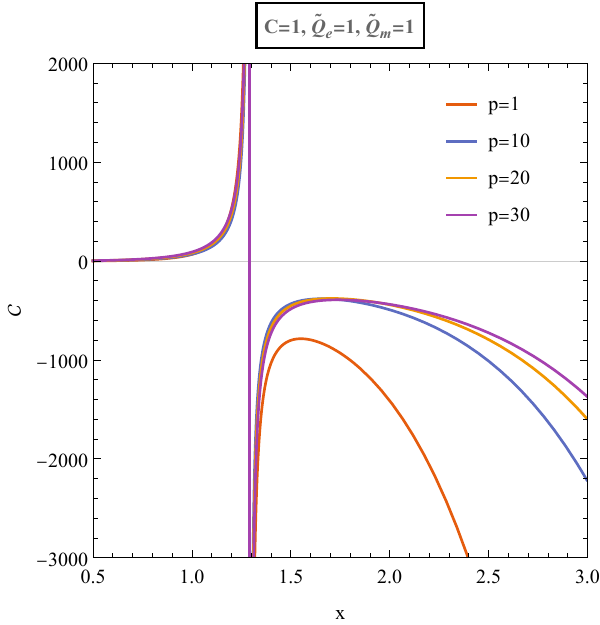}
\caption{}
\label{fig:plot1gg}
\end{subfigure}
\begin{subfigure}[b]{0.2\textwidth} 
\includegraphics[width=1\textwidth]{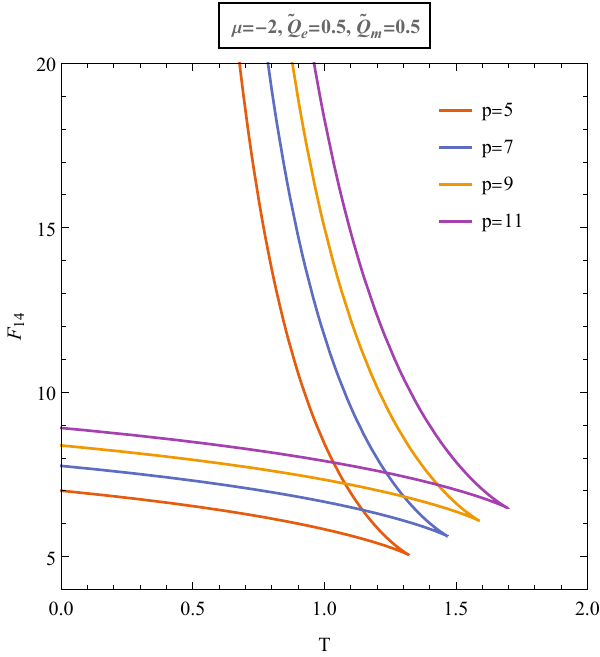}
\caption{}
\label{fig:plot8}
\end{subfigure}
\begin{subfigure}[b]{0.21\textwidth} 
\includegraphics[width=1\textwidth]{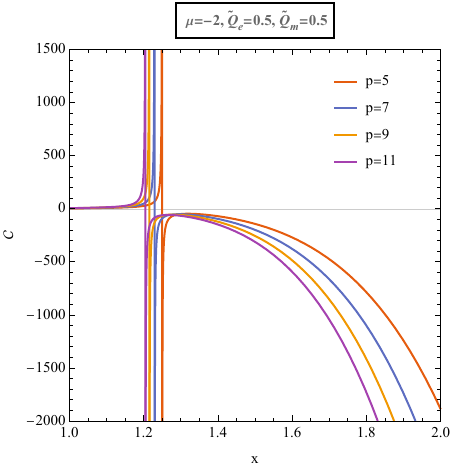}
\caption{}
\label{fig:plot1hh}
\end{subfigure}
    \caption{Free energy $F_{15}$ vs. temperature $T$ plot for the fixed $(\tilde{Q}_e,\tilde{Q}_m,R,C)$ ensemble in $(D=4/d=3)$.  }
 \label{fig:en:one}
\end{figure} 

\section{For the remaining ensembles }
\label{sec:other}
 We have studied in earlier sections all the seven ensembles in detail considering for fixed $(\tilde{\Phi}_e,\tilde{\Phi}_m,\mathcal{V},\mu)$ we get no free energy plots. There are eight more ensembles whose free energies are given in equation \eqref{free34}. Those free energies are namely $F_1$, $F_2$, $F_5$, $F_6$, $F_9$, $F_{10}$, $F_{13}$ $F_{14}$. In the Figure \ref{fig:en:one} given below, we have plot the graph of the free energy with respect to the temperatures well as the specific plots for all eight ensembles for different values of parameter. Putting a certain condition like the infinite volume and temperature scenario where $TR \rightarrow \infty$, we can see many free energy equations change drastically, namely:-

\begin{equation}
F_1=\tilde{\Phi}_m \tilde{Q}_m, \quad F_5=\tilde{\Phi}_e \tilde{Q}_e, \quad F_9=0, \quad F_{13}=\tilde{\Phi}_e \tilde{Q}_e+\tilde{\Phi}_m \tilde{Q}_m
\end{equation}
where $\mu C=-p \mathcal{V}$ in this boundary limit \cite{u}. Also, we see that the free energy of $F_1$ corresponds to the free energy of $F_4$ in the $(\tilde{Q}_m, \tilde{\Phi}_e, \mathcal{V}, \mu)$ and $F_5$ corresponds to $F_8$ in the $(\tilde{\Phi}_m, \tilde{Q}_e, \mathcal{V}, \mu)$ and lastly $F_{13}$ corresponds to the $F_{16}$ in the $(\tilde{Q}_m, \tilde{Q}_e,\mathcal{V}, \mu)$.
 In Figure \ref{fig:en:one}, we notice that some ensembles do show phase transitions namely the Davies type phase transition but no criticality. Hence, we can say there is no critical behaviour in the ensembles of the CFT containing fixed $p$. Moreover, out of the eight ensembles, we notice that seven ensembles namely at fixed  $(\tilde{Q}_m, \tilde{\Phi}_e, p, C)$, $(\tilde{Q}_m, \tilde{\Phi}_e,p,\mu)$, $(\tilde{\Phi}_m, \tilde{Q}_e, p, C)$, $(\tilde{\Phi}_m, \tilde{Q}_e, p, \mu)$,$(\tilde{Q}_m, \tilde{Q}_e, p, C)$, $(\tilde{Q}_m, \tilde{Q}_e, p, \mu)$, $(\tilde{\Phi}_e, \tilde{\Phi}_m,p,\mu)$ ensembles, shows Davies type phase transition having two branches. Here the down branch are analogous to a large black hole in the bulk which duals to a thermal state with high $S/C$, and the upper branch is analogous to the small AdS black hole which is dual to the thermal state with low $S/C$. The lower branch is more stable then the higher branch which is seen in the specific heat plots in Figure \ref{fig:en:one} and this phase dominates because it has a low free energy value. The last plot in Figure \ref{fig:plot6}, \ref{fig:plot1ff} shows a single branch which is unstable.

\section{Discussions}
 \label{sec:Discussions}

We explore the thermodynamic properties of dyonic conformal field theories (CFTs) in thermal states corresponding to charged Anti-de Sitter (AdS) black holes by utilizing the AdS/CFT correspondence. These conjugate thermodynamic variable pairs are connected to these thermal states: $(T, S)$, $(\tilde{\Phi}_e, \tilde{Q}_e)$, $(\tilde{\Phi}_m, \tilde{Q}_m)$, $(p, \mathcal{V})$ and $(\mu, C)$. Based on the thermodynamic properties of dyonic AdS black holes in the bulk, viewed through the holographic dictionary, each configuration has a corresponding dual description. We analysed each constructed ensemble and reviewed the resulting parametric plots. We only discovered critical behaviour in the ensembles $(\tilde{\Phi}_e, \tilde{Q}_m, \mathcal{V},C)$, $(\tilde{\Phi}_m, \tilde{Q}_e, \mathcal{V},C)$, $(\tilde{Q}_m, \tilde{Q}_e, \mathcal{V},C)$ and rest ensembles we see interesting phase transitions.\\
For the fixed ensemble $(\tilde{Q}_m, \tilde{\Phi}_e, \mathcal{V}, C)$ we graph for various values of $\tilde{Q}_m$ keeping rest parameters constant, we see Van der Waals type phase transition of order one for $\tilde{Q}_m<\tilde{Q}_m^{crit}$, where we see small (stable)-intermediate (unstable)-large (stable) entropy branches, a super-fluid $\lambda$ phase transition of second order for $(\tilde{Q}_m=\tilde{Q}_m^{crit})$ showing small (stable)-large (stable) entropy branches and a monotonous plot for $(\tilde{Q}_m>\tilde{Q}_m^{crit})$ showing stable branch. For values of $\tilde{\Phi}_e$, we see Van der Waals type phase transition of order one for $\tilde{\Phi}_e>\tilde{\Phi}_e^{crit}$, showing small(stable)-intermediate(unstable)-large(stable), a super-fluid $\lambda$ phase transition of order two for $(\tilde{\Phi}_e=\tilde{\Phi}_e^{crit})$ showing small(stable)-large(stable) and a monotonous plot for $(\tilde{\Phi}_e<\tilde{\Phi}_e^{crit})$ showing a stable entropy branch. Also at a supercritical value of $\tilde{\Phi}_m$ (green) we get a Davies type phase transition also. Finally, for various values of $C$ we see Van der Waals type phase transition for $C>C^{crit}$ showing small(stable)-intermediate(unstable)-large(stable), a super-fluid $\lambda$ phase transition of order of two for $C=C^{crit}$  and a monotonous stable branch plot for $(C<C^{crit})$.\\
For the fixed ensemble $(\tilde{Q}_m, \tilde{\Phi}_e, \mathcal{V}, \mu)$, we graph for numerous values of $\mu$. So when $\mu < 0$, we see only one phase, when $0\leq \mu<\mu_{coin}$ we see a change of phase in between the two entropy branches for the low and high values signing a Davies type phase transition. We do not get any plots for $\mu>\mu_{coin}$ In this interval the high entropy branch is more favoured as it has low free energy and hence stable. \\
For the fixed ensemble $(\tilde{Q}_e, \tilde{\Phi}_m, \mathcal{V}, C)$, we plot for various values of $\tilde{Q}_m$ where we see Van der Waals type phase transition of first order for $\tilde{Q}_e<\tilde{Q}_e^{crit}$ showing small(stable)-intermediate(unstable)-large(stable) entropy branches, a super-fluid $\lambda$ type phase transition of second order for $(\tilde{Q}_e=\tilde{Q}_e^{crit})$ showing small(stable)-large(stable) and a monotonous stable branch for $(\tilde{Q}_e>\tilde{Q}_e^{crit})$. For various values of $\tilde{\Phi}_m$, we see Van der Waals type phase transition of order one for $\tilde{\Phi}_m>\tilde{\Phi}_m^{crit}$ showing small(stable)-intermediate(unstable)-large(stable)branches. A super-fluid $\lambda$ type phase transition of second order for $(\tilde{\Phi}_m=\tilde{\Phi}_m^{crit})$ and a monotonous stable branch for $(\tilde{\Phi}_m>\tilde{\Phi}_m^{crit})$.At a supercritical value of $\tilde{\Phi}_m$ we see s Davies type phase transition also. For various  $C$ we see a Van der Waals type phase transition of order one for $C>C^{crit}$ showing small(stable)-internediate(unstable)-large(stable) branches, a super-fluid $\lambda$ type phase transition of order two for $(C=C^{crit})$ having small(stable)-large(stable) branches and a monotonous stable branch for $(C<C^{crit})$.\\
For the fixed ensemble $(\tilde{Q}_e, \tilde{\Phi}_m, \mathcal{V}, \mu)$, we plot for different values of $\mu$. When $\mu<0$, we see only one phase showing stable branch and when $0\leq \mu<\mu_{coin}$ we see a change of phase between two entropy branches having values both low and high where we see the Davies type phase transition showing small(unstable)-large(stable) entropy branches. We do not get plots for $\mu>\mu_{coin}$. \\
For the fixed ensemble $(\tilde{\Phi}_e, \tilde{\Phi}_m, \mathcal{V}, C)$, we graph for numerous instances of $\tilde{\Phi}_e$. We see a change of phase of the order one when the free energy changes sign at $F_{11}=0$. There is a "confined" state when $F_{11}>0$ and a "deconfined" state when $F_{11}<0$. We get the same results when we plot the graph for diverge range of values of $\tilde{\Phi}_m$. Also in Figure \ref{fig:en11}, in the red plot we also see Davies type phase transition in the specific heat plot on top of the $F_{11}-T$ plots.  
\\
\begin{table}[t!]
\begin{center}
\caption{Different ensembles and their phase transitions}
\label{tab:table1}
\begin{tabular}{|l|c|r|} 
\hline
  \textbf{Ensembles} & \textbf{Phase transitions shown}\\
      \hline
     $(\tilde{Q}_m, \tilde{\Phi}_e, p,C)$ & Davies type phase transition \\
      $(\tilde{Q}_m, \tilde{\Phi}_e, p,\mu)$ &  Davies type phase transition \\
      $(\tilde{Q}_m, \tilde{\Phi}_e, \mathcal{V},C)$ & Van der Waals, Super-fluid $\lambda$, Davies type phase transition and criticality \\
      $(\tilde{Q}_m, \tilde{\Phi}_e, \mathcal{V},\mu)$ & Davies type phase transition\\
      $(\tilde{\Phi}_m, \tilde{Q}_e,p, C)$ & Davies type phase transition \\
      $(\tilde{\Phi}_m, \tilde{Q}_e,p, \mu)$ & Davies type phase transition \\
      $(\tilde{\Phi}_m, \tilde{Q}_e,\mathcal{V}, C)$ & Van der Walls, Super-fluid $\lambda$, Davies type phase transitions and criticality \\
      $(\tilde{\Phi}_m, \tilde{Q}_e,\mathcal{V}, \mu)$ & Davies type phase transition \\
      $(\tilde{\Phi}_m, \tilde{\Phi}_e,p,C)$ &  No phase transition \\
      $(\tilde{\Phi}_m, \tilde{\Phi}_e,p,\mu)$ &  Davies type phase transition \\
      $(\tilde{\Phi}_m, \tilde{\Phi}_e,\mathcal{V},C)$ & Confined/deconfined and Davies type phase transitions \\
       $(\tilde{\Phi}_m, \tilde{\Phi}_e,\mathcal{V},\mu)$ & No phase transitions \\
      $(\tilde{Q}_m, \tilde{Q}_e,p,C)$ &  Davies type phase transition \\
      $(\tilde{Q}_m, \tilde{Q}_e,p,\mu)$ & Davies type phase transition \\
      $(\tilde{Q}_m, \tilde{Q}_e,\mathcal{V},C)$ & Van der Waals type, Super-fluid $\lambda$ phase transitions and criticality \\
      $(\tilde{Q}_m, \tilde{Q}_e,\mathcal{V},\mu)$ & Davies type phase transition \\
     \hline
    \end{tabular}
  \end{center}
\end{table}
For the fixed ensemble $(\tilde{Q}_e, \tilde{Q}_m, \mathcal{V}, C)$ we plot for various values of $\tilde{Q}_m$ and we see a Van der Waals type phase transition of order one for $\tilde{Q}_e<\tilde{Q}_e^{crit}$ showing small(stable)-intermediate(unstable)-large(stable) entropy branches, a superfluid $\lambda$ phase transition of order two for $(\tilde{Q}_e=\tilde{Q}_e^{crit})$ showing small(stable)-large(stable) branches and a monotonous stable branch for $(\tilde{Q}_e>\tilde{Q}_e^{crit})$. When we graph for numerous range of values of $\tilde{Q}_m$, we see a Van der Waals type phase transition of order one for $\tilde{Q}_m<\tilde{Q}_m^{crit}$ showing small(stable)-intermediate(unstable)-large(stable) branches, a superfluid $\lambda$ of order two for $(\tilde{Q}_m=\tilde{Q}_m^{crit})$ showing small(stable)-large(stable) and monotonous stable entropy branch for $(\tilde{Q}_m>\tilde{Q}_m^{crit})$. For a variety of values of $C$, we see a Van der Waals type phase transition of order one for $C>C^{crit}$, showing small(stable)-intermediate(unstable)-large(stable), asuperfluid $\lambda$ phase transition of order two for $(C=C^{crit})$ showing small(stable)-large(stable) and lastly a monotonous stable branch for  $(C<C^{crit})$.\\
For the fixed ensemble $(\tilde{Q}_e, \tilde{Q}_m, \mathcal{V}, \mu)$, we plot values of $\mu$ hence when $\mu \leq 0$, we see only one phase showing a stable entropy branch in the specific heat plot. When $0< \mu<\mu_{coin}$ we see change of phase between the entropy branches having  small (unstable)-large(stable)whic is the Davies type phase transition.\\
We also plot the free energy plot for various ensembles namely $(\tilde{Q}_m, \tilde{\Phi}_e, p, C)$, $(\tilde{Q}_m, \tilde{\Phi}_e, p, \mu)$, $(\tilde{Q}_e, \tilde{\Phi}_m, p, C)$, $(\tilde{Q}_e, \tilde{\Phi}_m, p, \mu)$, $(\tilde{\Phi}_m, \tilde{\Phi}_e, p, C)$, $(\tilde{\Phi}_m, \tilde{\Phi}_e, p, \mu)$, $(\tilde{Q}_m, \tilde{Q}_e, p, C)$, $(\tilde{Q}_m, \tilde{Q}_e, p, \mu)$. Except for ensemble $(\tilde{\Phi}_m, \tilde{\Phi}_e, p, C)$, rest all we get a Davies type phase transition but no criticality.\\
In summary, we aimed to elucidate the phase transitions and critical phenomena in Dyonic AdS black holes by exploring the impact of introducing $\mathcal{V}$ (CFT volume), $p$ (CFT pressure), and $C$ (central charge), both electric and magnetic charges $(\tilde{Q}_e, \tilde{Q}_m)$as extra parameters in the thermodynamic phase space. Our analysis covered 16 specific ensembles and demonstrated that incorporating these variables unveiled critical behaviours not previously observed in conventional thermodynamics or the extended phase space thermodynamic (EPST) framework. The results are summarized in \ref{tab:table1}.\\

Conclusively, the nature and order of phase transitions are profoundly influenced by the choice of ensemble and the complexity of variables involved. Our observations indicate that introducing an additional variable, such as magnetic charge $(\tilde{Q}_m)$ or magnetic potential $(\tilde{\Phi}_m)$, enriches the landscape of phase transitions. This underscores a broader spectrum of methodologies for studying phase transitions, paving the way for a more comprehensive understanding of these critical phenomena in black hole thermodynamics.

\section{Appendix}
\label{Appendix}

\begin{equation*}
 x_1=\frac{\sqrt[3]{6} \left(1-i \sqrt{3}\right) \left(A\right){}^{2/3}-2\ 2^{2/3} \sqrt[6]{3} \left(\sqrt{3}+3 i\right) R^4 \tilde{\Phi }_e^4+2\ 2^{2/3} \sqrt[6]{3} \left(\sqrt{3}+3 i\right) R^2 \tilde{\Phi }_e^2}{12 \left(R^2 \tilde{\Phi }_e^2-1\right) \sqrt[3]{A}}
 \end{equation*}
 
 \begin{equation*}
 x_2= \frac{\sqrt[3]{6} \left(1+i \sqrt{3}\right) \left(A\right){}^{2/3}-2\ 2^{2/3} \sqrt[6]{3} \left(\sqrt{3}-3 i\right) R^4 \tilde{\Phi }_e^4+2\ 2^{2/3} \sqrt[6]{3} \left(\sqrt{3}-3 i\right) R^2 \tilde{\Phi }_e^2}{12 \left(R^2 \tilde{\Phi }_e^2-1\right) \sqrt[3]{A}}
 \end{equation*}
 where the value of A in the above equation is:-
\begin{equation*}
A=\sqrt{3} \sqrt{R^2 \left(R^2 \tilde{\Phi }_e^2-1\right){}^3 \left(27 \mu ^2 \left(R^2 \tilde{\Phi }_e^2-1\right)+16 R^4 \tilde{\Phi }_e^6\right)}+9 \mu  R^5 \tilde{\Phi }_e^4-18 \mu  R^3 \tilde{\Phi }_e^2+9 \mu  R
\end{equation*} 

\begin{equation*}
\resizebox{1\hsize}{!}{$
A_1=243 \mu  \tilde{\Phi }_e^4 R^5-162 \mu  \tilde{\Phi }_e^2 R^3+27 \mu  R+\sqrt{729 R^2 \mu ^2 \left(1-3 R^2 \tilde{\Phi }_e^2\right){}^4+4 \left(-36 R^4 \tilde{\Phi }_e^4+21 R^2 \tilde{\Phi }_e^2-3\right){}^3}$}
\end{equation*}

\begin{equation*}
B_1=2^{2/3} \left(i \sqrt{3}+1\right) \left(A\right){}^{2/3}+6 \sqrt[3]{2} \left(1-i \sqrt{3}\right) \left(12 R^4 \tilde{\Phi }_e^4-7 R^2 \tilde{\Phi }_e^2+1\right)
\end{equation*}

\begin{equation*}
B_2=2^{2/3} \left(1-i \sqrt{3}\right) \left(A\right){}^{2/3}+12 \sqrt[3]{-2} \left(12 R^4 \tilde{\Phi }_e^4-7 R^2 \tilde{\Phi }_e^2+1\right)
\end{equation*}

\begin{equation}
A_4=2^{2/3} \left(1-i \sqrt{3}\right) \left(B_0\right){}^{2/3}+6 \sqrt[3]{2} \left(i \sqrt{3}+1\right) \left(12 R^4 \tilde{\Phi }_m^4-7 R^2 \tilde{\Phi }_m^2+1\right)
\end{equation}

\begin{equation}
\resizebox{1\hsize}{!}{$
B_4=12 \pi  R \left(1-3 R^2 \tilde{\Phi }_m^2\right) \sqrt[3]{B_0} \left(2^{2/3} \left(1-i \sqrt{3}\right) \left(B_0\right){}^{2/3}+6 \sqrt[3]{2} \left(i \sqrt{3}+1\right) \left(12 R^4 \tilde{\Phi }_m^4-7 R^2 \tilde{\Phi }_m^2+1\right)\right){}^2$}
\end{equation}

\begin{equation}
\resizebox{1\hsize}{!}{$
\begin{split}
&T_0=\frac{\left(R^2 \tilde{\Phi }_e^2-1\right)\left(R \tilde{\Phi }_e^2 \left(9 R \left(\sqrt{3} \sqrt{B_3} \mu -4 \sqrt[3]{6} \sqrt[3]{A_2} \mu \right)-4 \sqrt[3]{2} 3^{5/6} \sqrt[3]{A_2} \sqrt{B_3}+243 \mu ^2 R^2\right)-9 \mu  \left(\sqrt{3} \sqrt{B_3}+9 \mu  R\right)\right)}{2^{2/3} \pi  ^2 \left(-2 \sqrt[3]{6} R^4 \tilde{\Phi }_e^4+2 \sqrt[3]{6} R^2 \tilde{\Phi }_e^2+A_2^{2/3}\right) \sqrt[3]{3 \mu  R^5 \tilde{\Phi }_e^4-6 \mu  R^3 \tilde{\Phi }_e^2+\frac{\sqrt{B_3}}{\sqrt{3}}+3 \mu  R}}\\
&-\frac{\left(R^2 \tilde{\Phi }_e^2-1\right)\left(R^5 \tilde{\Phi }_e^6 \left(8\ 6^{2/3} A_2^{2/3}-36 \sqrt[3]{6} \sqrt[3]{A_2} \mu  R+81 \mu ^2 R^2\right)+R^3 \tilde{\Phi }_e^4 \left(8\ 6^{2/3} A_2^{2/3}-72 \sqrt[3]{6} \sqrt[3]{A_2} \mu  R+243 \mu ^2 R^2\right)\right)}{2^{2/3} \pi  ^2 \left(-2 \sqrt[3]{6} R^4 \tilde{\Phi }_e^4+2 \sqrt[3]{6} R^2 \tilde{\Phi }_e^2+A_2^{2/3}\right) \sqrt[3]{3 \mu  R^5 \tilde{\Phi }_e^4-6 \mu  R^3 \tilde{\Phi }_e^2+\frac{\sqrt{B_3}}{\sqrt{3}}+3 \mu  R}}
\end{split}$} 
 \end{equation}
 Where A and B are:-
 \begin{equation*}
 A_2=9 \mu  \tilde{\Phi }_e^4 R^5-18 \mu  \tilde{\Phi }_e^2 R^3+9 \mu  R+\sqrt{3} \sqrt{R^2 \left(R^2 \tilde{\Phi }_e^2-1\right){}^3 \left(16 R^4 \tilde{\Phi }_e^6+27 R^2 \mu ^2 \tilde{\Phi }_e^2-27 \mu ^2\right)}
 \end{equation*}
 \begin{equation*}
 B_3=R^2 \left(R^2 \tilde{\Phi }_e^2-1\right){}^3 \left(16 R^4 \tilde{\Phi }_e^6+27 R^2 \mu ^2 \tilde{\Phi }_e^2-27 \mu ^2\right)
 \end{equation*}

\begin{equation}
x_3=\frac{2^{2/3} \left(1-i \sqrt{3}\right) \left(A_3\right){}^{2/3}+6 \sqrt[3]{2} \left(1+i \sqrt{3}\right) \left(12 R^4 \tilde{\Phi }_m^4-7 R^2 \tilde{\Phi }_m^2+1\right)}{12 \left(1-3 R^2 \tilde{\Phi }_m^2\right) \sqrt[3]{A_3}}
\end{equation}
\begin{equation}
x_4=\frac{2^{2/3} \left(1+i \sqrt{3}\right) \left(A_3\right){}^{2/3}+6 \sqrt[3]{2} \left(1-i \sqrt{3}\right) \left(12 R^4 \tilde{\Phi }_m^4-7 R^2 \tilde{\Phi }_m^2+1\right)}{12 \left(1-3 R^2 \tilde{\Phi }_m^2\right) \sqrt[3]{A_3}}
\end{equation}
where A is given as:-
\begin{equation}
A_3=\sqrt{729 \mu ^2 R^2 \left(1-3 R^2 \tilde{\Phi }_m^2\right){}^4+4 \left(-36 R^4 \tilde{\Phi }_m^4+21 R^2 \tilde{\Phi }_m^2-3\right){}^3}+243 \mu  R^5 \tilde{\Phi }_m^4-162 \mu  R^3 \tilde{\Phi }_m^2+27 \mu  R
\end{equation}

\begin{equation}
A_5=2^{2/3} \left(i \sqrt{3}+1\right) \left(B_0\right){}^{2/3}+6 \sqrt[3]{2} \left(1-i \sqrt{3}\right) \left(12 R^4 \tilde{\Phi }_m^4-7 R^2 \tilde{\Phi }_m^2+1\right)
\end{equation}

\begin{equation}
\resizebox{1\hsize}{!}{$
B_5=12 \pi  R \left(1-3 R^2 \tilde{\Phi }_m^2\right) \sqrt[3]{B_0} \left(2^{2/3} \left(i \sqrt{3}+1\right) \left(B_0\right)^{2/3}+6 \sqrt[3]{2} \left(1-i \sqrt{3}\right) \left(12 R^4 \tilde{\Phi }_m^4-7 R^2 \tilde{\Phi }_m^2+1\right)\right)^2$}
\end{equation}

\begin{equation}
\resizebox{1\hsize}{!}{$
B_0=243 \mu  \tilde{\Phi }_m^4 R^5-162 \mu  \tilde{\Phi }_m^2 R^3+27 \mu  R+\sqrt{729 R^2 \mu ^2 \left(1-3 R^2 \tilde{\Phi }_m^2\right){}^4+4 \left(-36 R^4 \tilde{\Phi }_m^4+21 R^2 \tilde{\Phi }_m^2-3\right)^3}$}
\end{equation}

\begin{equation*}
\resizebox{1\hsize}{!}{$x_{01}= \frac{12 R^4 \tilde{\Phi }_m^4-12 R^2 \tilde{\Phi }_m^2}{3\ 2^{2/3} \left(R^2 \tilde{\Phi }_m^2-1\right) \sqrt[3]{108 \mu  R^5 \tilde{\Phi }_m^4-216 \mu  R^3 \tilde{\Phi }_m^2+\sqrt{\left(108 \mu  R^5 \tilde{\Phi }_m^4-216 \mu  R^3 \tilde{\Phi }_m^2+108 \mu  R\right){}^2+4 \left(12 R^4 \tilde{\Phi }_m^4-12 R^2 \tilde{\Phi }_m^2\right){}^3}+108 \mu  R}}$}
\end{equation*}
\begin{equation*} 
 \resizebox{1\hsize}{!}{$-\frac{\sqrt[3]{108 \mu  R^5 \tilde{\Phi }_m^4-216 \mu  R^3 \tilde{\Phi }_m^2+\sqrt{\left(108 \mu  R^5 \tilde{\Phi }_m^4-216 \mu  R^3 \tilde{\Phi }_m^2+108 \mu  R\right){}^2+4 \left(12 R^4 \tilde{\Phi }_m^4-12 R^2 \tilde{\Phi }_m^2\right){}^3}+108 \mu  R}}{6 \sqrt[3]{2} \left(R^2 \tilde{\Phi }_m^2-1\right)}$}	
 \end{equation*}
 \begin{equation}
\resizebox{1\hsize}{!}{$T_{01}=\frac{\left(R^2 \tilde{\Phi }_m^2-1\right) \left(R^5 \left(-36 A_3 +81 R^2 \mu ^2\right) \tilde{\Phi }_m^6-R^3 \left(-72 A_3+243 R^2 \mu ^2\right) \tilde{\Phi }_m^4+R \left(-4 \sqrt[3]{2} 3^{5/6} \sqrt{A_2} \sqrt[3]{A_1}+243 R^2 \mu ^2+9 R \left(\sqrt{3} \mu  \sqrt{A_2}-4 \sqrt[3]{6} \mu  \sqrt[3]{A_1}\right)\right) \tilde{\Phi }_m^2-9 \mu  \left(9 R \mu +\sqrt{3} \sqrt{A_2}\right)\right)}{B}$} 
 \end{equation}
where the value of $A_6$, $A_7$, $A_8$ and $B_6$ is given as
\begin{equation}
A_6=9 \mu  \tilde{\Phi }_m^4 R^5-18 \mu  \tilde{\Phi }_m^2 R^3+9 \mu  R+\sqrt{3} \sqrt{R^2 \left(R^2 \tilde{\Phi }_m^2-1\right){}^3 \left(16 R^4 \tilde{\Phi }_m^6+27 R^2 \mu ^2 \tilde{\Phi }_m^2-27 \mu ^2\right)}
\end{equation}

\begin{equation}
A_7=R^2 \left(R^2 \tilde{\Phi }_m^2-1\right){}^3 \left(16 R^4 \tilde{\Phi }_m^6+27 R^2 \mu ^2 \tilde{\Phi }_m^2-27 \mu ^2\right)
\end{equation}

\begin{equation}
A_8=\sqrt[3]{6} R \mu  \sqrt[3]{A_1}+8\ 6^{2/3} \left(9 \mu  \tilde{\Phi }_m^4 R^5-18 \mu  \tilde{\Phi }_m^2 R^3+9 \mu  R+\sqrt{3} \sqrt{A_2}\right){}^{2/3}
\end{equation}

\begin{equation}
\resizebox{1\hsize}{!}{$
B_6=^{2/3} \pi  \sqrt[3]{3 \mu  \tilde{\Phi }_m^4 R^5-6 \mu  \tilde{\Phi }_m^2 R^3+3 \mu  R+\frac{\sqrt{A_2}}{\sqrt{3}}} \left(\left(9 \mu  \tilde{\Phi }_m^4 R^5-18 \mu  \tilde{\Phi }_m^2 R^3+9 \mu  R+\sqrt{3} \sqrt{A_2}\right){}^{2/3}-2 \sqrt[3]{6} R^4 \tilde{\Phi }_m^4+2 \sqrt[3]{6} R^2 \tilde{\Phi }_m^2\right){}^2$}
\end{equation}

\begin{equation*}
x_5=\frac{i \sqrt[3]{2} \left(\sqrt{3}+i\right) \left(\sqrt{81 \mu ^2 R^2-12}-9 \mu  R\right)^{2/3}-2 \sqrt[3]{3} \left(1+i \sqrt{3}\right)}{2\ 6^{2/3} \sqrt[3]{\sqrt{81 \mu ^2 R^2-12}-9 \mu  R}}
\end{equation*}
\begin{equation}
x_6 =\frac{2 i \sqrt[3]{3} \left(\sqrt{3}+i\right)-\sqrt[3]{2} \left(1+i \sqrt{3}\right) \left(\sqrt{81 \mu ^2 R^2-12}-9 \mu  R\right)^{2/3}}{2\ 6^{2/3} \sqrt[3]{\sqrt{81 \mu ^2 R^2-12}-9 \mu  R}}
\end{equation}


\begin{thebibliography}{99}
\bibitem{a}
J.D. Berkenstein, \emph{Black holes and the second law}, \emph{Lett. Nuovo. Cim.} {\bf 4} (1972) 737-740

\bibitem{aa1}
J.D. Berkenstein, \emph{Black holes and entropy}, \emph{Phys. Rev. D} {\bf 7} (1973) 2333 [INSPIRE].

\bibitem{b}
S.W. Hawking, \emph{Particle creation by black holes}, \emph{Commun. math. Phys.} {\bf 43} (1975) 199.

\bibitem{bb1}
J. M. Bardeen, B. Carter, S. W. Hawking, \emph{The four laws of black hole mechanics}, \emph{Commun. Math. Phys. } {\bf 31} (1973) 161-170


\bibitem{0a}
 S. W. Hawking, \emph{Gravitational radiation from colliding black holes}, \emph{Phys. Rev. Lett.} 26,
 1344-1346 (1971) doi:10.1103/PhysRevLett.26.1344

\bibitem{0b}
 S. W. Hawking, \emph{Black hole explosions}, \emph{Nature} 248, 30-31 (1974) doi:10.1038/248030a0

\bibitem{0c}
 R. M. Wald, \emph{Entropy and black-hole thermodynamics}, \emph{Phys. Rev. D} 20, 1271-1282 (1979)
 doi:10.1103/PhysRevD.20.1271

\bibitem{0d}
J. D. Bekenstein. \emph{Black-hole thermodynamics}, \emph{Physics Today}, 33(1):24–31, 1980.

\bibitem{0e}
 R. M. Wald, \emph{The thermodynamics of black holes}, \emph{Living Rev. Rel.} 4, 6 (2001) doi:10.12942/lrr-2001-6 [arXiv:gr-qc/9912119 [gr-qc]]

\bibitem{1a}
 S.Carlip, \emph{Black Hole Thermodynamics}, \emph{Int. J. Mod. Phys. D} 23, 1430023 (2014) doi:10.1142/S0218271814300237 [arXiv:1410.1486 [gr-qc]]

\bibitem{1b}
A.C.Wall, \emph{A Survey of Black Hole Thermodynamics}, [arXiv:1804.10610 [gr-qc]].

\bibitem{1c}
P.Candelas and D.W.Sciama, \emph{Irreversible Thermodynamics of Black Holes}, \emph{Phys. Rev. Lett.}
 38, 1372-1375 (1977) doi:10.1103

\bibitem{2a}
 Chao Wang, Bin Wu, Zhen Ming Xu, Wen Li Yang, \emph{Thermodynamic geometry of the
 RN-AdS black hole and non-local observables},[arXiv:2210.08718 ]
 
\bibitem{2b}
 Yuchen Huang, Jun Tao, Peng Wang, Shuxuan Ying, \emph{Phase transitions and thermodynamic
 geometry of a Kerr-Newman black hole in a cavity}, \emph{Eur.Phys.J.Plus} 138, 265
 (2023)[https://doi.org/10.1140/epjp/s13360-023-03858-w]
 
\bibitem{2c} Wen-Xiang Chen, Yao-Guang Zheng, \emph{Thermodynamic geometric analysis of BTZ black hole
 under f(R) gravity}, [arXiv:2112.15032 ]

\bibitem{2d}
 Peng Wang, Feiyu Yao, \emph{Thermodynamic Geometry of Black Holes Enclosed by a Cavity in
 Extended Phase Space}, \emph{Nulc. Phys. B}, 976 (2022) 115715
 [https://doi.org/10.1016/j.nuclphysb.2022.115715]

\bibitem{2e}
 Amin Dehyadegari, Ahmad Sheykhi, \emph{Thermodynamic geometry and phase transition of
 spinning AdS black holes}, \emph{Phys. Rev. D}, 104, 104066, (2021)
 [doi:10.1103/PhysRevD.104.104066]

\bibitem{2f}
 Shao-Wen Wei, Yu-Xiao Liu \emph{General thermodynamic geometry approach for rotating Kerr
 anti-de Sitter black holes}, \emph{Phys. Rev. D}, {\bf 104}, 084087 (2021) [10.1103/PhysRevD.104.084087]– 26

\bibitem{2g}
 Seyed Ali Hosseini Mansoori, \emph{Thermodynamic geometry of the novel 4-D Gauss Bonnet AdS
 Black Hole}, Phys.Dark Univ., 31 (2021) 100776 [doi:10.1016/j.dark.2021.100776]

\bibitem{2h}
 Aritra Ghosh, Chandrasekhar Bhamidipati, \emph{Thermodynamic geometry for charged
 Gauss-Bonnet black holes in AdS spacetimes}, \emph{Phys. Rev. D} 101, 046005
 (2020),[doi:10.1103/PhysRevD.101.046005]

\bibitem{2i}
Panah, B.E., Rodrigues, M.E. \emph{Topological phantom AdS black holes in F(R) gravity}, \emph{Eur. Phys. J. C} {\bf 83}, 237 (2023). https://doi.org/10.1140/epjc/s10052-023-11402-4

\bibitem{2j}
B. Eslam Panah, \emph{Effects of energy dependent spacetime on geometrical thermodynamics and heat engine of black holes: gravity's rainbow} \emph{Phys. Lett. B} {\bf 787} (2018) 45. [ 10.1016/j.physletb.2018.10.042]

\bibitem{2k}
Kh. Jafarzade, J. Sadeghi, B. Eslam Panah, and S. H. Hendi, \emph{"Geometrical thermodynamics and P-V criticality of charged accelerating AdS black holes"
}, \emph{Annals of Physics} {\bf 432}, 168577 (2021).


\bibitem{3a}
 Shao-Wen Wei and Yu-Xiao Liu. \emph{Topology of black hole thermodynamics.}, \emph{Phys. Rev. D}
 105:104003, (2022).

\bibitem{3b}
 Shao-Wen Wei, Yu-Xiao Liu, and Robert B. Mann. \emph{Black hole solutions as topological
 thermodynamic defects}, \emph{Phys. Rev. Lett.}, 129:191101,(2022).

\bibitem{3c}
 B. Hazarika and P. Phukon, \emph{Thermodynamic topology of black Holes in f(R) gravity}, \emph{Progress
 of Theoretical and Experimental Physics}, {\bf 4}, (2024), https://doi.org/10.1093/ptep/ptae035
 [arXiv:2401.16756 [hep-th]].

\bibitem{3d}
 N. J. Gogoi and P. Phukon, \emph{Thermodynamic topology of 4D Euler-Heisenberg-AdS black hole in different ensembles},[arXiv:2312.13577 [hep-th]]

\bibitem{3e}
 B. Hazarika and P. Phukon, \emph{Thermodynamic Topology of D = 4,5 Horava Lifshitz Black
 Hole in Two Ensembles},[arXiv:2312.06324 [hep-th]].

\bibitem{3f}
 B. Hazarika, N. J. Gogoi and P. Phukon, \emph{Revisiting thermodynamic topology of
 Hawking-Page and Davies type phase transitions},[arXiv:2404.02526 [hep-th]].

\bibitem{3g}
Di Wu, \emph{Topological classes of rotating black holes}, \emph{Phys. Rev. D} {\bf 107} (2023) 024024, arXiv: 2211.15151.

\bibitem{3h}
 Di Wu, Shuang-Qing Wu, \emph{Topological classes of thermodynamics of rotating AdS black holes}, Phys. Rev. D 107 (2023) 084002, arXiv: 2301.03002;

\bibitem{3i}
 Di Wu, \emph{Consistent thermodynamics and topological classes for the four-dimensional Lorentzian charged Taub-NUT spacetimes}, \emph{Eur. Phys. J. C} {\bf 83} (2023) 589, arXiv: 2306.02324;
 
\bibitem{3j}
 Di Wu, \emph{Topological classes of thermodynamics of the four-dimensional static accelerating black holes}, \emph{Phys. Rev. D} {\bf 108} (2023) 084041, arXiv: 2307.02030;

\bibitem{3k}
 Di Wu, Shuang-Yong Gu, Xiao-Dan Zhu, Qing-Quan Jiang, Shu-Zheng Yang, \emph{Topological classes of thermodynamics of the static multi-charge AdS black holes in gauged supergravities: Novel temperature-dependent thermodynamic topological phase transition}, \emph{JHEP} {\bf 06} (2024) 213, arXiv: 2402.00106;

\bibitem{3l}
Behzad Eslam Panah, \emph{Analytic Electrically Charged Black Holes in F(R)-ModMax Theory}, \emph{Progress of Theoretical and Experimental Physics}, Volume 2024, Issue 2, February 2024, 023E01, https://doi.org/10.1093/ptep/ptae012

\bibitem{3m}
Bidyut Hazarika, B. Eslam Panah, Prabwal Phukon,\emph{Thermodynamic topology of topological charged dilatonic black holes} [ arXiv:2407.05325v1]

\bibitem{3n}
B.Eslam Panah, B.Hazarika, P.Phukon, \emph{Thermodynamic topology of topological black hole in F(R)-ModMax gravity's rainbow}, [https://doi.org/10.48550/arXiv.2405.20022]



\bibitem{c}
J.M. Maldacena, \emph{The large N limit of superconformal field theories and supergravity}, \emph{Int. J. Theor. Phys.} {\bf 43} (1999) 1113 [\emph{Adv. Theor. Math. Phys.} {\bf 43} (1998) 231] [hep-th/971120][INSPIRE].

\bibitem{d}
S.W. Hawking and D. N. Page, \emph{Thermodynamics of black holes in Anti-de Sitter space}, \emph{Commun. Math. Phys.} {\bf 87} (1983) 577 [INSPIRE].

\bibitem{i}
D. Kubiznak and R.B. Mann, \emph{P-V criticality of charged AdS black holes}, \emph{JHEP} {\bf 07} (2012)033 [arXiv:1205.0559] [INSPIRE].


\bibitem{x}
 B.P. Dolan, \emph{ Pressure and volume in the first law of black hole thermodynamics}, \emph{Class. Quant. Grav.} {\bf 28} (2011) 235017 [arXiv:1106.6260] [INSPIRE].
 
 \bibitem{y}
B.P. Dolan, \emph{ The cosmological constant and the black hole equation of state}, \emph{Class. Quant. Grav.} {\bf 28} (2011) 125020 [arXiv:1008.5023] [INSPIRE].

 \bibitem{yy}
 Dolan, \emph{Compressibility of rotating black holes}, \emph{Phys. Rev. D. } {\bf 84} (2011) 127503 [arXiv:1109.0198]
 
 \bibitem{yy1}
 R.-G. Cai, L.-M. Cao, L. Li, R.-Q. Yang, \emph{P-V criticality in the extended phase space of Gauss-Bonnet black holes in AdS space}, \emph{JHEP} (2013) 005 [arXiv:1306.6233]

\bibitem{yy2}
S. H. Hendi, S. Panahiyan, and B. Eslam Panah, \emph{P-V criticality and geometrothermodynamics of black holes with Born-Infeld type nonlinear electrodynamics
}, \emph{Int. J. Mod. Phys. D} {\bf 25} (2016) 1650010. [10.1142/S0218271816500103]

\bibitem{yy3}
S. H. Hendi, B. Eslam Panah, and S. Panahiyan, \emph{Einstein-Born-Infeld-Massive Gravity: adS-Black Hole Solutions and their Thermodynamical properties},  JHEP 11 (2015) 157.
ArXiv: arXiv:1508.01311

\bibitem{yy4}
S. H. Hendi, R. B. Mann, S. Panahiyan, and B. Eslam Panah, \emph{van der Waals like behaviour of topological AdS black holes in massive gravity}, \emph{Phys. Rev. D} {\bf 95}, 021501(R) (2017).
ArXiv: arXiv:1702.00432 



 \bibitem{j1}
   D. Kubiznak, R.B. Mann and M. Teo, \emph{Black hole chemistry: thermodynamics with Lambda}, \emph{Class. Quant. Grav. } {\bf 34} (2017) 063001 [arXiv:1608.06147] [INSPIRE].

\bibitem{p}
D. Kastor, S. Ray and J. Traschen, \emph{Enthalphy and the mechanics of AdS black holes}, \emph{Class. Quant. Grav.} {\bf 26} (2009) 195011 [arXiv:0904.2765] [INSPIRE].
\bibitem{e}
E. Witten, \emph{Anti-de Sitter space, thermal phase transition, and confinement in gauge theories}, \emph{Adv. Theor. Math. Phys.} {\bf 2} (1998) 505 [hep-th/9803131] [INSPIRE].


\bibitem{f}
A. Chamblin, R. Emparan, C. V. Jhonson and R. C. Myers, \emph{Charged AdS black holes and catastrophic holography}, \emph{Phys. Rev. D} {\bf 60} (1999) 064018 [hep-th/9902170] [INSPIRE].

\bibitem{g}
A. Chamblin, R. Emparan, C. V. Jhonson and R. C. Myers, \emph{Holography, thermodynamics and fluctuations of charged AdS black holes}, \emph{Phys. Rev. D} {\bf 60} (1999) 104026 [hep-th/9904197] [INSPIRE].

\bibitem{h}
M. Cvetič and S.S. Gubser, \emph{Phases of R charged black holes, spinning branes and strongly coupled gauge theories}, \emph{JHEP} {\bf 04} (1999)024 [hep-th/9902195] [INSPIRE].




\bibitem{j}
B.P. Dolan, A. Kostouki, D. Kubiznak and R.B. Mann, \emph{Isolated critical point from Lovelock gravity}, \emph{Class. Quant. Grav.} {\bf 31} (2014)242001 [arXiv:1407.4783] [INSPIRE].

\bibitem{o}
R.A. Hennigar, R.B. Mann and E. Tjoa, \emph{Superfluid black holes}, \emph{Phys. Rev. Lett.} {\bf 118} (2017) 021301 [arXiv:1402.2837] [INSPIRE].


\bibitem{m}
N. Altamirano, D. Kubiznak, R.B. Mann and Z. Sherkatghanad, \emph{Kerr-AdS analogue of triple point and solid/liquid/gas phase transition}, \emph{Class. Quant. Grav.} {\bf 31} (2014)042001 [arXiv:1308.2672] [INSPIRE].


\bibitem{n}
S.-W. Wei and Y.-X. Liu, \emph{Triple points and phase diagrams in the extended phase of charged Gauss-Bonnet black holes in AdS space}, \emph{Phys. Rev. D} {\bf 90} (2014)044057 [arXiv:1402:2837] [INSPIRE].

\bibitem{k}
N Altamirano, D. Kubiznak and R.B. Mann, \emph{Reentrant phase transition in rotating Anti-de Sitter black holes}, \emph{Phys. Rev. D} {\bf 88} (2013)101502 [arXiv:1306.5756] [INSPIRE].

\bibitem{l}
A.M. Frassino, D. Kubiznak, R.B. Mann and F. Simovic, \emph{Multiple reentrant phase transitions and triple points in Lovelock thermodynamics}, \emph{JHEP} {\bf 09} (2014)080 [arXiv:1406.7015] [INSPIRE].

 \bibitem{b1}
 C.V. Johnson, \emph{Holographic heat engines}, \emph{Class. Quant. Grav.} {\bf 31} (2014) 205002 [arXiv:1404.5982] [INSPIRE].
 
 \bibitem{bc1}
 H. Xu, Y. Sun, L. Zhao, \emph{Black hole thermodynamics and heat engines in conformal gravity}, \emph{Int. J. Mod. Phys. D} {\bf 26} (2017) no.13, 1750151 [arXiv:1706.06442]

  \bibitem{u}
 M.R. Visser, \emph{ Holographic thermodynamics requires a chemical potential for color}, \emph{Phys. Rev. D} {\bf 105} (2022) 106014 [arXiv:2101.04145] [INSPIRE].

 \bibitem{v}
 W. Cong, D. Kubiznak and R.B. Mann, \emph{Thermodynamics of AdS Black Holes: Central Charge Criticality}, \emph{Phys. Rev. Lett.} {\bf 127} (2021) 091301 [arXiv:2105.02223] [INSPIRE]
 
 \bibitem{vv}
 R. B. Alfaia, I. P. Lobo, L. C. T. Brito, \emph{Central charge criticality of charged AdS black hole surrounded by different fluids}, \emph{Eur. Phys. J. Plus} {\bf 137} (2022) 402 [arXiv:2109.06599 ] [hep-th]


 \bibitem{k2}
  W. Cong, D. Kubiznak, R.B. Mann and M.R. Visser, \emph{Holographic CFT phase transitions and critically for charged AdS black holes}, \emph{JHEP} {\bf 08} (2022) 174 [arXiv:2112.14848].
 
\bibitem{new}
 Gong, TF., Jiang, J., Zhang, M. \emph{Holographic thermodynamics of rotating black holes}, \emph{J. High Energ. Phys.} 2023, 105 (2023). https://doi.org/10.1007/JHEP06(2023)105
 
\bibitem{new1}
Moaathe Belhaj Ahmed, Wan Cong, David Kubiznak, Robert B. Mann, Manus R. Visser, \emph{Holographic CFT Phase Transitions and Criticality for Rotating AdS Black Holes} [
https://doi.org/10.48550/arXiv.2305.03161]
 
\bibitem{q}
 D. Kastor, S. Ray and J. Traschen, \emph{Smarr formula and an extended first law for Lovelock
gravity}, \emph{Class. Quant. Grav.} {\bf 27} (2010) 235014 [arXiv:1005.5053] [INSPIRE].


\bibitem{r}
 D. Kastor, S. Ray and J. Traschen, \emph{, Chemical potential in the first law for holographic entanglement entropy}, \emph{JHEP} {\bf 11} (2014) 120 [arXiv:1409.3521] [INSPIRE].
 
 \bibitem{s}
  A. Karch and B. Robinson, \emph{ Holographic black hole chemistry}, \emph{JHEP} {\bf 12} (2015) 073 [arXiv:1510.02472] [INSPIRE].

\bibitem{t}
  D. Sarkar and M. Visser, \emph{ The first law of differential entropy and holographic complexity}, \emph{JHEP} {\bf 11} (2020) 004 [arXiv:2008.12673] [INSPIRE].
  
\bibitem{rps}  
 Z. Gao,  L. Zhao, \emph{Restricted phase space thermodynamics for AdS black holes via holography}, \emph{Classical And Quantum Gravity.}, 39, 075019 (2022)  
  
\bibitem{rps1}  
  Gao, Z., Kong, X., Zhao, L. \emph{Thermodynamics of Kerr-AdS black holes in the restricted phase space}, \emph{Eur. Phys. J. C} 82, 112 (2022). https://doi.org/10.1140/epjc/s10052-022-10080-y
  
\bibitem{rps2}  
  Jafar Sadeghi, Mehdi Shokri, Saeed Noori Gashti, Mohammad Reza Alipour, \emph{RPS Thermodynamics of Taub-NUT AdS Black Holes in the Presence of Central Charge and the Weak Gravity Conjecture}, [
https://doi.org/10.48550/arXiv.2205.03648]

\bibitem{rps3}
Md Sabir Ali, Sushant G. Ghosh, Anzhong Wang, \emph{Thermodynamics of Kerr-Sen-AdS black holes in the restricted phase space} [https://doi.org/10.48550/arXiv.2308.00489]

\bibitem{rps4}
Mozib Bin Awal, Prabwal Phukon, \emph{Restricted Phase Space Thermodynamics of NED-AdS Black Holes}, [https://doi.org/10.48550/arXiv.2404.03261]

\bibitem{rps5}
Y. Ladghami, B. Asfour, A. Bouali, A. Errahmani, T. Ouali, \emph{4D-EGB black holes in RPS thermodynamics}, \emph{Physics of the Dark Universe},
 {\bf 41}, 2023, 101261, https://doi.org/10.1016/j.dark.2023.101261.

  \bibitem{l2}
  S. Dutta, A. Jain and R. Soni, \emph{Dyonic black hole and holography}, \emph{JHEP} {\bf 60} (2013) 07 [arXiv:1310.1748]


\bibitem{aa}
S. A. Hartnoll, C. P. Herzog and G. T. Horowitz, \emph{Building an AdS/CFT superconductor}, \emph{Phys. Rev. Lett.} {\bf 101} (2008) 031601 [arXiv:0803.3295] [hep-th]\\
S. A. Hartnoll, C. P. Herzog and G. T. Horowitz, \emph{Holographic Superconductors}, \emph{JHEP } {\bf 0812} (2008) 015 [arXiv:0810.1563] [hep-th]

\bibitem{ab}
S. A. Hartnoll, P. Kovtum, \emph{Hall conductivity from dyonic black holes}, \emph{Phys. Rev. D.} {\bf 76} (2007) 066001
[arXiv:0704.1160] [hep-th]

\bibitem{ac}
 M. M. Caldarelli, O. J. C. Dias and D. Klemm, \emph{Dyonic AdS black holes from magnetohydrodynamics}, \emph{JHEP} {\bf 0903} (2009) 025 [arXiv:0812.0801] [hep-th]
 
\bibitem{ac1}
Jeong, HS., Kim, KY., Sun, YW. \emph{Quasi-normal modes of dyonic black holes and magneto-hydrodynamics}, \emph{J. High Energ. Phys.}, {\bf 65} (2022). https://doi.org/10.1007/JHEP07(2022)065

\bibitem{ac2}
Ahn, Y.j., Baggioli, M., Huh, KB. et al. \emph{Holography and magnetohydrodynamics with dynamical gauge fields}, \emph{J. High Energ. Phys.}, {\bf 12} (2023). https://doi.org/10.1007/JHEP02(2023)012


\bibitem{ad}
S. A. Hartnoll, P. K. Kovtun, M. Muller, S. Sachdev, \emph{Theory of the Nernst effect near quantum phase transitions in condensed matter, and in dyonic black holes} , \emph{Phys. Rev. B.} {\bf 76} (2007) 144502 [arXiv:0706.3215] [cond-mat.str-el]


  
  
 \bibitem{z}
 M. Cvetič, G.W. Gibbons, D. Kubiznak and C.N. Pope, \emph{ Black hole enthalpy and an entropy inequality for the thermodynamic volume}, \emph{Phys. Rev. D} {\bf 84} (2011) 024037 [arXiv:1012.2888] [INSPIRE].
 
 \bibitem{a1}
 D. Kubiznak and R.B. Mann, \emph{Black hole chemistry}, \emph{Can. J.  Phys.} {\bf 93} (2015) 999 [arXiv:1404.2126] [INSPIRE].
 

 
 \bibitem{d1}
 J.-L. Zhang, R.-G. Cai and H. Yu, \emph{Phase transition and thermodynamical geometry for Schwarzschild AdS black hole in $AdS^5$ × $S_5$ spacetime}, \emph{JHEP} {\bf 02} (2015) 143 [arXiv:1409.5305] [INSPIRE].
 
 \bibitem{e1}
 J.-L. Zhang, R.-G. Cai and H. Yu, \emph{Phase transition and thermodynamical geometry of Reissner-Nordström-AdS black holes in extended phase space}, \emph{Phys. Rev. D} {\bf 91} (2015) 044028 [arXiv:1502.01428] [INSPIRE].
 
  \bibitem{f1}
 B.P. Dolan, \emph{Pressure and compressibility of conformal field theories from the AdS/CFT correspondence}, \emph{Entropy} {\bf 18} (2016) 169 [arXiv:1603.06279] [INSPIRE].
  \bibitem{c1}
B.P. Dolan, \emph{Bose condensation and branes}, \emph{JHEP} {\bf 10} (2014) 179 [arXiv:1406.7267] [INSPIRE].
 
  \bibitem{h1}
 J.P. Gauntlett, R.C. Myers and P.K. Townsend, \emph{Black holes of D = 5 supergravity}, \emph{Class. Quant. Grav.} {\bf 16} (1999) 1 [hep-th/9810204] [INSPIRE].
 
 \bibitem{i1}
  M.M. Caldarelli, G. Cognola and D. Klemm, \emph{Thermodynamics of Kerr-Newman-AdS black holes and conformal field theories}, \emph{Class. Quant. Grav. } {\bf 17} (2000) 3999 [hep-th/9908022] [INSPIRE].
 
  
   \bibitem{k1}
    E. Caceres, P.H. Nguyen and J.F. Pedraza, \emph{Holographic entanglement chemistry}, \emph{Phys. Rev. D } {\bf 95} (2017) 106015 [arXiv:1605.00595] [INSPIRE].
  
     \bibitem{l1}
  F. Rosso and A. Svesko, \emph{Novel aspects of the extended first law of entanglement}, \emph{JHEP} {\bf 08}
(2020) 008 [arXiv:2003.10462] [INSPIRE]

 \bibitem{m1}
M. Sinamuli and R.B. Mann, \emph{Higher order corrections to holographic black hole chemistry},
\emph{Phys. Rev. D} {\bf 96} (2017) 086008 [arXiv:1706.04259] [INSPIRE].


\bibitem{s1}
 S.S. Gubser, I.R. Klebanov and A.M. Polyakov, \emph{Gauge theory correlators from noncritical string theory}, \emph{Phys. Lett. B} {\bf 428} (1998) 105 [hep-th/9802109] [INSPIRE].

\bibitem{t1}
 E. Witten, \emph{Anti-de Sitter space and holography}, \emph{Adv. Theor. Math. Phys.} {\bf 2} (1998) 253 [hep-th/9802150] [INSPIRE].
 
 \bibitem{u1}
 I. Savonije and E.P. Verlinde, \emph{CFT and entropy on the brane}, \emph{Phys. Lett. B} {\bf 507} (2001) 305 [hep-th/0102042] [INSPIRE].
  
  \bibitem{v1}
  M. Henningson and K. Skenderis,\emph{ The holographic Weyl anomaly}, \emph{JHEP} {\bf 07} (1998) 023
[hep-th/9806087] [INSPIRE].

\bibitem{w1}
 V. Balasubramanian and P. Kraus, \emph{A Stress tensor for Anti-de Sitter gravity}, \emph{Commun. Math. Phys.} {\bf 208} (1999) 413 [hep-th/9902121] [INSPIRE].
 
 
\bibitem{e2}
 V. P. Maslov, \emph{Zeroth-order phase transitions}, \emph{Math. Notes} {\bf 76} (2004) 697.

  


 

 
\end{thebibliography}
\end{document}